\documentclass{aa}  
\usepackage{graphicx}
\usepackage{txfonts}
\usepackage{hyperref}
\usepackage[]{natbib}  
\usepackage{array}
\usepackage{slashbox}
\usepackage{colortbl}
\usepackage{multirow}
\usepackage{gensymb}
\usepackage{breqn}
\usepackage{fancyvrb}

\newcommand\rg{\textsc{rg}}
\newcommand\HR{Hertzsprung-Russell}
\newcommand\rgb{\textsc{rgb}}
\newcommand\sgb{\textsc{sgb}}
\newcommand\pms{\textsc{pms}}
\newcommand\ms{\textsc{ms}}

\newcommand\sg{\textsc{sg}}
\newcommand\agb{\textsc{agb}}
\newcommand\psd{\textsc{psd}}

\newcommand\kepler{\textsl{Kepler}}
\newcommand\numax{$\nu_\mathrm{max}$}

\newcommand\omegaenv{$\Omega_{\mathrm{env}}$}
\newcommand\omegacore{$\Omega_{\mathrm{core}}$}
\usepackage{booktabs}
\usepackage{amsmath}
\usepackage{siunitx}
\usepackage{subfig}
\newcommand{\iu}{{i\mkern1mu}}

\newcommand\bld{}

\begin{document}

\title{Magnetic signatures on mixed-mode frequencies}
   \subtitle{I. An axisymmetric fossil field inside the core of red giants.}
   \titlerunning{Magnetic signatures on mixed-mode frequencies I-Axisymmetric fossil field}

   \author{L.~Bugnet
          \inst{1,2}
          \and V.~Prat\inst{1}
          \and S.~Mathis\inst{1}
          \and A.~Astoul\inst{1}
          \and K.~Augustson\inst{1}
          \and R.~A.~Garc\'\i a\inst{1}
          \and S.~Mathur\inst{3,}\inst{4}
          \and L.~Amard\inst{5}\inst{} 
          \and C.~Neiner\inst{6} 
          }

\institute{AIM, CEA, CNRS, Université Paris-Saclay, Université Paris Diderot, Sorbonne Paris Cité, F-91191 Gif-sur-Yvette, France
\and {Flatiron Institute, Simons Foundation, 162 Fifth Ave, New York, NY 10010, USA},\\ \email{lbugnet@flatironinstitute.org}
\and
Instituto de Astrof\'{\i}sica de Canarias, E-38200, La Laguna, Tenerife, Spain
\and 
Universidad de La Laguna, Dpto. de Astrof\'{\i}sica, E-38205, La Laguna, Tenerife, Spain
\and University of Exeter, Department of Physics and Astronomy, Stoker Road, Devon, Exeter, EX4 4GL,  UK
\and LESIA, Observatoire de Paris, Université PSL, CNRS, Sorbonne Université, Université de Paris, 5 place Jules Janssen, F-92195 Meudon, France}

\authorrunning{L. Bugnet}
   \date{Received / Accepted}









\abstract{The discovery of the moderate differential rotation between the core and the envelope of evolved solar-like stars could be the signature of a strong magnetic field trapped inside the radiative interior. The population of intermediate-mass red giants presenting a surprisingly low-amplitude of their mixed modes (i.e. oscillation modes that behave as acoustic modes in their external envelope and as gravity modes in their core) could also arise from the effect of an internal magnetic field. Indeed, stars more massive than about 1.1 solar masses are known to develop a convective core during their main sequence. The field generated by the dynamo triggered by this convection could be the progenitor of a strong fossil magnetic field trapped inside the core of the star for the rest of its evolution.}{The observations of mixed modes can constitute an excellent probe of the deepest layers of evolved solar-like stars. Thus, magnetic fields in those regions can impact their propagation. The magnetic perturbation on mixed modes may thus be visible in asteroseismic data.
To unravel which constraints can be obtained from observations, we theoretically investigate the effects of a plausible mixed axisymmetric magnetic field with various amplitudes on the mixed-mode frequencies of evolved solar-like stars.}{The first-order frequency perturbations due to 
an axisymmetric magnetic field are computed for dipolar and quadrupolar mixed modes. These computations are carried out for a range of stellar ages, masses, and metallicities.}{We show that typical fossil-field strengths of $0.1-1$ MG, consistent with the presence of a dynamo in the convective core during the main sequence, provoke significant asymmetries on mixed-mode frequency multiplets during the red-giant branch. We provide constraints and methods for the detectability of such magnetic signatures. We show that these signatures may be detectable in asteroseismic data for field amplitudes small enough for the amplitude of the modes not to be affected by the conversion of gravity into Alfvén waves inside the magnetised interior. Finally, we infer an upper limit for the strength of the field, and the associated lower limit for the timescale of its action, to redistribute angular momentum in stellar interiors.}{}



  \maketitle

\section{Introduction}

    Oscillations on the surface of solar-type stars have been observed and studied across the Hertzsprung-Russel diagram thanks in large part to the data provided by CoRoT, \textsl{Kepler}, K2, and the TESS missions  \citep[e.g.][]{Michel2008, Chaplin2010, Lund2017, Huber2019a,Garcia2019, Chaplin2020}. While this list is hardly exhaustive, some key results are relevant to this paper and to the importance of the transport of angular momentum in the radiative regions of main-sequence (\ms) stars. Indeed, the radiative interior of the Sun seems to rotate as a solid body \bld{slightly slower than the equatorial surface rotation rate \citep{Garcia2007}} until 0.25 solar radius \citep[R$_\odot$, \bld{e.g.}][]{Thompson2003a,Couvidat2003}, and the nuclear core may rotate even faster \citep[][]{Garcia2007}. The study of some solar-like stars also shows that they present a nearly solid-body rotation \citep{Benomar2015a}. Likewise, in subgiant stars (\sg s) and red-giant stars (\rg s), some \bld{relevant} work are the discoveries of the unexpectedly slow rotation rate of their cores \citep[e.g.][]{Deheuvels2012a, Deheuvels2014a, Deheuvels2016, Mosser2012,Mosser2017, Gehan2018}, and the surprisingly low amplitude of dipolar mixed modes in some \rg s  \citep[][]{Garcia2014, Mosser2012, Mosser2017, Stello2016a}. 
    
    
    As of yet, there is no clear evolutionary model that yields internal rotation profiles akin to those observed \citep[e.g.][for the loss of angular momentum on the giant branch]{Eggenberger2012abis, Eggenberger2017bis, Eggenberger2019b, Ceillier2013, Marques2013a}, nor a robust explanation for the observed dipole mode amplitude suppression \citep{Fuller2015,Cantiello2016, Lecoanet2017b, Mosser2017, Loi2018}.
    
    Some magneto-hydrodynamic fluid behaviors can strongly impact the rotation profile of the star both on secular time scales \citep{Eggenberger2005, Cantiello2014a, fuller2014, fuller2019a} and even on their dynamical time scales \citep{Brun2005, Featherstone2009, Augustson2016}. Moreover, the transport of chemical species, energy, and angular momentum by internal waves can also play a role \citep[e.g.][]{Belkacem2015, Pincon2017}.
    However, none of these cited solutions totally explains the angular momentum transport from the subgiant stage until the end of the red-giant branch (\rgb). This raises the issue of the incompleteness or inaccuracy of the included set of physical processes driving the internal dynamics in stellar models at each step of the evolution. Magnetic fields are not considered for instance inside the radiative interior of solar-type stars along their evolution in models \citep[apart from a few studies that do include the effect of the Tayler-Spruit dynamo, e.g.][]{Cantiello2014a}, while we do know that at least weak fields must be present, resulting from the relaxation of past dynamo events \citep[e.g.][]{Braithwaite2004, Braithwaite2008a,Duez2010}. Such internal magnetism may prevent the differential rotation inside the radiative interior, a configuration that has been observed inside solar-like stars and the Sun \citep{Garcia2007, Benomar2015a, Fossat2017a}.

    The internal structure of \sg s and \rg s allows acoustic and gravity modes to couple to form mixed modes. As they probe the deepest layers of the star, they are of great interest for the understanding of physical processes taking place from the deepest layers of the radiative interior towards the surface of the star. For instance, they are known to provide estimates of mean rotation rate of the core of \sg s and \rg s \citep[e.g.][]{Deheuvels2012a, Deheuvels2014a, Deheuvels2016, Mosser2012, Mosser2017, Gehan2018}. \cite{Fuller2015}, \cite{Lecoanet2017b}, and \cite{loi2017} suggested via different magnetic conversion mechanisms that the presence of a strong magnetic field inside the core of a RG can convert magneto-gravity waves into Alfvén waves. It results in a loss of energy of the observed modes, which are no longer mixed modes and only present an acoustic signature. Their amplitude is thus diminished in the power spectrum density (PSD). Even if these theories are controversial \citep[e.g.][]{Mosser2017}, they point out the potentially large effect of magnetic fields on mixed-mode amplitudes. The impact of buried pure toroidal or poloidal magnetic fields on acoustic mode frequencies was studied in the context of the SOlar and Heliospheric Observatory \citep[SOHO,][]{Domingo1995} mission (Solar Oscillation Imager/Michelson Doppler Imager \citep[SOI/MDI,][]{Scherrer1995} and the Global Oscillations at Low Frequencies \citep[GOLF,][]{Gabriel1995} instruments) for the observation of the Sun \citep[e.g.][]{gough1984, Gough1990a, dziembowski1985, dziembowski1989, Takata1994, Kiefer2018} but no trace of such fields has been found in the solar acoustic data. \cite{Rashba2007} developed the equivalent theory for the Sun's $g$ modes, whose detection can still be controversial \citep[][]{Garcia2007, Appourchaux2010, Fossat2017a, Fossat2018,Schunker2018, Appourchaux2018, Scherrer2019}. More recent studies \citep[e.g.][]{Hasan2005a, prat2019, VanBeeck2020} focused on the impact of stable fields (with a magnetic configuration given by \cite{Duez2010} that consists in mixed poloidal and toroidal fields) on slowly pulsating B and $\gamma$-Doradus stars showing pure gravity modes. The effect of moderate amplitude magnetic fields on mixed-mode frequencies of evolved stars has been theoretically studied by \cite{Loi2020}, in the case of a non-rotating star. That study provides analytical (non-perturbative) expressions of the impact of magnetic fields on mixed mode frequencies that are complementary to the perturbative analysis presented in this paper.
\\


    Led by all these previous theoretical studies, we focus our understanding on the core dynamics of \sg s and \rg s. We investigate the impact of a realistic axisymmetric fossil magnetic field buried inside the core of evolved \bld{low-mass ($M_\star\lesssim 1.3$ solar masses, $\mathrm{M}_\odot$) and intermediate-mass ($1.3 \mathrm{M}_\odot \lesssim M_\star \lesssim 7 \mathrm{M}_\odot$) solar-like pulsators (i.e. subgiant and red giant stars)} on their observable mixed-mode frequencies. In our study, as opposed to the work by \cite{Loi2020}, we consider the magnetic field amplitude to be small enough for its effects on the mixed-mode frequencies to play as a first-order perturbation, along with the first order effects resulting from the slow differential rotation of the star \citep[e.g.][]{Deheuvels2014a, Gehan2018}. This approximation is motivated by the non-discovery to date of any magnetic signatures on the mixed-mode frequency pattern of observed \rg s. 
    
    For the magnetic field, we use a realistic mixed poloidal and toroidal configuration \citep{Braithwaite2004, Braithwaite2006a, duez2010a, duez2010b}, and evaluate its impact on typical mixed modes computed by using the stellar evolution code MESA \citep{Paxton2011} and stellar oscillation code GYRE \citep{townsend2013}. After describing the magnetic-field configuration and its potential origin and evolution in Sect.~\ref{sec:field}, we provide in Sect.~\ref{sec:firstorder} the first-order perturbative analysis leading to the magnetic and rotational shift of mixed-mode frequencies. Section~\ref{sec:typical_RG} investigates the shifting of the $\ell=1$ and $\ell=2$ mixed mode frequencies in the case of a $M_\star=1.5\mathrm{M_\odot}$, $Z=0.02$ star along its evolution on the \rgb. This star is massive enough to develop a convective core during the main sequence, which can lead to the intense production of magnetic energy. We provide values of the critical field strength associated to this axisymmetric topology above which the effect on mode frequencies should be visible in real data, and conclude on the validity of the perturbative approach depending on the magnetic-field strength and the evolutionary stage of the star. In Sect.~\ref{sec:SG}, we then follow the same approach as we did for the \rg{} branch for the \sg{} stage. During the \sg{} branch (\sgb{}), the nature of most mixed modes is transitioning from acoustic- toward gravity-dominated modes. Section~\ref{sec:AM} discusses the consequences of the presence of fossil magnetic fields inside evolved solar-like stars on angular momentum transport. In Sect.~\ref{sec:dependency} we investigate the stellar mass and metallicity dependence of the magnetic splitting of mixed modes.  Section~\ref{sec:depressed} focuses on comparing the magnetic field amplitude needed for $g$-mode conversion into Alfvén modes with those needed for magnetic splitting to be detectable in the data. Finally, we conclude on the large potential of this approach for the future detection of magnetic fields from the inversion of magnetic-frequency splitting from real data.
\\

\begin{figure}[t]
    \centering
    \includegraphics[width=0.49\textwidth]{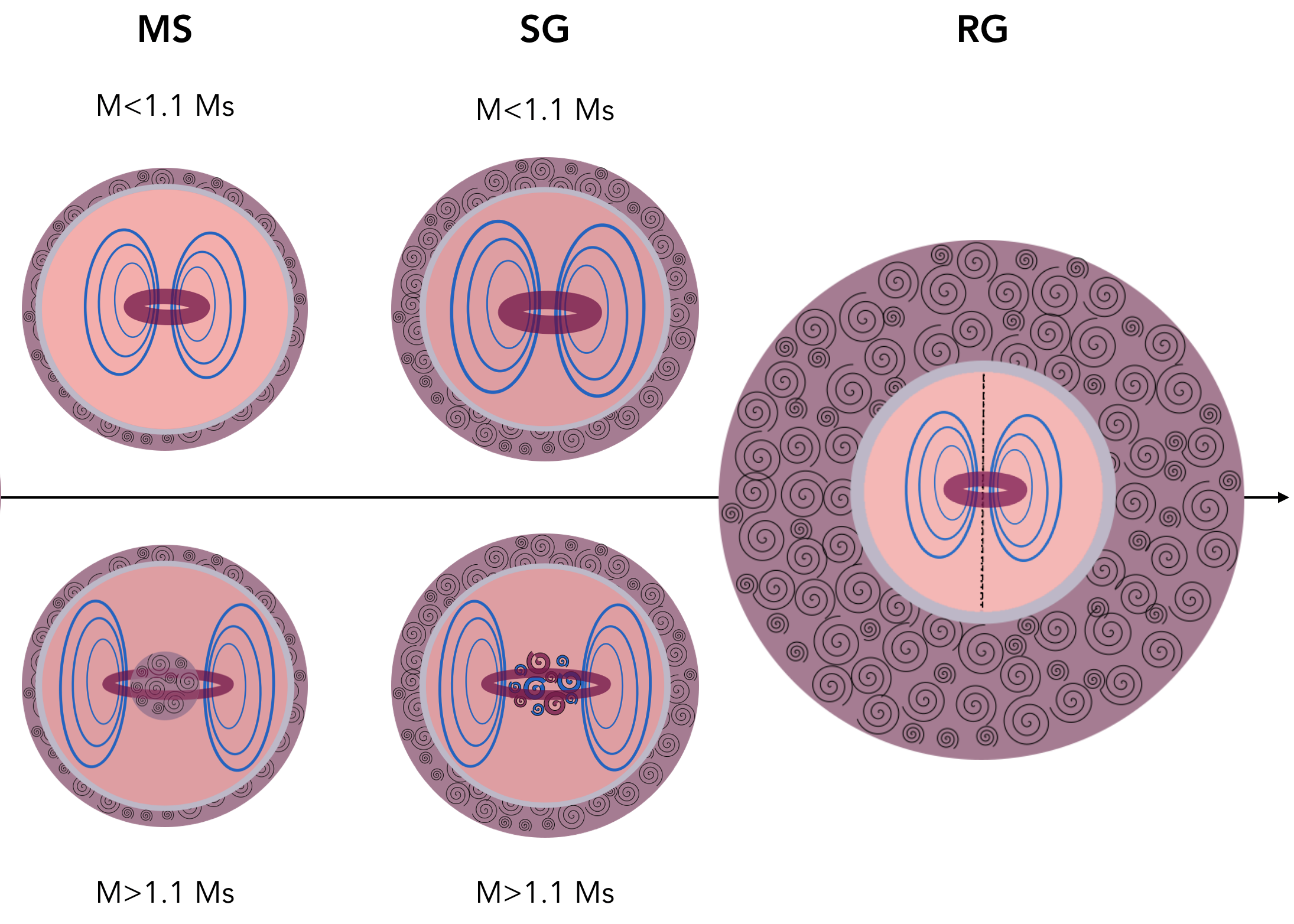}
    \caption{Magnetic configuration schemes (not to scale) following the evolution of low- and intermediate-mass stars from the main sequence. As the star evolves on the \ms, its radiative interior possesses a relaxed fossil field resulting from the convective dynamo taking place during the \pms. On the \ms, either the star is of low mass ($M_\star\leq 1.1\mathrm{M_\odot}$) and the convective region simply reduces to a surface layer, or for ($M_\star\geq1.1\mathrm{M_\odot}$) the convective region reduces to a surface layer while a convective core is formed. During the \sg{} phase the interior is completely radiative, and a fossil field is formed, also present during the \rgb. The field is located inside the radiative region of the star, for which the extent varies with the evolution. The represented configuration corresponds to the field topology considered in our study. It presents both a poloidal (blue lines) and a toroidal (red lines) component, with the axis of symmetry aligned with the rotation axis of the star.}
    \label{fig:axisymmetric}
\end{figure}
\section{Internal magnetic fields along the evolution}
\label{sec:field}

Magnetism in the depths of stars is very difficult to probe. Spectropolarimetry, which provides most measurements of stellar magnetic fields through the Zeeman effect, only provides information on magnetism emerging from the external layers of the star \citep[e.g.][]{Donati1997}. Probing internal magnetism in evolved solar-like stars could be game changing, for magnetism is known to enhance angular momentum transport \citep{Mestel1987, Charbonneau1993, Gough1998, Spruit1999, spruit2002, Mathis2005, Fuller2014a, fuller2019a,  Eggenberger2020a}. Thus, strong magnetic fields within subgiants and red giants could potentially explain their low core-to-envelope rotation-rate ratio.

One powerful mechanism strengthening and sustaining magnetic fields is a convective dynamo, which can create strong magnetism from a weak initial field \cite[e.g.][for the Sun's dynamo]{Dikpati1999}. Such mechanism is likely to be the origin of the solar surface magnetic cycle \citep{Brun2004, Brun2017a}, and is probably active in any convective rotating region inside stars \citep[e.g.][]{Brun2005, Browning2008, Brown2010, Brown2011, Augustson2012, Augustson2015, Augustson2016}. However, internal layers of low-mass evolved stars are radiative during the subgiant and red-giant stages: no convective dynamo can take place in their depths. For a magnetic field to be present inside the radiative interior of evolved solar-like stars, either a dynamo-originated stochastic field has been preserved inside the radiative cavity since the last dynamo episode (fossil field scenario), or the field is transient and currently generated by magnetohydrodynamics instabilities \citep{spruit2002, Zahn2007,fuller2019a}. \cite{spruit2002} and \cite{fuller2019a} propose a solution to generate magnetic energy from the combination of the Tayler instability \citep{Tayler1973} and radial differential rotation. This rotational instability may result in large-scale internal magnetic fields, but is out of the scope of this paper. We focus here on the so-called fossil field scenario, resulting from the relaxation of magnetic fields \bld{originating} from a past convective episode, as described in \cite{Braithwaite2008a}, \cite{Duez2010}, and \cite{Mathis2010b}. In the case of giant stars, internal fields may form from two previous convective dynamo episodes, depending on the mass of the star.
During the pre-\ms, the star is fully convective: a similar dynamo process as the solar dynamo could take place inside the whole star. When the convective region reduces to a thin surface layer on the \ms, the dynamo-originated magnetism may start to relax inside the radiative interior \citep{Arlt2013, Emeriau-Viard2017, Villebrun2019}. Main-sequence stars with $M_\star \gtrsim 1.1 \mathrm{M_\odot}$ additionally develop a convective core, due to the change in the hydrogen fusion mechanisms. Such a core likely also convectively generates a magnetic field via a dynamo (see Fig.~\ref{fig:axisymmetric}). This field could thus enhance the potentially already present relaxed magnetic field inside the radiative interior of the star.

\subsection{The fossil field scenario}

  In the fossil field scenario, the relaxing magnetic field may be preserved inside the radiative interior when the convection ends. If no disruptive processes occur inside the radiative interior from the \ms{} towards the \rgb, such as strong differential rotation \citep{Auriere2014, Auriere2015,Gaurat2015}, the stochastic field will eventually stabilise into a fossil equilibrium configuration, reached when the Lorentz force balances hydrodynamics forces \citep{Chandrasekhar1958}. In that case, and by assuming a constant magnetic diffusivity, the induction equation is written as: 

\begin{equation}
    \frac{\partial \boldsymbol{B}}{\partial t} = - \eta \boldsymbol{\nabla} \wedge \left( \boldsymbol{\nabla} \wedge \boldsymbol{B} \right),
    \label{eq:diffusivity}
\end{equation}
with $\eta$ the magnetic diffusivity. If there is no turbulence in the region, the associated Ohmic relaxation time due to atomic processes is $\tau=R^2/\eta$.  For a "fossil field", this is comparable to the lifetime of the star itself, being $10^{10}$ years in the case of a field inside the radiative interior of the Sun \citep[e.g.,][]{Cowling1945}. Thus, if ever generated and never disrupted by hydrodynamical or other MHD processes, the dynamo-originated field should remain trapped for the rest of the star's evolution as a fossil field, and still be present in its radiative interior during the RGB \citep{stello2016d}. The intense magnetism discovered in some white dwarfs \citep{angel1981, Putnet1999, schmidt2001} could then result from such a \pms-\ms\  dynamo field, surviving as a fossil field during the succeeding evolutionary stages. The primary challenges regarding these equilibrium magnetic fields are the determination of their 3D configuration and the estimation of their amplitude. 

\begin{figure*}[t]
    \centering
    \includegraphics[width=1\textwidth]{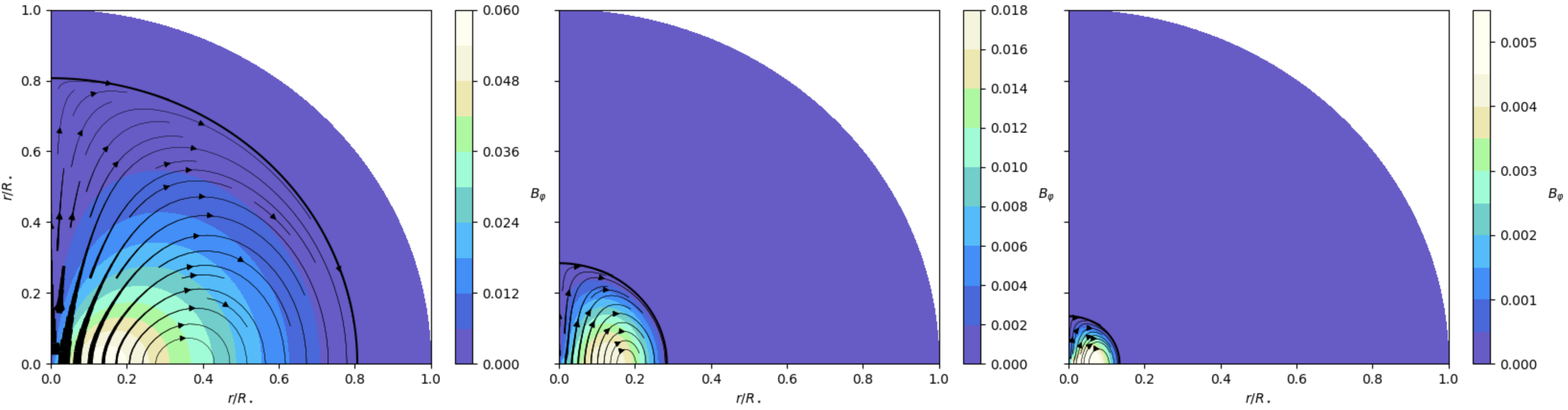}
    \caption{Mixed  poloidal (black lines) and toroidal (color scale) magnetic field modelled using the formalism of \cite{Duez2010}, see Eq.~\eqref{eq:compu}. Values are normalised by the maximum radial field amplitude. The field is confined inside the radiative region of the star, for which the extent is given by the corresponding evolution model MESA \bld{and indicated by the black circle.} Left to right: cases of a 1.5 M$_\odot$, Z=0.02 star from the \sg{} phase to \rgb{}, with ages of resp. $2.65$, $2.75$, and $2.80$ Gy, and effective temperatures of resp. $5967$, $5075$, $4900$ K .}
    \label{fig:field}
\end{figure*}

\subsection{Fossil field topology}
\label{sec:field_topo}
\interfootnotelinepenalty=10000
Purely toroidal and purely poloidal magnetic configurations are known to be unstable \citep[e.g.][]{Tayler1973, Markey1973, Braithwaite2006, Braithwaite2007a}. The stability of mixed configurations with both poloidal and toroidal components is demonstrated in \citet{Tayler1980}, and the relative energy contained in the toroidal and poloidal components for the field to be stable has been evaluated numerically \citep{Braithwaite2008a} and semi-analytically \citep{Akgun2013}. In a similar vein, \cite{Braithwaite2004} simulated the relaxation of stochastic fields in a non-rotating radiative medium. Those simulations show that a stochastic field representing a dynamo-generated field inside a stably stratified region relaxes into a stable, larger-scale, mixed poloidal and toroidal magnetic field. \cite{Duez2010} give the closest semi-analytic description of such stable mixed toroidal and poloidal fossil fields. As opposed to previous studies \citep[e.g.][]{Gough1990a} that considered purely toroidal or purely poloidal field topologies, we use this stable mixed formalism to investigate the effect on mixed-mode frequencies of a hypothetical fossil field trapped inside the radiative interior of size $R_{\mathrm{rad}}$ of evolved stars (see Fig.~\ref{fig:axisymmetric}). We do not consider in our study non-perturbative boundary conditions associated to more realistic magnetic field configurations \citep[e.g.][]{Roberts1983, Campbell1986, Dziembowski1996, Bigot2000}. The mixed toroidal and poloidal expression from \cite{Duez2010} that minimizes the energy of the system is dipolar. Such magnetism confined inside the radiative interior of the star is written as:


\begin{equation}
    \boldsymbol{B}=\begin{cases}
    \displaystyle
    \frac{1}{r \sin{\theta}}\left( \boldsymbol{\nabla} \psi(r,\theta) \wedge \boldsymbol{e_\varphi} + \lambda \frac{\psi(r,\theta)}{R_{\mathrm{rad}}} \boldsymbol{e_\varphi}\right) \mathrm{\,if\,} r<R_{\mathrm{rad}}\, ,\\
    \displaystyle
    0 \mathrm{\,if\,} r>R_{\mathrm{rad}}
    \end{cases}
    \label{eq:compu}
\end{equation}

\noindent where $\psi$ is the stream function satisfying 
\begin{equation}
    \psi(r,\theta)=\mu_0 \alpha \lambda \frac{A(r)}{R_{\mathrm{rad}}}\sin^2{\theta}\, ,
\end{equation} with $\mu_0$ the vacuum magnetic permeability expressed as $4\pi$ in cgs units, $\alpha$ a normalisation constant fixed by the chosen magnetic-field amplitude, $\lambda$ the eigenvalue of the problem to be determined, $R_{\mathrm{rad}}$ the radius of the radiative interior, and 
\begin{multline}
A(r)=-r j_1\left(\lambda \frac{r}{R_{\mathrm{rad}}} \right) \int_r^{R_{\mathrm{rad}}} y_1\left(\lambda \frac{x}{R_{\mathrm{rad}}} \right)\rho x^3 \textrm{d}x\\
-r y_1\left(\lambda \frac{r}{R_{\mathrm{rad}}} \right) \int_0^r j_1\left(\lambda \frac{x}{R_{\mathrm{rad}}} \right)\rho x^3 \textrm{d}x,
\label{eq:A}
\end{multline}
\noindent with $j_1$ (resp. $y_1$) the first-order spherical Bessel function of the first (resp. second) kind \citep{Abramowitz1972} and $\rho$=$\rho(r)$ the density of the star.

In our study, we consider the field to be aligned with the rotation axis of the star; hence, non-aligned and non-fossil field configurations are out of the scope of this paper. Finally, the axisymmetric poloidal and toroidal magnetic field used in our analytical calculations can be expressed by 
\begin{equation}
    \boldsymbol{B}=B_0\left[ b_r(r) \cos{\theta}, b_\theta(r) \sin{\theta}, b_\phi(r) \sin{\theta}\right],
    \label{ana}
\end{equation}
\noindent with the following correspondences to match the \cite{Duez2010} formalism:
\begin{align}
    B_0&=\frac{\mu_0 \alpha \lambda}{R_{\mathrm{rad}}}\\
    b_r(r)&=\frac{2A}{r^2} \label{eq:br}\\ 
    b_\theta(r)&=-\frac{A'}{r} \label{eq:btheta}\\ 
    b_\varphi(r)&=\frac{\lambda A}{r R_{\mathrm{rad}}}, \label{eq:bphi}
\end{align}
\noindent with the prime symbol denoting the radial derivative.

 In order to confine the field inside the radiative area we set $\lambda$ as the smallest positive constant for $\boldsymbol{B}$ to vanish at the radiative/convective boundary (see Fig.~\ref{fig:field}). As opposed to massive stars, red giants present a strong internal density gradient. This property prevents $\int_0^{R_{\mathrm{rad}}} j_1\left(\lambda \frac{x}{R_{\mathrm{rad}}} \right)\rho x^3 \textrm{d}x$ to go to zero for any value of $\lambda$. As a result, we could not simultaneously cancel the $b_r$ and $b_\theta$ components of the field at the radiative/convective boundary. \bld{To ensure that the field is confined into the radiative interior}, we choose to eliminate only $b_r$. It leads to \bld{an azimuthal current sheet that might create instabilities \citep[see][for a detailed description]{Duez2010}.}  This is accomplished by searching for zeros of $y_1(\lambda)$ instead \bld{(a more detailed description is given in Appendix~\ref{appendix:mag_topo}).} We find that the corresponding smallest value for $\lambda$ is about $2.80$. This eigenvalue corresponds to the first zero of the function $y_1$, providing the most stable parametrisation that eliminates $b_r$ at $R_{\mathrm{rad}}$. This rather small value of $\lambda$ leads to poloidally-dominated fields as represented in Fig.~\ref{fig:field}.

\subsection{Estimating the evolving magnetic field amplitude}
\label{sec:B_amp}

By considering an ideal fossil-field scenario, one can estimate the amplitude of the field along the evolution of the star by considering magnetic-flux conservation from the end of the more recent dynamo episode, without considering any Ohmic loss of energy due to the reconnection of field lines during successive relaxations. \bld{Therefore, we here adopt the approach of giving an upper-limit of the field amplitude by first computing the rate of magnetic energy produced by stellar dynamos and transmitting its amplitude to the fossil field, a simple approach that has been currently used in studies in asteroseismology \citep[e.g.][]{Fuller2015,Cantiello2016}. However, we know that during the relaxation of fossil fields, loss of magnetic energy occurs at small scales. This has been observed in numerical simulations \citep[e.g.][]{Braithwaite2008a,Duez2011,Emeriau-Viard2017}. How to properly quantify this Ohmic loss of energy and the corresponding turbulent scales all along the evolution of stars is an open and difficult question because a coherent theoretical modeling should be provided \citep[e.g.][]{Moffat2015,Hotta2017} while our current MHD simulations assume magnetic diffusivities (and diffusivities in general) that are in general higher than in stellar interiors. This question will be examined in detail in a forthcoming separate article.}

We consider the extent of the fossil field configuration to match the size of the radiative interior of the star during the \sg{} and \rg{} phases (see Fig.~\ref{fig:axisymmetric}). It implies that no reconnections with the envelope field are considered in our simplified configuration. \bld{In our fossil field scenario,} the last recorded internal dynamo is either the convective core dynamo for stars more massive than $1.1\mathrm{M_\odot}$ \bld{as presented on Fig.~\ref{fig:axisymmetric}} or from the global convective dynamo during the pre-main sequence (PMS) for low-mass stars. 

As in \cite{Fuller2015} and \cite{Cantiello2016}, we estimate the magnetic amplitude resulting from magnetic flux conservation from the end of the last convective dynamo episode, towards the current giant stage, through:

\begin{equation}
    B_{0\star}\simeq B_{\mathrm{0,init}}\frac{R^2_{\mathrm{init}}}{R^2_{\mathrm{rad, \star}}},
    \label{eq:flux_conservation}
\end{equation}
\noindent with $R_{\mathrm{init}}$ the radius of the convective cavity of the last internal dynamo phase, and $R_{\mathrm{rad, \star}}$ the size of the current radiative sphere inside which the field is supposed to be confined. This flux conservation equation ensures that the whole magnetic energy is conserved inside the radiative interior, with no exchanges with the surrounding convective zone, and with no disruption of the magnetic field by an external mechanism such as differential rotation. $R_{\mathrm{init}}$ and $R_{\mathrm{rad, \star}}$ are evaluated along the evolution by using the MESA stellar evolution code \citep{Paxton2011}. The $B_{\mathrm{0,init}}$ field amplitude is estimated during the recession phase of the last internal dynamo by the use of rotating stellar models computed with STAREVOL \cite[STAREVOL is used instead of MESA to take into account rotation during the early stages of the evolution of the star needed in order to estimate the magnetic field amplitude resulting from magnetostrophy and buoyancy regimes, see][for details]{Amard2019, Astoul2019}. The field is evaluated following \cite{Augustson2019} and \cite{Astoul2019}, by considering the magnetostrophic, equipartition, and buoyancy driven regimes (see Appendix~\ref{sec:appendix_magnetostrophy} for details about how the different regimes are constructed). The turbulent equipartition regime corresponds to the state where the convective kinetic energy density of the fluid is fully converted into a magnetic energy density (superequipartition regime thus refers to the fluid's magnetic energy greater than its kinetic energy). An equipartition or subequipartition regime is invoked for the Sun's dynamo inside the convective envelope \citep{Brun2017}. The magnetostrophic regime is reached when Coriolis acceleration balances Lorentz force in the momentum equation. It corresponds to a maximum amplitude estimate of the magnetic field, and can be a superequipartition regime. It is suspected by \cite{Augustson2019} to be applicable for stellar dynamos. The buoyancy dynamo regime corresponds to the case where Coriolis acceleration, buoyancy, and Lorentz forces all have the same order of magnitude. This regime is usually considered for rapidly-rotating low-mass stars and planets with strong density gradient \citep{Christensen2009}.

\subsubsection{Case of $M_\star\lesssim 1.1 \mathrm{M_\odot}$}

\begin{figure}[t]
    \centering
    \includegraphics[width=0.49\textwidth]{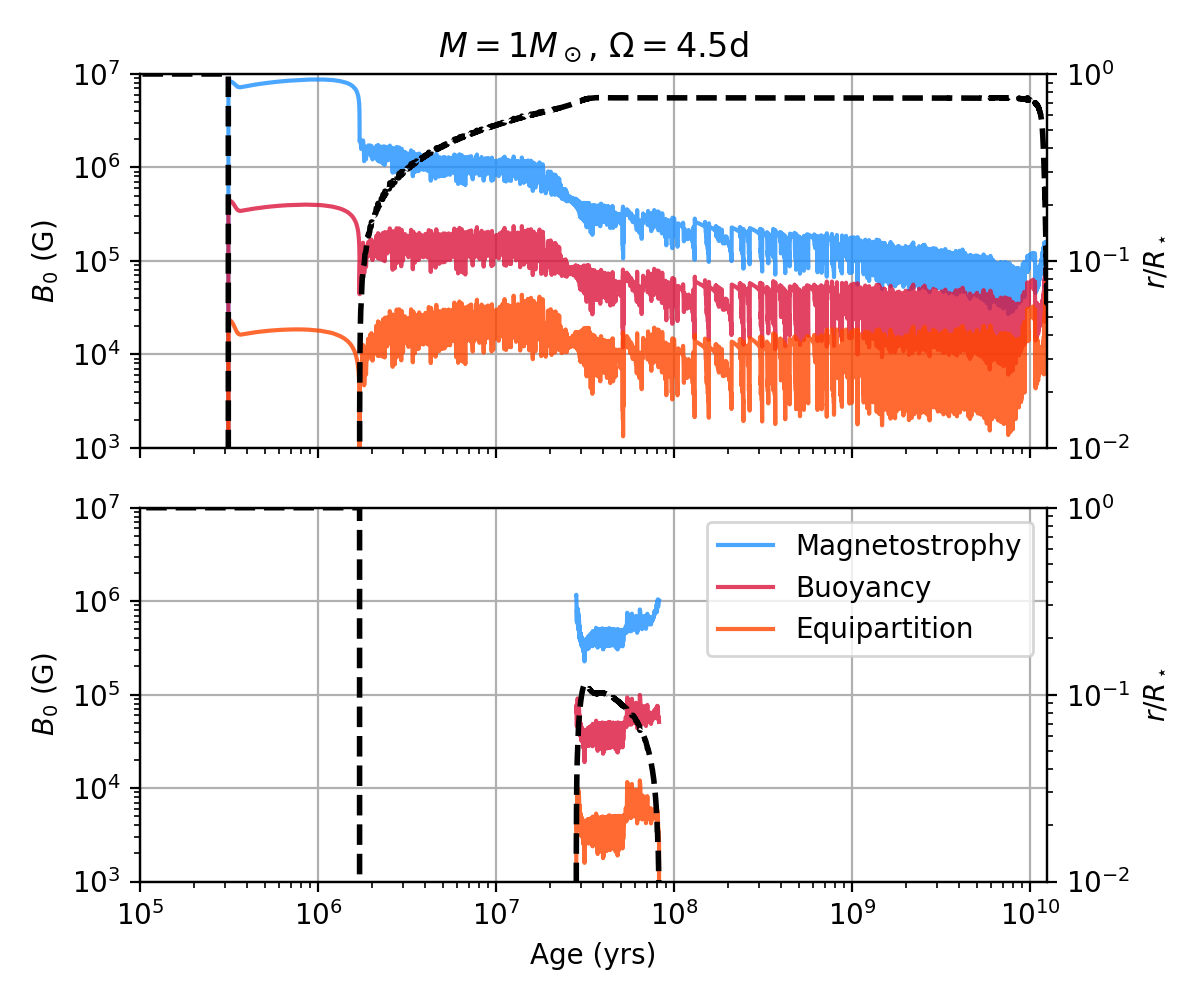}
    \caption{\textsl{Top:} Magnetic field amplitude at the base of the convective envelope along the evolution of a $M_\star=1\mathrm{M_\odot}$ star, with solar metallicity (Z=0.0134), uniformly rotating with an initial rotation rate during the disk-locking of the \pms{} of $\Omega=2.5\,$\si{\micro\hertz}. The blue, red, and orange lines respectively represent the field estimated by considering the magnetostrophic, buoyancy, and equipartition regimes. The black dashed line represents the location of the limit radius between the internal radiative layer and the convective envelope. \textsl{Bottom:} Same as top panel at the top of the convective interior along the evolution of the star. The black dashed line represents the location of the limit radius between the convective core (when existing) and the radiative layer.}
    \label{fig:1p0_aurel}
\end{figure}

Low-mass stars do not develop a convective core during the main-sequence, apart from a small convective interior that arises from the energy released during the CNO cycle before the temperature drops (down to $\sim5600$K) leading to the p-p fusion mechanism. Such stars however possess a convective envelope, schematically represented on Fig.~\ref{fig:axisymmetric}. Top panel of Fig.~\ref{fig:1p0_aurel} represents the amplitude of the magnetic field generated at the bottom of the convective envelope while the star evolves. 
When the radiative interior replaces the convection at the beginning of the main sequence ($\sim 2\times 10^7$ years), the amplitude of the expected magnetic field that relaxes into the radiative interior at the boundary with the convective envelope is of the order of magnitude of respectively $10^6$, $10^5$, and $10^4$ Gauss for each of the magnetostrophic, buoyancy, and equipartition regimes.  \cite{Astoul2019} concluded that in order to reproduce the magnetic amplitude observed at the surface of rotating \pms{} and \ms{} stars, the magnetostrophic regime should be considered. For the remainder of this paper we thus use the amplitude of the dynamo field estimated by considering the magnetostrophic regime as an upper estimate of the magnetic amplitude inside the radiative interior. The equipartition approximation that is used by \cite{Cantiello2016} and \cite{Fuller2015} provides much smaller estimates of the field amplitude, as seen on Fig.~\ref{fig:1p0_aurel}: we consider these estimates as the lowest possible value of the fossil field in the remainder of the article.

We shall also discuss the quick appearance of a convective core at the beginning of the main sequence. The characteristic timescale for a dynamo to emerge from convection and differential rotation is of the order of a year (by considering a typical solar internal rotation rate). Therefore, despite its comparatively short lifespan, this early convective core should still produce a dynamo field. Its amplitude would be of the order of magnitude represented on the bottom panel of Fig.~\ref{fig:1p0_aurel}. This newly generated field has an amplitude equivalent to the one of the original field resulting from the dynamo in the convective envelope during the \pms: the presence of this convective core right after the end of the \pms{} should not modify the amplitude of the field already relaxing. We ignore by this statement the interaction between the dynamo field and the surrounding fossil field as shown by \cite{Featherstone2009} in the case of A-type stars. Such interaction may enhance the dynamo process, and may lead to stronger core field amplitudes. For low-mass stars with no persistent convective core during the \ms, we thus consider the relaxation and stabilisation of the field to occur at the end of the \pms{} inside the newly formed radiative interior, ignoring the early convective core, where $B_{\mathrm{init}}=B_{\mathrm{PMS}}$ with values ranging from $10^4$ to $10^6$ Gauss, depending on the chosen dynamo force balance regime.

\subsubsection{Case of $M_\star\gtrsim 1.1 \mathrm{M_\odot}$}
\label{sec:rotation}
If the star is massive enough for a sustainable convective core to form during the MS, we then consider this convective core to be the location of the last internal dynamo episode from which the fossil field relaxed. In that case, $R_{\mathrm{init}}$ is equal to the radius corresponding to the maximum extend of the convective core during the \ms{} ($R_{\mathrm{MS}}$), that is estimated with evolution models from MESA (see appendix~\ref{sec:MESA_inlist}). Figure~\ref{fig:1p4_aurel} represents the same diagrams as Fig.~\ref{fig:1p0_aurel} for a $M_\star=1.4\mathrm{M_\odot}$ star. The top panel indicates the relaxation of the \pms{} magnetic field with the same order of magnitude than the $B_{\mathrm{PMS}}$ field estimated for low-mass stars represented on Fig.~\ref{fig:1p0_aurel}. In the bottom panel of Fig.~\ref{fig:1p4_aurel} we observe the formation of the convective core around $10$ My, that lasts until the end of the \ms. The corresponding field amplitudes are of the same order of magnitude as the different fields estimated from the relaxation of the magnetic field resulting from the \pms{} dynamo (see top panel of Fig.~\ref{fig:1p4_aurel}). In that case, we consider the relaxation to a fossil configuration of the field to occur inside the radiative interior at the end of the \ms: $B_{\mathrm{init}}=B_{\mathrm{MS}}$, with values also ranging from $10^4$ to $10^6$ Gauss. This new relaxing field eventually couples with the surrounding relaxed magnetic field from the \pms\  that has an amplitude of the same order of magnitude.

\begin{figure}[h]
    \centering
    \includegraphics[width=0.49\textwidth]{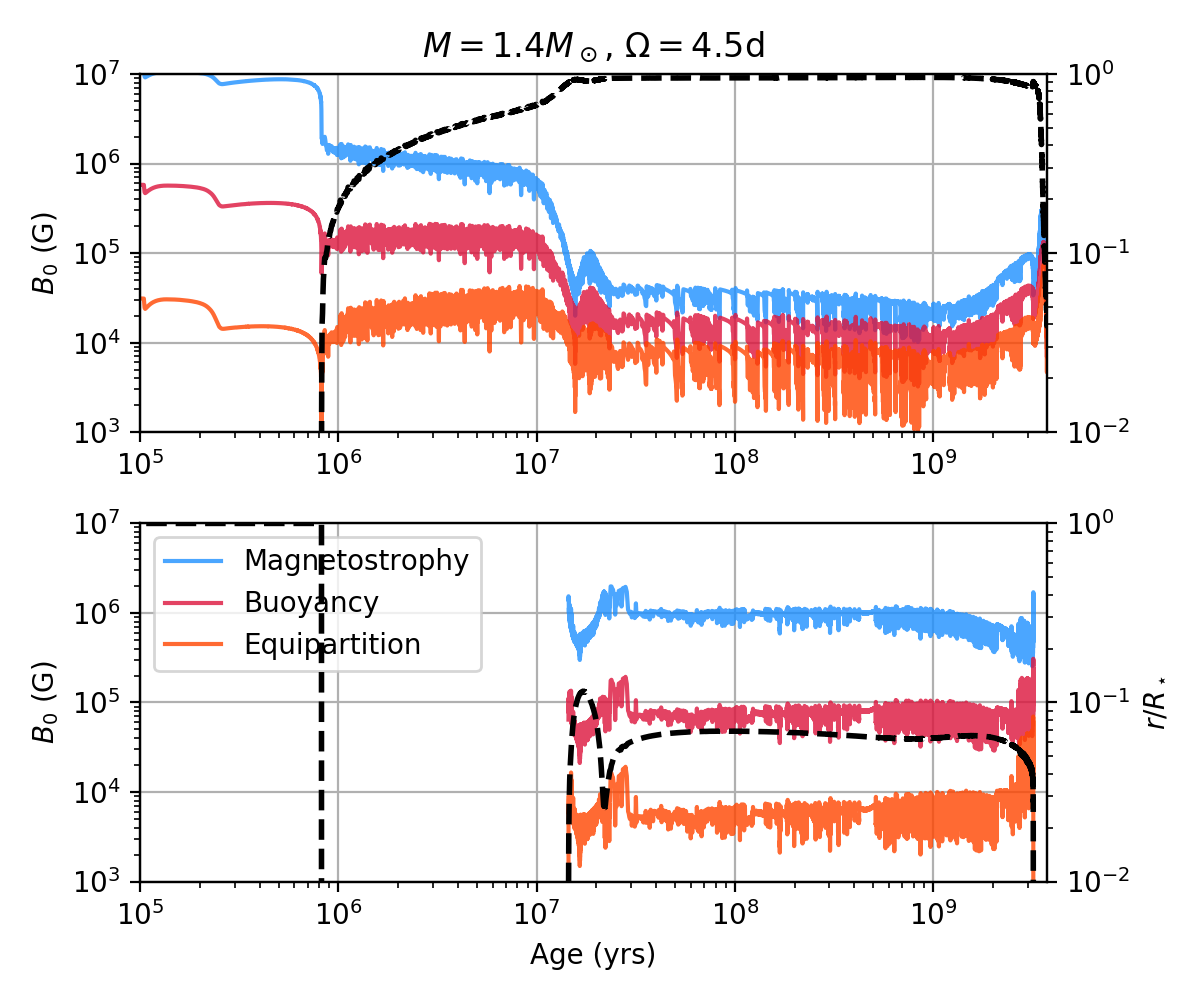}
    \caption{Same as  Fig.~\ref{fig:1p0_aurel} but for a $M_\star=1.4\mathrm{M_\odot}$ star.}
    \label{fig:1p4_aurel}
\end{figure}


\bld{As shown in figs.~\ref{fig:1p0_aurel}~and~\ref{fig:1p4_aurel}, the expected magnetic-field amplitude is very similar in the convective star during the \pms{} and inside the convective core during the \ms. Indeed, $B_{\mathrm{init}}$ is in both cases evaluated to be of the same order of magnitude (between $10^4$ and $10^6$G). As a result, it appears from our study that there is no correlation between the development of a convective core during the main sequence and the fossil magnetic field amplitude on the \rgb{}. This would imply that the discrepancy between low- and intermediate-mass stars dipolar mode amplitudes on the \rgb{} \citep{Stello2016a} might not be provoked by the fossil magnetic field resulting from the core dynamo action on the \ms. 
The main difference of field amplitude during the \rgb{} in our scenario is then due to the contraction of the radiative interior, as massive-star interiors contract more efficiently than low-mass star's after the \ms, resulting in larger magnetic field amplitudes during the \rgb{} from Eq.~(\ref{eq:flux_conservation}).}\\ 

\bld{We consider the dissipation of the fossil field to take place during the Ohmic timescale computed with its relaxed equilibrium global scale. Therefore, we neglect the losses of energy induced by its relaxation from the previous dynamo field during the conversion of convective layers into stably stratified radiative ones along the evolution of stars. This is a strong approximation, as already pointed out before, as the reconnection of small-scale magnetic structures into a larger global structure would decrease the fossil-field initial amplitude \citep[e.g.][]{Braithwaite2008a,Duez2011,Moffat2015,Emeriau-Viard2017}. This reconnection process could result into amplitude discrepancy between low- and intermediate-mass stars fossil fields.\\
In addition, our estimation of magnetic field amplitudes along the evolution is based on scaling laws and flux conservation only; we did not take into account any interaction between the fossil field in the radiative shell and the dynamo action inside the convective core of intermediate-mass stars on the main sequence. Such interaction could enhance the dynamo action inside the convective core of intermediate-mass stars \citep[e.g.][]{Featherstone2009} and thus result into stronger fossil-field amplitudes inside intermediate-mass stars on the \rgb{} than inside low-mass stars. Such interactions between stable and dynamo fields during the \ms{} could thus support the magnetic scenario for the amplitude of dipolar modes discrepancy as presented by \citep{Stello2016a}.}

\section{First-order frequency perturbation of mixed modes}
\label{sec:firstorder}

In this section we develop the first-order perturbation of the eigenfrequency of the oscillation modes, due to the stellar rotation and internal magnetism.
\subsection{Oscillations in non-rotating, non-magnetised stars}
The general linearised equation of motion of the fluid inside a non-rotating, non-magnetised star can be written in the inertial frame as: 

\begin{equation}
\rho \frac{d^2\boldsymbol{\xi}}{dt^2}=\boldsymbol{F}(\boldsymbol{\xi}),
\end{equation} 

\noindent where $\boldsymbol{\xi}$ denotes eigenfunctions of the modes propagating inside the star. In the case of evolved solar-like oscillators, they correspond to the superposition of mixed acoustic and gravity waves \citep{Beck2011, Bedding2011}. $\boldsymbol{F}$ represents all the applied forces in the inertial frame, and is composed of the effect of pressure and density gradient for a non-rotating, non-magnetised star (where the effects of the centrifugal acceleration and of the Lorentz force are neglected when computing the structure of the star).\\

\noindent We suppose a periodic Lagrangian displacement of the fluid inside the star due to oscillation modes $\boldsymbol{\xi}=\boldsymbol{\tilde{\xi}}(r, \theta, \varphi)e^{-i\omega t}$, with $\omega$ the eigenfrequencies of the oscillations. The non-perturbative equilibrium state refers to the non-rotating, non-magnetic oscillating star, for which the equation of motion can be written: 
\begin{equation}
\omega_0^2\boldsymbol{\xi}_0= -\boldsymbol{F}_0(\boldsymbol{\xi}_0),
\label{order0}
\end{equation} 

\noindent with $\omega_0$ the unperturbed eigenfrequencies, $\boldsymbol{\xi}_0$ the unperturbed eigenmodes and $\boldsymbol{F}_0$ reduced to pressure and buoyancy forces. Unperturbed eigenmodes for slowly or non-rotating stars can be written as a function of the azimuthal order $m$ and the degree $\ell$ of the mode \citep{Unno1989}:

\begin{equation}
    \boldsymbol{\xi}_0= \left[\xi_{r}(r) Y_\ell^m(\theta),  \xi_{h}(r) \frac{\partial Y_\ell^m(\theta)}{\partial \theta},  \frac{im}{\sin{\theta}}\xi_{h}(r) Y_\ell^m(\theta)\right] e^{i(m\varphi-\omega t)},
\end{equation}\\
\noindent with $(r, \theta, \varphi)$ the usual spherical coordinates, and $Y_\ell^m$ the spherical harmonics of degree $\ell$ and azimuthal order $m$.

\subsection{First-order perturbative forces}

\subsubsection{Rotational inertial forces} 

In order to connect our theory to observations, we approximate the star as a slowly rotating object with two layers, where each layer is rotating as a solid body. Let \omegacore{} be the angular rotation rate in the radiative interior and \omegaenv{} be the angular rotation rate in the convective envelope. This approximation is justified since asteroseismology generally provides only one measurement inside the deep radiative interior of the star and one at the surface (due to the limitation of the number of measured modes and due to the spatial structure of the kernels or eigenfunctions). These rotation rates are considered small enough so that the resulting terms in $\Omega(r)$ only come into play as a first-order perturbation \citep[e.g.][]{Gough1990a}. Observational studies of rotational splitting of mixed modes provide averaged values of surface and internal rotation for \sg s and \rg s. Specifically, we fix $\Omega_{\mathrm{core}}\simeq 0.5$\,\si{\micro\hertz} and $\Omega_{\mathrm{env}}\simeq  \Omega_{\mathrm{core}}$/10 \citep{Deheuvels2014a, Gehan2018}. 
Finally, with $R_{\mathrm{rad}}$ being the size of the radiative interior,
\begin{align}
\Omega(r, \theta)&=\Omega(r)\\&=
    \begin{cases}
     \Omega_{\mathrm{core}} &\mathrm{\,if\,} r<R_{\mathrm{rad}}\\
     \Omega_{\mathrm{env}} &\mathrm{\,if\,} r>R_{\mathrm{rad}} 
    \end{cases}
.
\end{align}

\noindent \bld{The system is thus affected in the inertial frame at first order in $\Omega$ by the Coriolis acceleration ($\boldsymbol{F_{c}}$, Eq.~(\ref{eq:coriolis})) and by the advection linked to the rotating motion of the star relatively to an observer in an inertial frame ($\boldsymbol{F_{f}}$, Eq.~(\ref{eq:frame})). In the following, we make use for rotational effects of the decompositions $\omega= \omega_0 + \epsilon_\Omega \omega_1$, $\boldsymbol{\xi}= \boldsymbol{\xi}_0 + \epsilon_\Omega \boldsymbol{\xi}_1$ and $\boldsymbol{F}= \boldsymbol{F}_0 + \epsilon_\Omega \boldsymbol{F_1}$, with $\epsilon_\Omega \ll 1$ the dimensionless parameter $\epsilon_{\Omega}={\Omega}/{\Omega_C}$ with $\Omega_C=\left(GM/R^3\right)^{-1/2}$ the Keplerian critical angular velocity \citep[see][]{Gough1990a}. The Coriolis operator stands for: }

\begin{equation}
    \boldsymbol{F_{c}}(\boldsymbol{\xi}_0)=2\iu\omega_0\boldsymbol{\Omega} \wedge \boldsymbol{\xi}_0,
    \label{eq:coriolis}
\end{equation}
\bld{with $\boldsymbol{\Omega}=\Omega(r)(\cos{\theta}\boldsymbol{e_r}-\sin{\theta}\boldsymbol{e}_{\theta})$ the rotation vector (whose amplitude is small enough in the core and in the envelope for the effect of the Coriolis acceleration to be considered as a first-order perturbation). Under the same approximation, the advection operator associated with the change of frame is written at first order as}
\begin{equation}
    \boldsymbol{F_{f}}(\boldsymbol{\xi}_0)= -2 m\omega_0\Omega\boldsymbol{\xi}_0. 
    \label{eq:frame}
\end{equation}

\subsubsection{The Lorentz force associated to magnetism} 

For \rg{} with a moderate internal magnetic field, we consider the magnetic field energy to be weak enough for the effects of the unperturbed Lorentz force on the system to be negligible compared to the gravitational force \citep[e.g.][Augustson \& Mathis, \textsl{in prep}]{Augustson2018}.\\

\bld{The system is thus also affected in the inertial frame at first order by the perturbed Lorentz force ($\boldsymbol{\delta F_L}$,~Eq.~(\ref{eq:lorentz})). We again make use of decompositions, associated with the perturbed Lorentz force: $\omega= \omega_0 + \epsilon_B \omega_1$, $\boldsymbol{\xi}= \boldsymbol{\xi}_0 + \epsilon_B \boldsymbol{\xi}_1$ and $\boldsymbol{F}= \boldsymbol{F}_0 + \epsilon_B \boldsymbol{F_1}$, with $\epsilon_B \ll 1$ the dimensionless parameter $\epsilon_B=B_0\left(\mu_0GM^2/R^4\right)^{-1/2}$ that compares the magnetic and gravific forces \citep{Gough1990a}. Finally, the perturbed linearised magnetic operator is given by}
\begin{equation}
    \boldsymbol{\delta}\boldsymbol{F_{L}}(\boldsymbol{\xi}_0) = \boldsymbol{\delta}\boldsymbol{F_{L, j+t}}(\boldsymbol{\xi}_0)
    +\boldsymbol{\delta}\boldsymbol{F_{L, c}}(\boldsymbol{\xi}_0),
    \label{eq:lorentz}
\end{equation}

\noindent \bld{where $\boldsymbol{\delta}\boldsymbol{F_{L, j+t}}$ represents the sum of the current and tension terms:}
\begin{equation}
    \boldsymbol{\delta}\boldsymbol{F_{L, j+t}}(\boldsymbol{\xi}_0) =\frac{1}{\mu_0}[(\boldsymbol{\nabla} \wedge \boldsymbol{B}) \wedge \boldsymbol{\delta B} + (\boldsymbol{\nabla} \wedge \boldsymbol{\delta B}) \wedge \boldsymbol{B}] 
    \label{eq:lorentz_j+t}
\end{equation}

    \noindent with the fluctuation of the magnetic field that comes from the linearised induction equation:\footnote{\bld{By combining the Maxwell Faraday and the Maxwell Ampère equations with the Ohm's law and by considering an ideal plasma with infinite electric conductivity (low Ohmic diffusivity $\eta$, see Eq.~\ref{eq:diffusivity}, as usually verified inside stars), the induction equation reduces to ${\partial \boldsymbol{B}}/{\partial t}=\boldsymbol{\nabla}\wedge\left(\boldsymbol{u} \wedge \boldsymbol{B}\right)$, with $\boldsymbol{u}$ the velocity field.}}
    \begin{equation}
\boldsymbol{\delta B}=\boldsymbol{\nabla} \wedge(\boldsymbol{\xi}_0 \wedge \boldsymbol{B})
\label{eq:induction}
\end{equation}
\noindent \bld{and $\boldsymbol{\delta}\boldsymbol{F_{L, c}}$ is the compression term associated with the compressibility of the mode \cite{Gough1990a}:}
\begin{equation}
    \boldsymbol{\delta}\boldsymbol{F_{L, c}}(\boldsymbol{\xi}_0) =\frac{\boldsymbol{\nabla}\cdot {\left(\rho\boldsymbol{\xi_0}\right)}}{\rho}\left(\boldsymbol{\nabla}\wedge \boldsymbol{B}\right)\wedge\boldsymbol{B},
    \label{eq:lorentz_c}
\end{equation}


\subsection{First-order perturbation equations}
The first-order momentum equation is thus written in the inertial frame as

\begin{equation}
    2 \omega_1 \omega_0 \boldsymbol{\xi}_0 + \omega_0^2\boldsymbol{\xi}_1= -\boldsymbol{F}_0(\boldsymbol{\xi}_1) - \boldsymbol{\delta F_L}(\boldsymbol{\xi}_0) /\rho- \boldsymbol{F_c}(\boldsymbol{\xi}_0) - \boldsymbol{F_f}(\boldsymbol{\xi}_0),
    \label{eqw1}
\end{equation}

\noindent with $\boldsymbol{\xi}_1$ the perturbed eigenfunctions of the modes.
We then apply the scalar product $\langle \boldsymbol{\xi}_0,\square \rangle = \int_V \rho \boldsymbol{\xi}_0^* \centerdot \square \textnormal{d}V$ to Eq.~(\ref{eqw1}). As $\boldsymbol{F}_0$ is Hermitian, Eq.~(\ref{eqw1}) simplifies and leads to the expression of the frequency shift at first order in the inertial frame:


\begin{equation}
    \omega_1=-\frac{\langle \boldsymbol{\xi}_0,\boldsymbol{\delta F_L}(\boldsymbol{\xi}_0)/\rho \rangle + \langle \boldsymbol{\xi}_0,\boldsymbol{F_c}(\boldsymbol{\xi}_0)\rangle + \langle \boldsymbol{\xi}_0,\boldsymbol{F_f}(\boldsymbol{\xi}_0) \rangle}{2 \omega_0 \langle \boldsymbol{\xi}_0, \boldsymbol{\xi}_0 \rangle}.
    \label{eq:w1}
\end{equation}

\subsection{Analytical development of frequency shifts}

We directly estimate the inertia of the modes, along with the contribution of the Coriolis acceleration and of the advection term associated to the change of frame to the splitting in the inertial frame. Equation~\eqref{eq:massmode} is the inertia of the modes, independent of the azimuthal order $m$:

\begin{equation}
\langle \boldsymbol{\xi}_0,\boldsymbol{\xi}_0 \rangle = \int_0^R\rho r^2 \left(|\xi_r^2+\Lambda|\xi_h|^2\right)\textrm{d}r, \, \text{with } \Lambda=\ell(\ell+1).
\label{eq:massmode}
\end{equation}

The effect of the Coriolis acceleration and of the change of frame are usually written together as the global rotational perturbation \citep{Aerts2010}. For clarity we chose to detail the two terms in Eq.~\eqref{eq:corioliseffect}~and~\eqref{eq:frameeffect}.
\begin{dmath}
\langle \boldsymbol{\xi}_0,\boldsymbol{F_c}(\boldsymbol{\xi}_0) \rangle = 4 m \omega_0  \int_0^R \rho r^2 \Omega(r) |\xi_h|^2\textrm{d}r \\ 
+ 8m\omega_0 \int_0^R \rho r^2 \Omega(r)\xi_r^* \xi_h \textrm{d}r
\label{eq:corioliseffect}
\end{dmath}


\begin{equation}
\langle \boldsymbol{\xi}_0,\boldsymbol{F_f}(\boldsymbol{\xi}_0) \rangle~=~{-2} m \omega_0 
\left(\int_0^{R} \rho r^2 \left(|\xi_r|^2+\Lambda|\xi_h|^2\right) \Omega(r) \textrm{d}r \right)
\label{eq:frameeffect}
\end{equation}

We verify that in the case of solid rotation $\Omega_s$, the change of frame can as usual be written as 
\begin{equation}
-\frac{\langle \boldsymbol{\xi}_0,\boldsymbol{F_f}(\boldsymbol{\xi}_0) \rangle}{2 \omega_0 \langle \boldsymbol{\xi}_0,\boldsymbol{\xi}_0 \rangle }= + m \Omega_s.
\end{equation}
The calculation of $\langle \boldsymbol{\xi}_0,\delta \boldsymbol{F_L}(\boldsymbol{\xi}_0)/\rho \rangle$ involves \bld{50} terms \bld{for which the angular integral does not cancel by geometrical combination of the spherical harmonics}. They are reported in Appendix~\ref{sec:terms}. Depending on the nature of the mixed modes, dominant terms can be extracted from this massive expression \bld{\citep[see for instance works by ][for high radial order pure gravity modes]{Hasan2005a, Rashba2007}}. In the case of subgiant stars, mixed-mode patterns are strongly dominated by peaks located at the acoustic mode eigenfrequencies. They are denoted as $p-m$ modes in the following study. On the contrary, modes located at internal gravity-mode frequencies have very small acoustic signatures. They dominate the spectrum of red giants, and are denoted $g-m$ modes. \bld{The dominant nature of the mode strongly influences the value of each of the $50$ terms; we refer to sections \ref{sec:asympt_g} and \ref{sec:asympt_p} for the extraction of dominant terms and to \cite{Mathis2020} for asymptotic formulations in the case of g-dominated modes and p-dominated modes.}

\subsection{Validity of the perturbative analysis}
\label{sec:critic}
In our scenario the magnetic field perturbation on mixed-mode frequencies plays a role at first order. For this approximation to be valid, the magnetic field amplitude should be small when compared to the zeroth-order processes. Thus, the frequency of the $g-m$ modes should be larger than the characteristic Alfvén frequency $\omega_A$, usually written as
\begin{equation}
    \omega_A=\frac{\boldsymbol{B} . \boldsymbol{k}}{\sqrt{\mu_o \rho}},
\end{equation}
\noindent with $\boldsymbol{k}$ the wave vector. For high-order radial modes, $\boldsymbol{B}.\boldsymbol{k} \simeq B_0 b_r k_{\mathrm{r}}$, with $b_r$ close to unity, $k_{\mathrm{r}}$ \bld{the vertical wavenumber} scaling for $g-m$ modes as $\frac{N_{\mathrm{max}}}{\omega_0}\frac{\sqrt{\Lambda}}{R_{\mathrm{rad}}}$, the critical field amplitude value for the perturbative study to be valid can be expressed as: 
    
\begin{equation}
    B_{c,g} \simeq \frac{\omega_0^2 R_{\mathrm{rad}} \sqrt{\mu_0 \rho_c}}{\sqrt{\Lambda}N_{\mathrm{max}}},
    \label{eq:Bc_g}
    \end{equation}
with $\rho_c$ the central density, and $N_{\mathrm{max}}$ the maximum value of the Brunt-V\"ais\"al\"a angular frequency profile \bld{defined as:
\begin{equation}
N^2=g\left(\frac{1}{\Gamma_{1}}\frac{{\rm d}\ln P}{{\rm dr}}-\frac{{\rm d}\ln\rho}{{\rm d}r}\right),
\label{eq:brunt_nu}
\end{equation}
\noindent where $g$ is the gravity of the hydrostatic background.} In order to ensure that the frequency of the mode is much larger than the Alfvén frequency, we set the limit of validity of the perturbative analysis for $g-m$ modes to $B_{c,g}/1000$.

In the case of high-radial order $p-m$ modes, the Alfvén speed
\begin{equation}
    v_A=B_0/\sqrt{\mu_0\rho}
    \label{eq:Alfvén_speed}
\end{equation}
should be compared to the sound speed of the gas ($c_s$) 
\bld{through $S_l$ the Lamb frequency defined as:
\begin{equation}
S_l^2=\frac{\Lambda c_s^2}{r^2}=k_{h}^{2}\,c_s^2,\quad\hbox{where}\quad k_{h}=\frac{\sqrt{\Lambda}}{r}.
\label{eq:lamb}
\end{equation}}
\noindent For high-radial-order $p-m$ modes, the minimum field above which the perturbative analysis may not be valid, considering the minimum value of the Lamb angular frequency $S_{l,\mathrm{min}}$ inside the radiative core, is written as

\begin{equation}
    B_{c,p}\simeq \sqrt{\frac{\mu_0\rho_c}{\Lambda}}\times S_{l,{\mathrm{ min}}}\times R_{\mathrm{rad}}.
    \label{eq:Bc_p}
\end{equation}

In the rest of our study, we set the limit of validity for the perturbative analysis for $p-m$ modes to $B_{c,p}/1000$.

\begin{figure*}[t]
    \centering
    \includegraphics[width=1\textwidth]{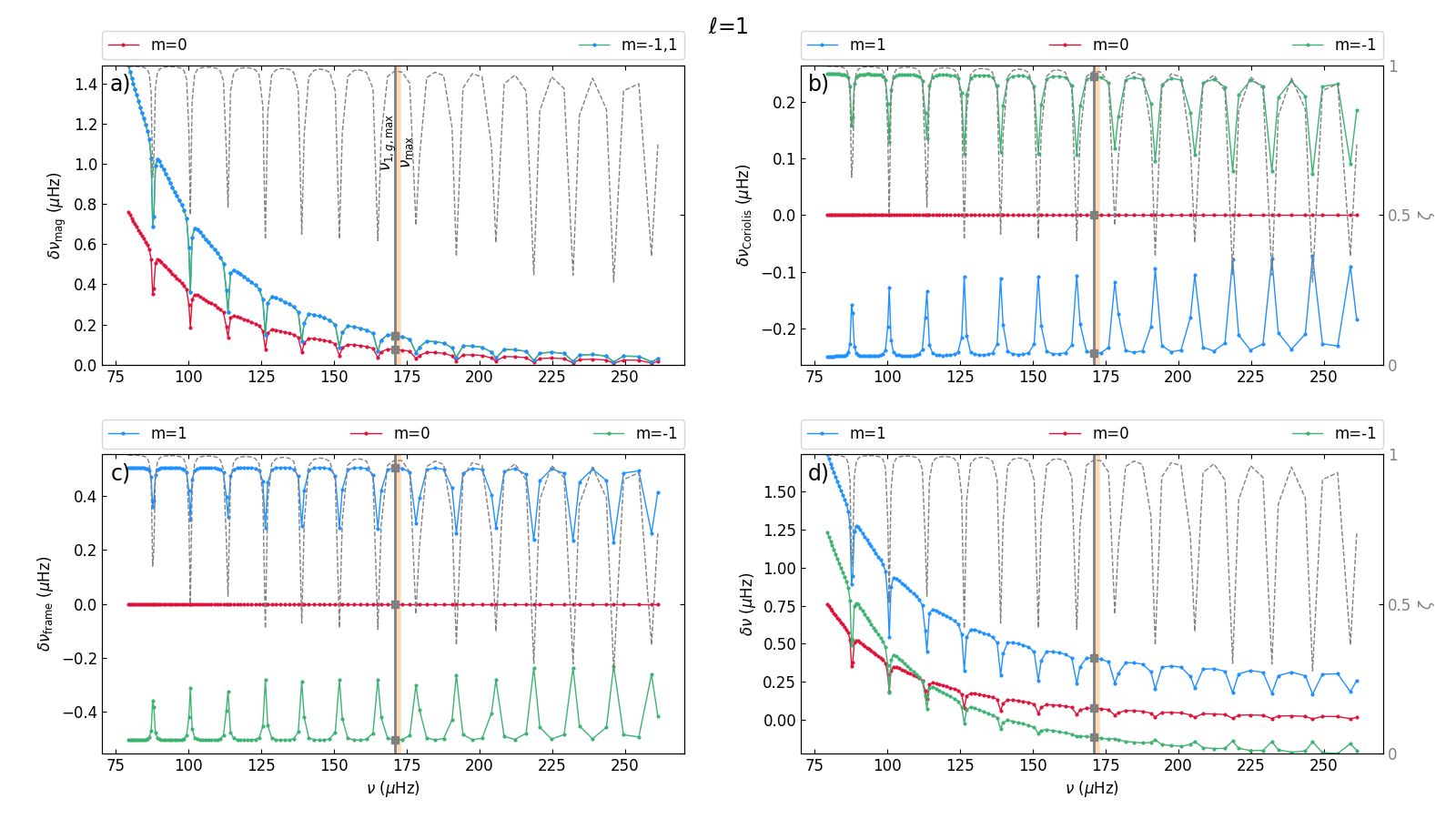}
     \caption{Frequency splittings ($\delta\nu$) calculated for mixed modes frequencies of a $M_\star=1.5 \mathrm{M_\odot}$, $Z=0.02$, \numax$\simeq 172$\,\si{\micro\hertz} red giant, corresponding to $a)$ Magnetic splitting, $b)$ Coriolis effect, $c)$ change from co-rotating to inertial frame, and $d)$ total frequency shifts. Red lines and points represent splittings for the $m=0$ component. The blue and green lines represent the magnetic splitting for the $m=1$ and $m=-1$ components respectively, that overlap on panel $a)$. The grey dashed line represents the $\zeta$ function as defined by \cite{Goupil2013a}. The beige vertical line indicates the position of \numax, and the grey vertical line the frequency of the $\ell=1$ g-dominated mode closest to \numax, $\nu_{\mathrm{1,g, max}}$, with squares indicating the corresponding splitting values associated with the different orders $m$.}
    \label{fig:nu_1g_def}
\end{figure*}

\section{Investigation of the magnetic impact on mixed-mode frequencies of red giants}
\label{sec:typical_RG}
To evaluate the different contributions from Eq.~(\ref{eq:w1}) to the frequency perturbation of mixed modes, we model stellar structures with the MESA evolutionary code \citep[][]{Paxton2011}, and we compute the associated oscillation modes' frequency by using GYRE pulsation code \citep[][]{townsend2013}. In this section, we estimate frequency perturbations from rotation and magnetism on mixed dipolar ($\ell=1$) and quadrupolar ($\ell=2$) modes for the reference star with $M_\star=1.5\mathrm{M_\odot}$ and $Z=0.02$ along the \rgb. This star is characteristic of intermediate-mass stars observed on the \rgb{} \citep[e.g.][]{Yu2016}. In Sect.~\ref{sec:SG} the effect of internal magnetism on mixed modes during the \sg{} stage will also be investigated.

 \subsection{Expected fossil magnetic field signature on mixed-mode frequencies during the \rgb}
 \label{sec:example}
 
 From the conservation of the magnetic flux from the end of the \ms, we evaluate the magnetic amplitude range expected for a typical red giant with $M_\star=1.5\mathrm{M_\odot}$ and $Z=0.02$. The star has oscillation modes centered around the frequency of maximum mode power \numax{} of about $172$\,\si{\micro\hertz}. For such a star, the initial radiative radius $R_{\mathrm{init}}$ is about \bld{$92\%$ of the star's total radius at the beginning of the subgiant phase ($R_\star=1.8\times 10^{11}$cm, $R_{\mathrm{init}}=1.6\times 10^{11}$cm).} It reduces to about \bld{$13\%$ of the total radius for the considered evolutionary stage star at the middle of the \rgb{} ($R_\star=3.7\times 10^{11}$cm, $R_{\mathrm{rad}}=5.0\times 10^{10}$cm)}. The equipartition regime, which is used to estimate fields amplitudes in the study of \cite{Cantiello2016} leads to $B_0\sim0.7$ MG, while the magnetostrophic regime sets the upper boundary of the field amplitude to $B_0\sim7$ MG (these boundaries will be extended along the evolutionary path of the reference star for the analysis of the impact of magnetism along the evolution of the star in Sect.~\ref{sec:colormap}). In the next paragraph, we investigate the effect of the change of internal structure of the star along its evolutionary path on the \rgb{} on the effect of magnetism on mixed-mode frequencies. We choose to use a magnetic-field amplitude value of $1$ MG, consistent with the presence of a convective core dynamo during the \ms.
 
As a first step, we verify that our perturbative approach can be used for this typical $M_\star=1.5\mathrm{M_\odot}$, $Z=0.02$, \numax$\simeq 172$\,\si{\micro\hertz} red giant. By fixing the validity limit of the perturbative analysis to a reasonable value of $B_0=B_{c,g}/1000$, the critical value for the magnetic-field amplitude at \numax{} is $B_0\simeq 3$ MG from Eq.~(\ref{eq:Bc_g}). Regarding the relatively large value of this limit, we conclude that the typical magnetic frequency-shift amplitudes of $g-m$ modes around \numax{} represented on Fig.~\ref{fig:nu_1g_def} belong to the valid frequency range of the perturbative analysis. However, a similar calculation for a frequency of $100$\,\si{\micro\hertz} leads to a critical field value (divided by $1000$) of $B_0\sim1$ MG, equal to the value used to build Fig.~\ref{fig:nu_1g_def}. The MG order of magnitude is too high to ensure that the perturbative analysis is valid for evolved red giants with \numax{} lower than $\sim 100$\,\si{\micro\hertz}. This upper boundary will further be discussed along the evolution of the reference star in Sect.~\ref{sec:colormap}. The reasonable limiting value of $B_0=B_{c,p}/1000$ for the pertubative analysis to be valid for $p-m$ modes leads to field amplitudes of about $B_0\simeq10$ MG for typical $S_{l,\mathrm{min}}\simeq200$\,\si{\micro\hertz} for the considered red giant by using Eq.~(\ref{eq:Bc_p}). The following study with a field amplitude of $1$ MG is therefore consistent with a perturbative regime for $p-m$ modes.  
 
\subsection{Results for a typical $M_\star=1.5\mathrm{M_\odot}$, $Z=0.02$, \numax$\sim172 $\,\si{\micro\hertz} star on the \rgb}

\begin{figure*}[h]
    \centering
    \includegraphics[width=1\textwidth]{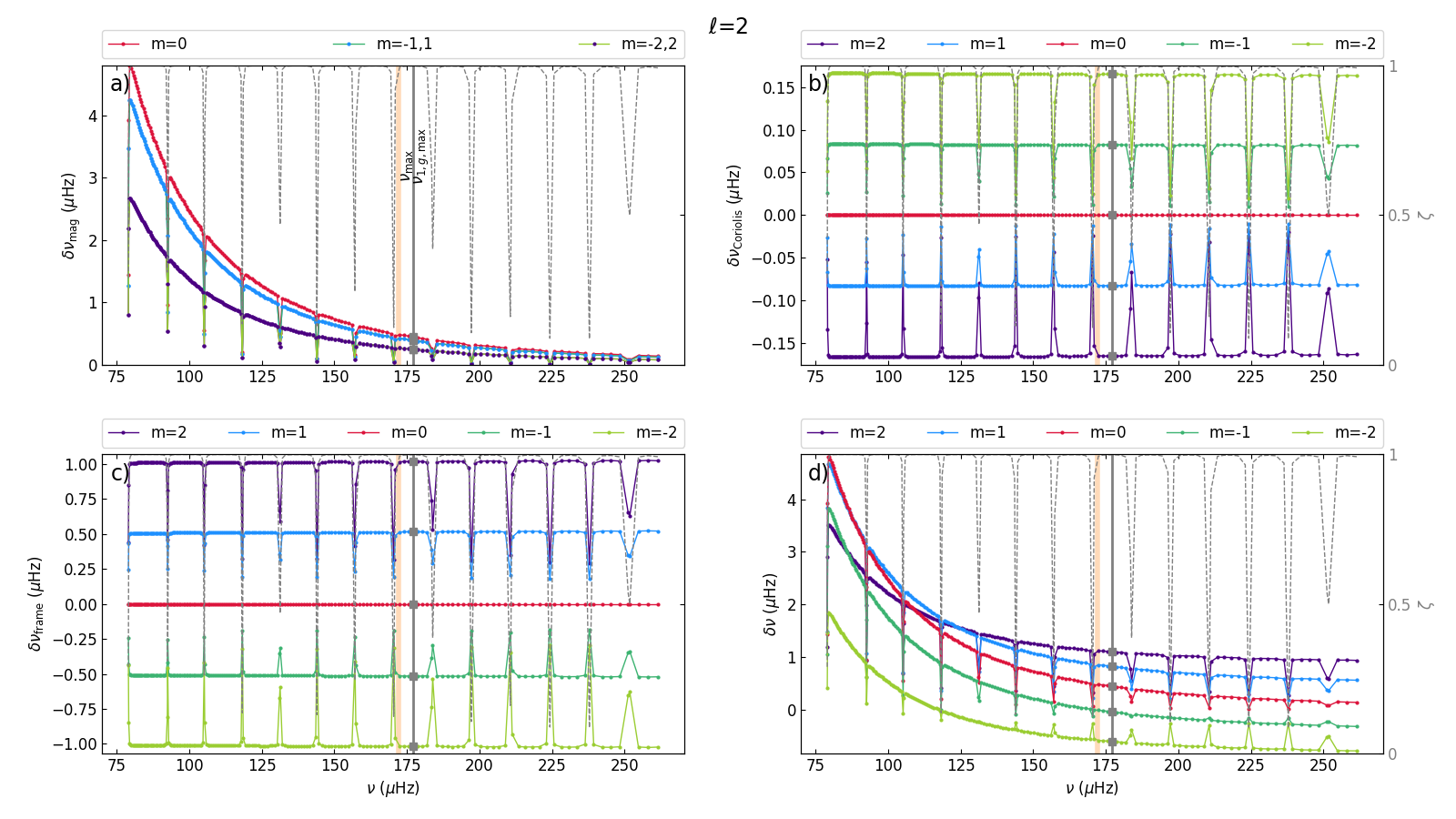}
    \caption{Same diagrams as on Fig.~\ref{fig:nu_1g_def} for $\ell=2$ mixed  modes. The red line represents the $m=0$ mode, the green line with blue points the $m=-1,1$ components, and the green line with purple points the $m=-2,2$ components.}
    \label{fig:splitting2}
\end{figure*}

The different terms constituting Eq.~(\ref{eq:w1}) are evaluated for mixed-modes $\ell=1$ and $\ell=2$ frequencies within the [\numax-7$\Delta\nu$:\numax+7$\Delta\nu$] interval, which ensures that detectable solar-like-oscillation modes are inside this region for most stars.
The unperturbed frequencies of mixed modes computed with GYRE \citep{townsend2013} are degenerated, due to the choice of a non-rotating and non-magnetic equilibrium. Each unperturbed mode is thus expected to be split into $2\ell+1$ components due to both rotational and magnetic effects when applying Eq.~(\ref{eq:w1}). Both magnetic first-order terms (reported in Appendix \ref{sec:terms}) and the mode inertia (see Eq.~(\ref{eq:w1})) depend on $m^2$ or are independent of $m$; thus, we expect the magnetic-frequency splittings of the $\pm m$ modes to be equal.\\

Panel $a)$ of Fig.~\ref{fig:nu_1g_def} represents magnetic shifts on $\ell=1$ mixed-mode frequencies $\left(\delta\nu_{\mathrm{mag, core},m}= -{\langle \boldsymbol{\xi}_0,\boldsymbol{F_L}(\boldsymbol{\xi}_0) \rangle}/(2 \omega_0 \langle \boldsymbol{\xi}_0,\boldsymbol{\xi}_0 \rangle)\right)$, resulting from a field of amplitude $B_0=1$ MG affecting the reference star on the \rgb. The magnetic field components $b_r$, $b_\theta$, and $b_\varphi$ expressed by Eqs.~\eqref{eq:br}~to~\eqref{eq:bphi} describe the stable axisymmetric fossil field from {the configuration given by} Eq.~(\ref{eq:compu}). The magnetic splittings of the $m=-1$ and $m=1$ modes are equal as expected ($C_{1,[1,-1]}$), and are larger than the splitting for $m=0$ modes ($C_{1,0}$) by a factor $C_{1,[1,-1]}/C_{1,0}=2$, as calculated by \cite{Hasan2005a}. A characteristic pattern in the magnetic splittings can be noticed: the global trend follows a power law in $\nu$, with deviations toward lower splitting values equivalently spaced in frequency. \bld{Such deviations are associated with the $\zeta$ function, a measure of the $g$ nature of the mode through the ratio of the mode inertia in the $g$ cavity over the total mode inertia \citep{Deheuvels2012a, Goupil2013a}:} 

\begin{equation}
    \zeta=\frac{\mathcal{I_{\mathrm{core}}}}{\mathcal{I}}=  \frac{\int_0^{R_{\mathrm{rad}}} \left(\xi_r^2+\Lambda\xi_h^2\right) r^2 dr}{\int_0^{R_\star} \left(\xi_r^2+\Lambda\xi_h^2\right) r^2 dr},
    \label{eq:zeta}
\end{equation}

This characteristic pattern is very similar to the pattern of mixed modes split by rotation as studied by \cite{Goupil2013a} and as represented by the grey dashed line on Fig.~\ref{fig:nu_1g_def}. Panel $b)$ of Fig.~\ref{fig:nu_1g_def} represents the frequency splittings due to the Coriolis acceleration only on $\ell=1$ mixed-mode frequencies, whereas panel $c)$ shows the effect of the change of frame on mixed mode frequencies due to the rotation of the star. These two rotational components are evaluated by considering typical red-giant rotation rates from \cite{Gehan2018} of $\Omega_{\mathrm{core}}/(2\pi)=0.5 $\,\si{\micro\hertz}, and $\Omega_{\mathrm{env}}/(2\pi)=0.05 $\,\si{\micro\hertz}. Combined together, these terms are well known from previous studies \citep[e.g.][]{Aerts2010, Mosser2015, Vrard2015}\bld{, and can as well be expressed as a function of the $\zeta$ function.}

\begin{figure*}[t]
    \centering
    \includegraphics[width=0.9\textwidth]{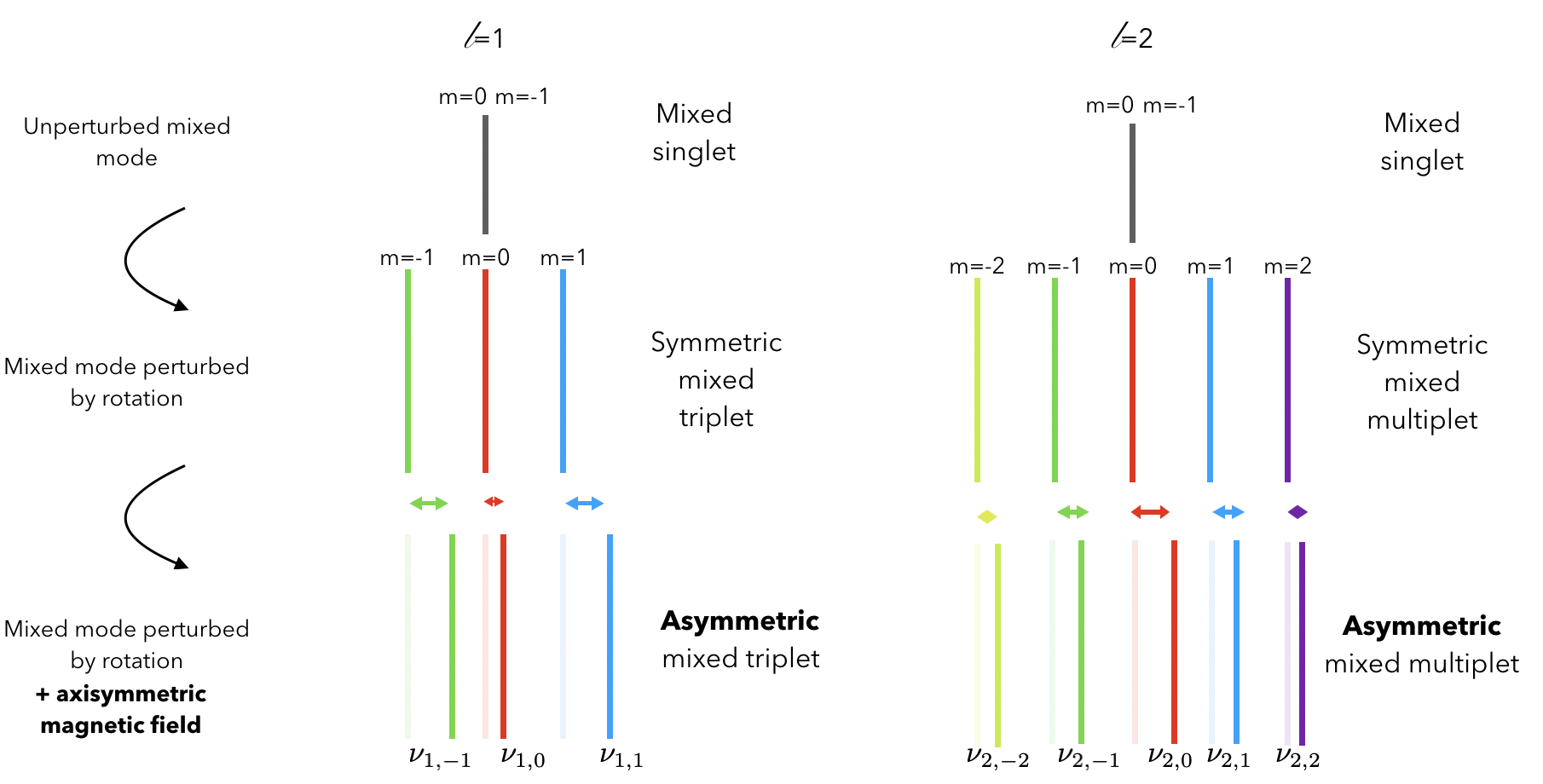}
    \caption{Sketch representing the effect of rotation and magnetism on a single unperturbed mixed mode frequency of degree $\ell=1$ and $\ell=2$.}
    \label{fig:schema}
\end{figure*}

For comparison, we represent on each panel the $\zeta$ function (grey dashed line). \bld{Dips} in the $\zeta$ function indicate the location of $p-m$ modes whereas the value of $\zeta$ tends towards 1 for $g-m$ modes. As the rotation rate of evolved solar-like stars is much higher in the depths than at the surface, the $g$-mode cavity is more affected by rotation. It leads to a global rotation splitting (including both $b)$ and $c)$ components) that are larger for $g-m$ than $p-m$ modes, which also probe the slowly rotating envelope. As a consequence of the confinement of the magnetic field inside the g-mode cavity, $g$-dominated mixed modes are also more affected by the magnetic field than $p$-dominated mixed modes. Indeed, the minimum values of $\zeta$ correlate with the minimum in the magnetic splittings on panel $a)$. On all panels, the frequency of maximum power (\numax) is marked by the beige vertical line, and the frequency of the closest $g-m$ mode ($\nu_{\mathrm{1,g, max}}$) is located by the grey vertical line. Grey squares indicate the typical frequency location ($\nu_{\mathrm{1,g, max}}$) and associated shifts for the $g-m$ mode closest to \numax,  defined as the frequency corresponding to the location of the maximum of the $\zeta$ function closest to \numax.

Finally, the bottom right panel provides the global splitting due to rotation and magnetism under the previously detailed rotational and magnetic configurations. Unperturbed frequencies are \bld{generally shifted towards higher frequencies for modes around \numax}, with a greater shift for prograde ($m=1$) and retrograde ($m=-1$) components than for the zonal modes ($m=0$), which are only affected by magnetism. As a consequence of this global frequency change, magnetism leads to a break of symmetry of the usually studied rotational \bld{g-dominated} triplet, as seen by the non-regular vertical spacing between two consecutive grey squares. \bld{Rotation plays against magnetism for the shifting of $m=-1$, $p-m$ modes (at frequencies corresponding to dips of the $\zeta$ function), resulting in upward peaks in the bottom right diagram at high frequencies.}

In Fig.~\ref{fig:splitting2} the same approach is taken for quadrupolar $\ell=2$ mixed modes. On panel $a)$ magnetic splits are represented as a function of the frequency of the modes, depending on the value of $m$. As opposed to the $\ell=1$ case, zonal modes are more affected by magnetism than the $m=1,-1$ and even more so than the $m=2,-2$ mixed modes, by factors of $C_{2,0}/C_{2,[1,-1]}=1.125$ and $C_{2,0}/C_{2,[2,-2]}=1.8$ respectively, as in \cite{Hasan2005a}. Panels $c)$ and $d)$ represent the contribution from the Coriolis acceleration and from the change of frame. Panel $d)$ shows the global frequency shifts for $m \in \{-2,-1,0,1,2\}$ in the inertial frame, due to magnetism and rotation. The $\zeta$ function and the location of $\nu_{\mathrm{1,g, max}}$ are reported on each panel for reference. As for $\ell=1$ mixed modes, $p-m$ modes are less affected by both rotation and magnetism than $g-m$ modes. Each perturbed quintuplet corresponding to one given unperturbed mixed mode is expected to have non-constant spacings between its components, as shown by the non-constant vertical spacing between the quintuplet components on panel $d)$. The asymmetry of $\ell=1$ and $\ell=2$ mixed-mode multiplets are schematically represented on Fig.~\ref{fig:schema}, in which we can observe the simplified behaviour of dipolar- and quadrupolar-mode frequency patterns due to internal magnetism.

\subsection{Asymptotic expressions of the magnetic perturbations on $g-m$ mode frequencies}
\label{sec:asympt_g}

In order to interpret the previously presented asymptotic patterns of magnetic splittings, we investigate the frequency dependency of the magnetic splitting at low ($\omega_0 \ll N$) and high ($\omega_0  \gg S_l$) frequency, \bld{with $N$ the angular Brunt-V\"ais\"al\"a frequency (Eq.~\ref{eq:brunt_nu}) and $S_l$ the Lamb angular frequency (Eq.~\ref{eq:lamb}) in $\mathrm{rad\, s}^{-1}$.}

In both cases, the oscillation vertical wavelength is much smaller than the characteristic distance of \bld{change} of the equilibrium state describing the surrounding fluid. In the Cowling approximation, and for high-order modes, the equation of non-radial oscillations can be written \citep[e.g.][]{Aerts2010}:

\begin{equation}
    \frac{d^2 \xi_r}{dr^2}=\frac{\omega_0^2}{c^2}\left(1-\frac{N^2}{\omega_0^2} \right)\left(\frac{S_l^2}{\omega_0^2}-1\right)\xi_r.
    \label{eq:oscillations}
\end{equation}

In the approximation of high-order mixed modes, the local radial displacement is dominated by the horizontal one ($\xi_r \ll \xi_h$) in the radiative region. The high-radial-order modes have a small vertical wavelength, allowing us to perform an asymptotic Jeffreys-Wentzel-Kramers-Brillouin (JWKB) analysis \citep{Hasan2005a, prat2019}.
Terms with high-order $\xi_h$ derivatives dominates because $\xi_h'\propto ik_{\mathrm{r}}\xi_h, \xi_h''\propto-k_{\mathrm{r}}^2\xi_h$, and $k_{\mathrm{r}}\gg1$. 



These approximations allow us to estimate the dominant terms composing Eq.~\eqref{eq:w1} in the case of low-frequency $g-m$ modes: 
\begin{equation}
\langle \boldsymbol{\xi}_0,\boldsymbol{\xi}_0 \rangle_{g} \simeq \int_0^R \rho r^2 \Lambda |\xi_h|^2 \textrm{d}r, 
\label{eq:massmode_g}
\end{equation}

\begin{equation}
\langle \boldsymbol{\xi}_0,\boldsymbol{F_c}(\boldsymbol{\xi}_0) \rangle_{g} \simeq 4 m \omega_0 \int_0^R \rho r^2 \Omega |\xi_h|^2\textrm{d}r, 
\label{eq:corioliseffect_g}
\end{equation}

\begin{equation}
\langle \boldsymbol{\xi}_0,\boldsymbol{F_f}(\boldsymbol{\xi}_0) \rangle_{g}~\simeq~{-2}  m \omega_0
\int_0^R \rho r^2 \Omega \Lambda |\xi_h|^2\textrm{d}r, 
\label{eq:frameeffect_g}
\end{equation}

\begin{dmath}
\langle \boldsymbol{\xi}_0,\boldsymbol{\delta F_L}(\boldsymbol{\xi}_0)/\rho \rangle_{g} \simeq 2 \pi B_0^2 \int_0^R r \xi_h^* b_r \left(r\xi_h b_r\right)'' \textrm{d}r \\
\int_0^\pi\left(\left(\frac{mY_l^m}{\sin{\theta}} \right)^2 + \left(\partial_\theta Y_l^m\right)^2 \right)\cos^2{\theta} \sin{\theta}\textrm{d}\theta.
\label{eq:mag_g}
\end{dmath}

From Eq.~\eqref{eq:w1}, the magnetic contribution to the frequency perturbation is
\begin{equation}
    {\delta \omega_{\mathrm{mag,g}}} \propto \frac{B_0^2}{\omega_0}\times \mathcal{I}_{g},
\end{equation}

\noindent with, in the case of high-order gravity modes,
\begin{equation}
    \mathcal{I}_{g}= \frac{\int_0^R{|(r b_r \xi_h)'|^2 dr}}{\int_0^R{{\rho} r^2 |\xi_h|^2  dr}},
\end{equation}
\noindent where we used Eqs.~\eqref{eq:massmode_g}~and~\eqref{eq:mag_g}. In the low-frequency regime, where $\omega_0 \ll N\ll S_l$, Eq.~\eqref{eq:oscillations} becomes 
\begin{equation}
    \frac{d^2 \xi_r}{dr^2}=-\frac{N^2}{\omega_0^2}\frac{\Lambda}{r^2}\xi_r = -k_{\mathrm{r}}^2\xi_r.
\end{equation}

In the Cowling approximation, we obtain a set of radial equations of momentum \citep[see for instance][]{Alvan2013}, with $p_1$ the perturbation associated with the gas pressure:
\begin{equation}
\xi_h=\frac{1}{r \omega_0^2}\frac{1}{\rho_0}p_1,
\end{equation}
\noindent and 
\begin{equation}
\frac{d p_1}{d r}\simeq -\rho_0 N^2\xi_r. 
\end{equation} 

\noindent Finally, we have that
\begin{equation}
    \xi_h=\frac{i}{r}\left(\frac{N^2}{\omega_0^2} \right)\frac{\xi_r}{k_{\mathrm{r}}}.
\end{equation}

We write $N=N_{\mathrm{max}}\times f(r)$, with $N_{\mathrm{max}}$ the maximum of the Brunt-V\"ais\"al\"a frequency inside the radiative interior and $f(r)$ containing all the radial dependence of the $N$ profile. As we look for the frequency dependence of $\mathcal{I}_{g}$, and by the use of the JWKB solution $\xi_r \propto \frac{1}{\sqrt{k_{\mathrm{r}}}}e^{i\int{k_{\mathrm{r}} dr}}$ \citep{Froman2005}, we obtain
\begin{equation}
    \xi_h=\left(\frac{N_{\mathrm{max}}^2}{\omega_0^2}\right)^{1/4}\times F(r),
\end{equation}
with $F(r)$ containing all factors that are independent of $\omega$. Considering high-order $g-m$ modes, we evaluate 

\begin{equation}
    |(r^2 b_r \xi_h)'|^2 \propto \left(\frac{N_{\mathrm{max}}^2}{\omega_0^2}\right)^{3/2},
\end{equation}
and 
\begin{equation}
    |\xi_h|^2 \propto \left(\frac{N_{\mathrm{max}}^2}{\omega_0^2}\right)^{1/2}.
\end{equation}
This leads to
\begin{equation}
    \mathcal{I}_{g}\propto \frac{N_{\mathrm{max}}^2}{\omega_0^2},\, \mathrm{and}\, \left.{\delta \omega}\right._{g} \propto \frac{B_0^2}{\omega_0^3}N^2_{\mathrm{max}}.
    \label{eq:power_law_g}
\end{equation}

\begin{figure}[t]
    \centering
    \includegraphics[width=.5\textwidth]{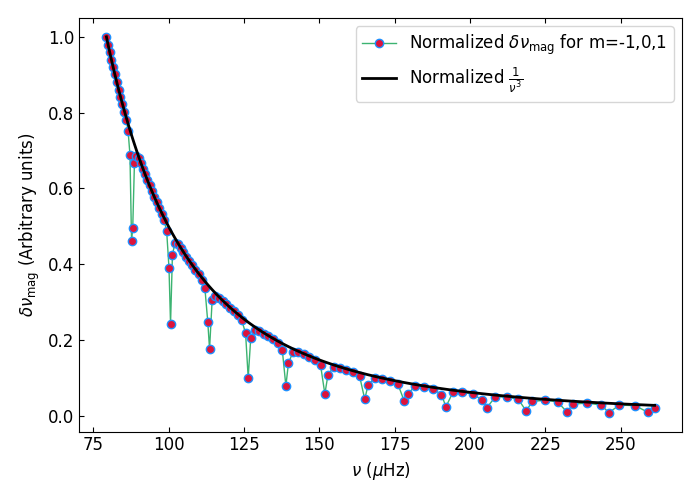}

    \caption{Normalised magnetic splittings versus mixed-mode frequencies computed for a simulated $M_\star=1.5\mathrm{M_\odot}$, Z=0.02, \numax$\simeq172$\,\si{\micro\hertz} red giant. The normalised power law following $1/\nu^3$ is superimposed in black.}
    \label{fig:asympt_g}
\end{figure}

The magnetic splitting of $g-m$ modes is thus proportional to $\omega_0^{-3}$ \bld{to leading order. In} Fig.~\ref{fig:asympt_g}, we represent the magnetic splittings $\delta\nu_{\mathrm{mag}}=\delta\nu_{\mathrm{mag, core},m}$ normalised to one of $m=-1, 0, 1$ mixed modes for the reference red giant from panel a) of Fig.~\ref{fig:splitting_color}. The black line indicates the normalized frequency power law from Eq.~\eqref{eq:power_law_g}. As a result, g-m modes perfectly follow the frequency power law, as infered from the JWKB analysis. However, \emph{p-m} modes that are less affected by magnetism are not reproduced by this analysis. For a complete theoretical description of asymptotic behaviours, including the origin of the $p-m$ dips through the $\zeta$ function and the theoretical estimation of angular integrals setting the amplitude of $\delta \nu_{\mathrm{mag}}$, we refer to our paper Mathis et al. (submitted).

\begin{figure*}[t]
    \centering
    \includegraphics[width=1\textwidth]{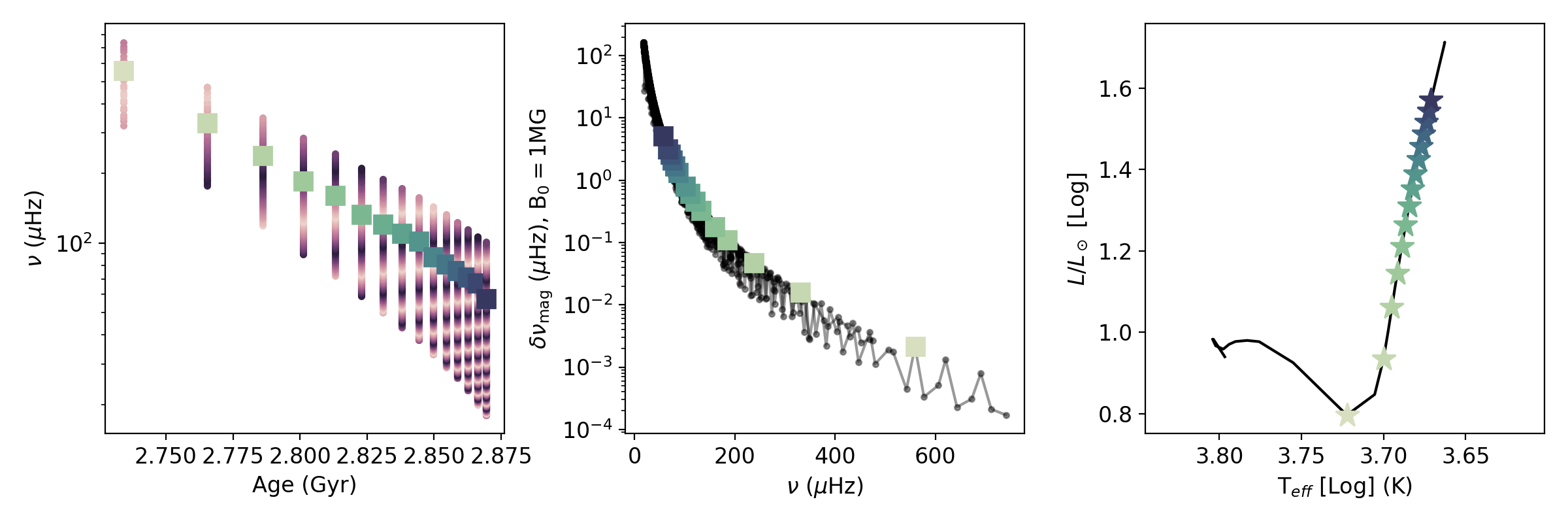}
    \caption{\textsl{Left:} Mixed-mode frequencies at different evolutionary stages along the \rgb{} indicated on the right panel. Purple-coloured dots represent the individual frequencies of mixed modes, with the shade of the dots indicating the value of the mixed-mode order $n_{pg}$ (the order is higher at high frequencies). Squares represent for each evolutionary stage the frequency of the $\ell=1$ g-dominated mode closest to \numax{} (see Fig.~\ref{fig:splitting_color} for the determination of this frequency).
    \textsl{Middle:} The $m=1$ magnetic splitting of $\ell=1$ mixed modes located inside $[\nu_{\mathrm{max}}-7\Delta\nu:\nu_{\mathrm{max}}+7\Delta\nu]$ for each of the evolutionary stages denoted on the right panel. Squares represent for each evolutionary stage the frequency of the $\ell=1$ g-dominated mode closest to \numax{} ($\nu_{\mathrm{1,g,max}}$, see Fig.~\ref{fig:nu_1g_def})
    \textsl{Right:} Hertzsprung-Russell diagram of the $M_\star=1.5\mathrm{M_\odot}$, $Z=0.02$ star ascending the \rgb{}. Coloured stars indicate the different evolutionary stages represented on the other panels.}
    \label{fig:evol}
\end{figure*}
\subsection{Characterisation of the multiplet asymmetry}
\label{sec:asymmetry}
Regardless of the degree $\ell$ of the mixed modes, the $g-m$ multiplet formed by the lifting of degeneracy by first-order perturbations linked to rotation and magnetism is asymmetric. This asymmetry can be quantified for each $|m|$ through the calculation of an asymmetry measure $\delta_{\ell, m}$ \citep[similar formalism as in][adapted for the consideration of magnetism in addition to rotation]{Deheuvels2017}:

\begin{equation}
    \delta_{\ell, m}=\frac{\nu_{\ell,m}+\nu_{\ell,-m}-2\nu_{\ell,0}}{\nu_{\ell,m}-\nu_{\ell-m}}=\frac{\delta \nu_{\ell,m}+\delta\nu_{\ell,-m}-2\delta\nu_{\ell,0}}{\delta\nu_{\ell,m}-\delta\nu_{\ell,-m}}.
    \label{eq:asymmetric_degree}
\end{equation}

This measure is equal to zero when the $m$ and $-m$ components of the multiplet are perfectly symmetric around the $m=0$ component, and reaches $1$ or $-1$ when one of the $|m|$ components overlaps with the $m=0$ mode. For the reference red giant studied in this section, with the considered $\Omega_{\mathrm{core}}=10\times \Omega_{\mathrm{env}}=0.5$\,\si{\micro\hertz}, $B_0=1$ MG, total frequency perturbations are estimated at $\nu_{1,g,\mathrm{max}}$: $\delta\nu_{1,1}\simeq0.41$\,\si{\micro\hertz}, $\delta\nu_{1,-1}\simeq-0.076$\,\si{\micro\hertz}, $\delta\nu_{1,0}\simeq-0.11$\,\si{\micro\hertz}, $\delta\nu_{2,2}\simeq0.93$\,\si{\micro\hertz}, $\delta\nu_{2,-2}\simeq-0.74$\,\si{\micro\hertz}, $\delta\nu_{2,1}\simeq0.56$\,\si{\micro\hertz}, $\delta\nu_{2,-1}\simeq-0.27$\,\si{\micro\hertz}, and $\delta\nu_{2,0}\simeq0.17$\,\si{\micro\hertz} from $d)$ panels of Figs.~\ref{fig:nu_1g_def}~and~\ref{fig:splitting2}. These values lead to asymmetry measures of $\delta_{1, 1}\simeq0.28$, $\delta_{2, 2}\simeq-0.09$, and $\delta_{2, 1}\simeq-0.06$. We notice the change of sign of the measure from $\ell=1$ to $\ell=2$ modes, where the $m=0$ mode is closer to the $m=-1$ mode for $\ell=1$, and closer to the $m=1$ mode for $\ell=2$. The absolute value of the asymmetry measure is higher for $\ell=1$ modes than for $\ell=2$, meaning that the impact of magnetism compared to rotation is stronger for $\ell=1$ modes. This is not surprising considering that the \bld{acoustic} dipolar modes \bld{couple with internal gravity modes, which are sensitive to the magnetic field,} deeper inside the radiative interior than $\ell=2$ modes for a given frequency.

\begin{figure*}[t]
    \centering
    \includegraphics[width=1.\textwidth]{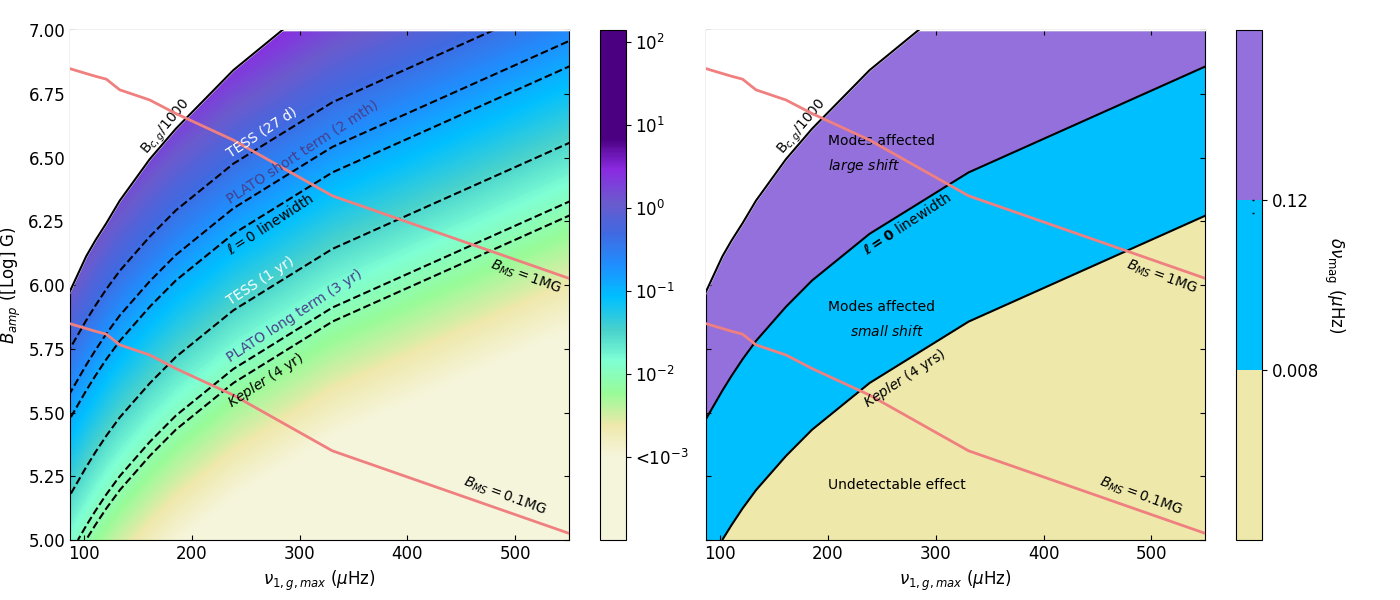}
    \caption{\textsl{Left:} $m=1$ mixed-mode mean frequency perturbation due to magnetism only ($\delta \nu_{\mathrm{mag}}$ in \si{\micro\hertz}) as a function of mixed-mode frequencies ($\nu_{\mathrm{1, g, max}}$ in \si{\micro\hertz}) and the magnetic field amplitude in Gauss for a $M_\star=1.5$~M$_\odot$, $Z=0.02$. Typical limiting frequencies are indicated: the black dashed lines represent the frequency resolution in the \psd{} corresponding to 4 years of \kepler{} continuous observations ($\sim8$ \,\si{\nano\hertz}), 3 years of PLATO two main fields ($\sim 11$ \,\si{\nano\hertz}), 1 year of TESS continuous viewing zone data ($\sim 30$ \,\si{\nano\hertz})), a typical linewidth of radial modes, which are an upper limit for the linewidth of dipolar mixed modes ($\sim 0.12$ \,\si{\micro\hertz}), 2 months of PLATO stare-and-step observations ($0.19$ \,\si{\micro\hertz}), and 27 days of TESS individual sectors ($0.37$ \,\si{\micro\hertz}). The coloured area is limited by the critical magnetic-field amplitude range for the validity of the perturbative analysis. We choose to limit the results at $B_{c,g}/1000$, as defined in Sect.~\ref{sec:critic}. Red lines represent the expected magnetic-field amplitude by considering magnetic-flux conservation from the end of the main-sequence, with original fossil field amplitudes of $0.1$ and $1$ MG, as detailed in Sect.~\ref{sec:B_amp}. \textsl{Right:} Same as left panel, but simplified with the color map replaced by typical limiting frequency values corresponding to \kepler{} data. The yellow area corresponds to the combination of $\nu_{\mathrm{1, g, max}}$ and magnetic amplitude for which magnetic splittings cannot be visible in \textsl{Kepler} observations. The blue area corresponds to small magnetic shifts that should be visible in \textsl{Kepler} observations. The purple area corresponds to very large magnetic shifts, larger than the typical linewidth of radial modes.}

    \label{fig:splitting_color}
\end{figure*}

Figure \ref{fig:evol} depicts the changes in the magnetic splittings along the evolutionary track of the reference star from the base of the \rgb, during which mixed modes can be approximated as $g-m$ modes. 
Squares indicate for each evolutionary stage the $\nu_{\mathrm{1,g, max}}$ frequency defined in the previous section, for each of the colour-coded evolutionary stages reported on the \HR{} diagram in the right-hand side panel of Fig.~\ref{fig:evol}. The middle panel is composed of the superposition of the magnetic splittings (equivalent to panel $a)$ in Fig.~\ref{fig:nu_1g_def}) corresponding to a field of amplitude $B_0=1$ MG for the $m=1$ mixed modes at the given evolutionary stages, as a function of their unperturbed frequencies. We notice that for stars ascending the \rgb{} the value of $\nu_{1,g, max}$ is globally monotonously decreasing. As a consequence, the magnetic splitting value of $g-m$ modes increases as the star evolves on the \rgb{} (the base of the \rgb{} corresponds to $\nu_{1,g, max} \lesssim 550$\,\si{\micro\hertz} for the considered reference star). The $p-m$ modes, which are less affected by magnetism generally, are also more split as they evolve along the \rgb{} as seen by the minima on the curve of the middle panel. The increase of $\delta\nu_{\mathrm{mag, core},m}$ with the evolution of the star on the \rgb{} at a given magnetic field amplitude should make the detection of the magnetic effect on mixed-mode ($g-m$ and $p-m$) frequencies easier for evolved red giants than younger ones on the \rgb.\\ 

Similar results are obtained for different magnetic field amplitudes $B_0$, ranging from $0.1$ to $10$ MG, which may arise from past dynamo events (see Sect.~\ref{sec:B_amp}). As a summary, the colormap on the left panel of Fig.~\ref{fig:splitting_color} represents values of $m=1$, $g-m$-mode-frequency splittings due to magnetism only, as function of the frequency $\nu_{\mathrm{1,g, max}}$, which is a proxy for the evolutionary stage along the \rgb, and of the applied magnetic field amplitude. The critical value of the magnetic field $B_c/1000$ estimated by Eq.~\eqref{eq:Bc_g} for $g-m$ modes at all points along a star's evolution delimits the top of the colormap.  Specifically, results above this line are beyond the domain of applicability of the perturbative analysis and have been removed. The magnetic signature increases as the star evolves, as deduced from Fig.~\ref{fig:evol}. The magnetic signature also increases with the magnetic field amplitude, as expected from the $B_0^2$ dependency in the magnetic splitting expression, and as already deduced from Fig.~\ref{fig:evol}.

\subsection{Detectability of the magnetic signature}
\label{sec:colormap}

For comparison, we represent the position in the diagram of different typical observational frequencies: the frequency resolutions of the \kepler, TESS and PLATO data, and the typical value of $\ell=0$ mode linewidths as estimated by \cite{Vrard2017} and \cite{Mosser2018}. The linewidth of $\ell=1$ mixed modes of normal amplitude was shown by \cite{Benomar2014} to be linked to that of $\ell=0$ modes by
\begin{equation}
    \Gamma_1=\Gamma_0(1-\zeta).
    \label{eq:linewidth}
\end{equation}
Here, this radial mode linewidth is considered to be an upper limit for mixed-modes linewidth values. All these typical frequencies provide constraints on the minimum field amplitude needed at each evolutionary stage in order for magnetic effects to be visible in asteroseismic observations. When considering a star from the continuous viewing zone (CVZ) of the TESS satellite so that the star has been observed for about $1$ year, with $\nu_{\mathrm{max}}\simeq172$\,\si{\micro\hertz} , the lower bound value for the magnetic field amplitude to have detectable signatures is about $0.4$ MG. On the right panel, the same diagram is shown but simplified, emphasing the ($B_0$, $\nu_{\mathrm{1,g, max}}$) combinations for which the magnetic signature should be either easily detectable (purple area), detectable (blue area), or undetectable (yellow area) based on  $4$ years of observations, such as done by $\kepler$. 

For simplification, we consider the magnetic signature to be observable when $\delta \nu_{\mathrm{mag}}$ is larger than the data frequency resolution ($\delta f$). However, this threshold should be discussed and considered carefully as described by the following criteria:  
\begin{enumerate}
    \item It usually takes a few $\delta f$ for a signal to be detectable within observational data, due to the spreading of the signal over several frequency bins. However, typical linewidths of $g-m$ mixed modes are very small, as measured from Eq.~\eqref{eq:linewidth} where $\zeta\rightarrow1$. For the typical $g-m$ mode located at $\nu_{\mathrm{1,g,max}}$, we estimate $\zeta\simeq0.98$ from Fig.~\ref{fig:nu_1g_def}, which leads to an estimation of $g-m$ mode linewidths of about $2.4$\,\si{\nano\hertz} by considering the typical linewidth of radial modes of $120$\,\si{\nano\hertz} \citep[e.g.][]{Mosser2018}. The minimal frequency resolution of the \kepler{} satellite being greater than $\sim7.9$\,\si{\nano\hertz}, $g-m$ mixed modes are not resolved, and a dilution factor must be considered \citep{Dupret2009, Mosser2018}. It results in a lower limit for $g-m$ modes linewidths of $2\delta f/\pi$, evaluated at $5$\,\si{\nano\hertz} for the \kepler{} 4-year observations \citep[see also][for the estimation of $\ell=1$-mode linewidths for a typical red giant]{Mosser2018}. We conclude that the linewidth of $g-m$ modes is of the order of magnitude of the resolution of the data. Therefore, the detection limit of magnetic signature at $\delta\nu_{\mathrm{mag, core},m}\simeq \delta f$ is consistent. In contrast, even though $p-m$ modes have a larger linewidth of about $70$\,\si{\nano\hertz}, the magnetic effect has a much smaller amplitude and are therefore much more difficult to detect. We keep the detection limit at $\delta\nu_{\mathrm{mag, core},m}\gtrsim \delta f$ that is pertinent for $g-m$ modes, but one should use this lower boundary with care especially when looking at p-dominated mixed modes.
    
    \item In addition to the magnetic splitting, one should also add the shift due to the rotation of the star from Eqs.~\eqref{eq:coriolis}~and~\eqref{eq:frame}, whose measurements come with their own uncertainties. Thereby, the magnetic effect should be large enough for its signature to be easily distinguishable from rotational pattern adjustment errors. These errors have been estimated by \cite{Mosser2018} at about $\Delta\nu/200$, corresponding to $\sim 1.3$ times the \kepler{} typical frequency resolution. Once again, the lower limit of detectability evaluated at $\delta f$ may be too small, especially in the case of noisy data. 
    
    \item When adding rotational perturbations, the magnetic signature can only be measured through the use of the asymmetry measure $\delta_{\ell,m}$ defined in Eq.~\eqref{eq:asymmetric_degree}, as magnetic and rotational effects add up. For the  effect of magnetism to be detectable, the criterion is changed from $\delta\nu_{\mathrm{mag, core},m}\gtrsim \delta f$ to
    \begin{equation}
        \nu_{\ell,m}+\nu_{\ell,-m}-2\nu_{\ell,0}\gtrsim \delta f
        \label{eq:new_criterion0}
    \end{equation} 
    in the presence of rotation. From a perfectly symmetric rotational $\ell=1$ triplet, at given frequencies $[-\delta\nu_{\mathrm{rot},1}$,~$0$,~$\delta\nu_{\mathrm{rot},1}]$, perturbations by magnetism produces the shifts $[\delta\nu_{\mathrm{mag},1}$,~$\delta\nu_{\mathrm{mag},1}/2$,~$\delta\nu_{\mathrm{mag},1}]$ (see Sect.~\ref{sec:example}). The criterion from Eq.~\eqref{eq:new_criterion0} may then be rewritten as 
    
        \begin{multline}
    \underbrace{{\nu_{0,\ell,0 }}+\delta\nu_{\mathrm{mag},\ell,m}+ \delta\nu_{\mathrm{rot},\ell, m}}_{\text{{$\nu_{\ell,m}$}}}+ \underbrace{{\nu_{0,\ell,0}}+\delta\nu_{\mathrm{mag},\ell, m}-\delta\nu_{\mathrm{rot},\ell,m}}_{\text{{$\nu_{\ell,-m}$}}}\\
    -2(\underbrace{{\nu_{0,\ell,0}}+\delta\nu_{\mathrm{mag},\ell,m}/2}_{\text{{$\nu_{\ell,0}$}}})\gtrsim \delta f,
    \label{eq:new_criterion}
  \end{multline} 
  
 \noindent with $\nu_{0,\ell,0}$ the unperturbed frequency of the $m=0$ component. Simplifying Eq.~\eqref{eq:new_criterion} leads back to the equation $\delta\nu_{\mathrm{mag, core},m}\gtrsim \delta f$. The chosen minimum limit of detection is thus conserved when the star rotates.
  
    \item For a few red giants, a departure of the symmetric rotational triplet due to buoyancy glitches can be observed, which is caused by strong chemical gradients generated by the first dredge-up and left behind by the retreating envelope \citep[][]{Cunha2015, Cunha2019, Jiang2020}.  \cite{Mosser2018} shows that KIC3216736 is the only red giant among the $200$ studied that exhibits buoyancy glitches, with only its $m=0$ component visible. As deduced analytically by \cite{Cunha2015, Cunha2019}, buoyancy glitches are very rare on the \rgb, and glitch-induced oscillation variation occur only at the luminosity bump. As a consequence, glitches will be neglected in the rest of our study.

    \item If the star rotates fast enough, second-order and higher-order asymmetric rotational perturbations of the centrifugal and Coriolis accelerations can affect the symmetric rotational frequency pattern \citep{Dziembowski1992, Suarez2006}. Such second-order perturbations should affect both $\ell=1$ and $\ell=2$ mixed modes. However, \cite{Deheuvels2017} emphasises that a measured core rotation rate of $\Omega_c\simeq710$ \,\si{\nano\hertz} along with an envelope rotation rate $5$ times smaller are much too low for second-order rotational effects to significantly contribute to rotational splittings. Thereby, second-order rotational effects should not produce significant asymmetric perturbations in the spectrum of red giants nor subgiants. Moreover, we consider only rotation as a perturbation in the case of red giants as they are considered as slow rotators. Indeed, \cite{Ouazzani2013} show that the effect of rotation can be expressed by perturbative calculations when $\Omega/\left(2\pi\right)/(\Delta P/P^2)\lesssim 2$, which is the case for red giants according to core rotation measurements by \cite{Gehan2018}. For the study of rapid rotators with non-perturbative developments, we refer to \cite[][]{prat2019} and \cite[][]{VanBeeck2020} (gravity modes), and to \cite[][]{Reese2006a} (acoustic modes).
    
    \item Latitudinal differential rotation may also induce asymmetries, but as evaluated through non-perturbative calculations with the Adiabatic Code of Oscillation including Rotation (ACOR) by \cite{Deheuvels2017}, typical latitudinal differential rotation profiles lead to a very small asymmetry measure of $\sim10^{-3}$. It corresponds to $\delta\nu_{\mathrm{mag, core},m}\simeq 2\times 10^{-3}\delta\nu_{\mathrm{rot, core},m}$, evaluated at $\delta\nu_{\mathrm{mag, core},m}\simeq0.5$\,\si{\nano\hertz} for the rotational splittings associated with core rotation of $0.5$\,\si{\micro\hertz}. Given the frequency resolution of asteroseismic data ranging from $\sim7.9$ (\kepler{} 4 years data) to $\sim380$\,\si{\nano\hertz} (TESS 1 month data), and considering typical latitudinal differential rotation inside solar-like stars, the effect of latitudinal differential rotation on the symmetry of the mixed-mode pattern is therefore negligible.
    
    \item Near-degeneracy effects occur by the combination of rotation and mode mixing: when two mixed modes with the same ($\ell, m$) combination have frequencies too close to each other (i.e. the frequency spacing between the two mixed modes is smaller than the rotation rate), their frequencies are perturbed \citep{Dziembowski1992, Suarez2006}. We refer to the complete study of \cite{Deheuvels2017} for the theoretical development of near-degeneracy effects on the asymmetry of rotational multiplets. We emphasise the fact that near-degeneracy effects produce increasing asymmetry measure $\delta_{\ell,m}$ when $\ell$ increases ($\delta\nu_{\mathrm{degeneracy, \ell=1}}\ll \delta\nu_{\mathrm{degeneracy, \ell=2}}$). It can be interpreted as the fact that the frequency separation between two consecutive $\ell=1$ mixed modes is much larger than the separation between \bld{two} consecutive $\ell=2$ mixed modes. In the case of KIC7341231 studied by \cite{Deheuvels2017}, no asymmetries were found in the $\ell=1$ triplet whereas $\ell=2$ multiplet asymmetries are $|\delta_{2,2}|\gtrsim 0.14$. In order to disentangle near-degeneracy from magnetic effects, measures of $\ell=1$ and $\ell=2$ mixed mode asymmetries are effective. More precisely, an asymmetry measure such that $\delta\nu_{\ell=1}\gtrsim \delta\nu_{\ell=2}$ is a clear indicator that the magnetic effects are larger than those of near-degeneracy.
\end{enumerate}

In conclusion, ($B_0, \nu_{\mathrm{1,g,max}}$) areas delimited in the right panel of Fig.~\ref{fig:splitting_color} should be used with care. In addition to the characteristic frequency positions, expected magnetic-field amplitudes at given evolutionary stages from Eq.~\eqref{eq:Bc_g} are represented by red lines on each panel, considering amplitudes at the end of the \ms{} of $B_{\mathrm{ms}}=0.1$ and $1$ MG. \bld{The maximum expected magnetic amplitude is represented by the $B_{\mathrm{ms}}=1$ MG upper-red line on the diagram (magnetic field amplitude along the evolution. We recall that it results from the conservation of the magnetic field flux from the end of the main sequence as explained in Sect.\ref{sec:B_amp}). Note that the value of $B_{\mathrm{ms}}=0.01$ MG corresponding to the equipartition regime leads to magnetic field amplitude values too low on the \rgb{} to be represented on the diagram. From} Fig.~\ref{fig:splitting_color} we conclude that the more evolved the star the easier it is to detect its internal magnetic signature. More specifically, magnetic fields of $B_{\mathrm{MS}}\simeq1$ MG at the end of the \ms{} produces frequency perturbations that are too small to detect inside young \rg{} with $\nu_{\mathrm{max}}\gtrsim 475$\,\si{\micro\hertz} and that are too large for the perturbative analysis to be valid for older \rg{} with $\nu_{\mathrm{max}}\lesssim 190$\,\si{\micro\hertz}. Globally, the expected range for the magnetic-field amplitude to reach large enough values for the magnetic signature to be observable depends upon the duration of the observing campaign.  For instance, the red lines crossing the different \kepler{} boundaries are shown in the right panel of Fig.~\ref{fig:splitting_color}.

\subsection{Stretched spectrum}
\begin{figure*}[t]
    \centering
    \includegraphics[width=1.\textwidth]{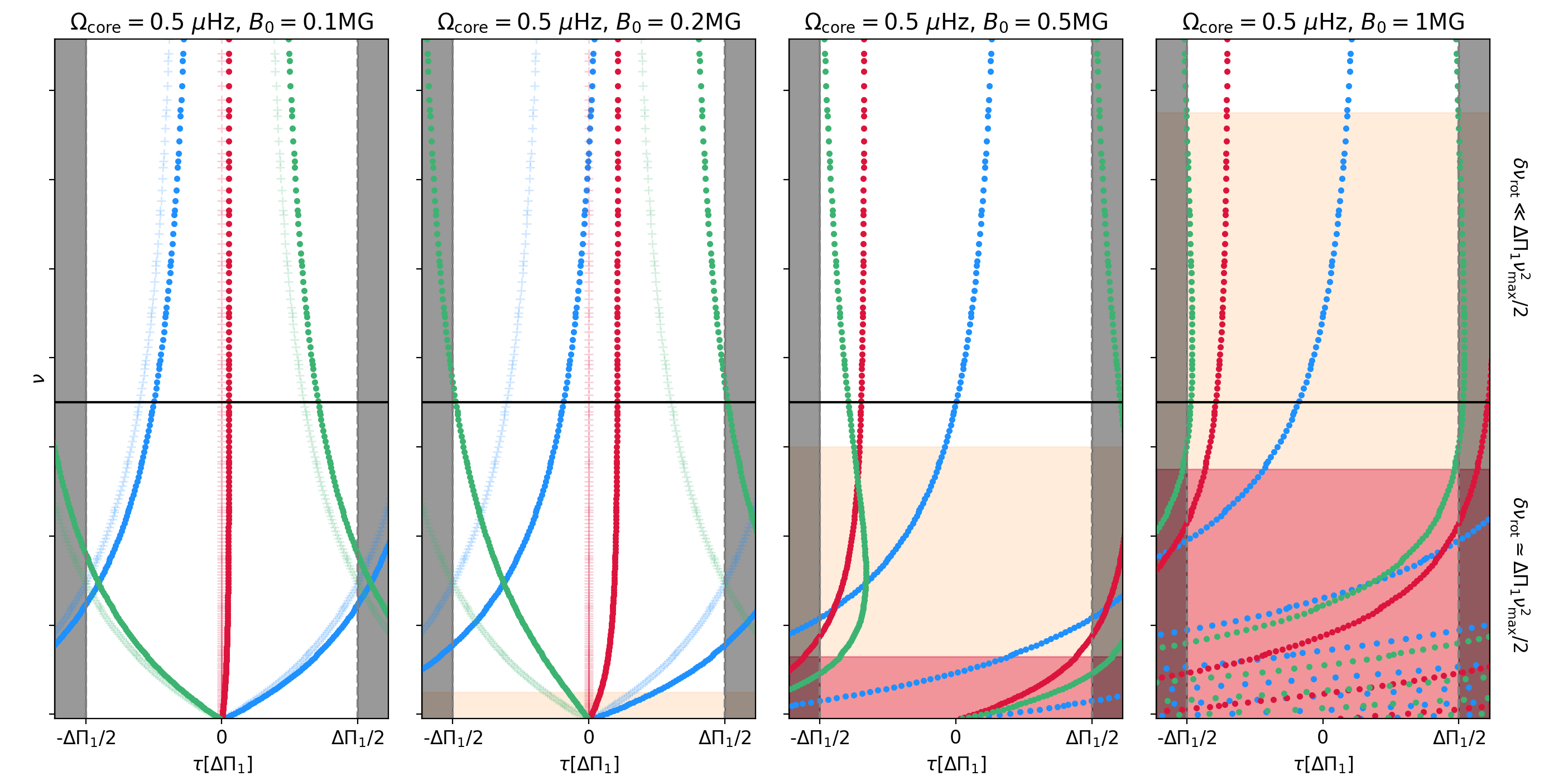}
    \caption{Synthetic stretched period echelle diagram built from Eq.~\eqref{eq:differential_tau}, showing modelled mixed mode frequencies affected by rotation with $\Omega_{\mathrm{core}}=0.5 $\,\si{\micro\hertz} and magnetic-field strengths increasing left to right with $B_0=0.1$ MG, $B_0=0.2$ MG, $B_0=0.5$ MG, and $B_0=1$ MG in the radiative interior. Purely rotationally split components are given for reference by background crosses on the first two panels. Coloured dots indicate the positions of the $\ell=1$ mixed modes of azimuthal order $m \in \{1, 0, -1\}$, respectively blue, red, and green dots. \bld{The horizontal black line delimits domains for which the rotational splitting is small compared to $\Delta\Pi_1 \nu_{\mathrm{max}}^2/2$ \citep[see][for more details]{Gehan2018}. Synthetic patterns above this line are typically representative of patterns of mixed modes for subgiants and early giants, and those below are associated with evolved red giants.} The magnetic effect is small compared to rotational effects in white areas, of the same order of magnitude in orange areas, and larger than $\delta\nu_{\mathrm{rot, core},m}$ in red areas.}
    \label{fig:stretched}
\end{figure*}

As a way of summarising previous results concerning the asymmetry of the perturbed rotational triplet and the amplitude of this perturbation by magnetism, we illustrate the stretched spectrum corresponding to perturbed frequencies of mixed modes in Fig. \ref{fig:stretched}. This visualisation technique was initially developed by \cite{Vrard2015} to estimate the observational period spacing ($\Delta \Pi_1$) of mixed modes, and it is quite often employed to estimate the internal rotation rate of red giants \citep[e.g.][]{Gehan2018}. In these diagrams (see Fig.~\ref{fig:stretched}), $\ell=1$ mixed-mode frequencies are represented by coloured dots (green: $m=-1$ red: $m=0$, blue: $m=1$) as a function of the corrected period $\tau$ modulo $\Delta\Pi_1$, defined via the differential equation:

\begin{equation}
    \mathrm{d} \tau_m=\frac{1}{\zeta}\frac{\mathrm{d}\nu}{\nu^2},
    \label{eq:differential_tau}
\end{equation}
with $\zeta$ defined by Eq.~\eqref{eq:zeta}, and $\nu$ the observational frequency of the mode. For a rotating star without magnetism, the period separation between two mixed-modes of the same azimuthal order $m$ is given by integrating Eq.~\eqref{eq:differential_tau}:

\begin{equation}
    \Delta \tau_m=\Delta\Pi_1\left(1+2\zeta \frac{\delta\nu_{\mathrm{rot, core},m}}{\nu}\right),
    \label{eq:previous_delta_tau}
\end{equation}
with $\delta\nu_{\mathrm{rot, core},m}$ the rotational perturbation of the mode due to the core rotation. We show in Appendix~\ref{sec:delta_tau_with_magnetism_appendix} that this stretched period spacing can be rewritten in the presence of magnetism as
\begin{equation}
    \Delta \tau_m=\Delta\Pi_1\left(1+2\zeta \frac{\delta\nu_{\mathrm{rot, core},m}+\delta\nu_{\mathrm{mag, core},m}}{\nu}\right),
    \label{eq:delta_tau}
\end{equation}
with $\delta\nu_{\mathrm{mag, core},m}$ the magnetic perturbation in the core of the star. The method described by \cite{Gehan2018} to estimate the rotation period of the star can therefore still be applied to magnetised stars due to the similarities between Eqs.~\eqref{eq:previous_delta_tau}~and~\eqref{eq:delta_tau}.  However, the resulting value is no longer an estimate of $\delta\nu_{\mathrm{rot,core}}$ but rather of $\left(\delta\nu_{\mathrm{rot,core}}+\delta\nu_{\mathrm{mag,core}}\right)$. A second analysis step is necessary in order to separate the rotational and magnetism signatures. This is accomplished through the measurement of the asymmetry of the multiplet described in Sect.~\ref{sec:asymmetry}.

In Fig.~\ref{fig:stretched} we represent four stretched spectra. From left to right, we increase the core magnetic field amplitude ($B_0\in\{0.1, 0.2, 0.5, 1\}$ MG). On the first two panels, the positions of non-magnetised modes are represented by faded crosses for comparison. They are constructed following the method described in \cite{Gehan2018}. As in the case of non-magnetised rotating stars, we obtain three ridges, corresponding to the $m\in\{-1,0,1\}$ components of the mixed modes. We can see on the left panel that for a weak magnetic field the rotational ridges are nearly identical to the magnetically influenced ones, with the triplet being slightly shifted towards higher frequencies as expected given our discussion in Sect.~\ref{sec:asymmetry}. The horizontal black line separates two rotational regimes. Below this line the effect of rotation is moderate, where the resulting rotational splitting of the modes is of the order of $\Delta\Pi_1\nu_{\mathrm{max}}^2/2$. This regime is associated with crossing of the three ridges in the stretched spectrum, wherein the rotational splitting leads to the crossing of mixed modes with different $n_g$. In contrast, above the black line, the core rotation rate is small enough for the individual multiplets not to overlap with each other. 

\begin{figure*}[t]
    \centering
    \includegraphics[width=1\textwidth]{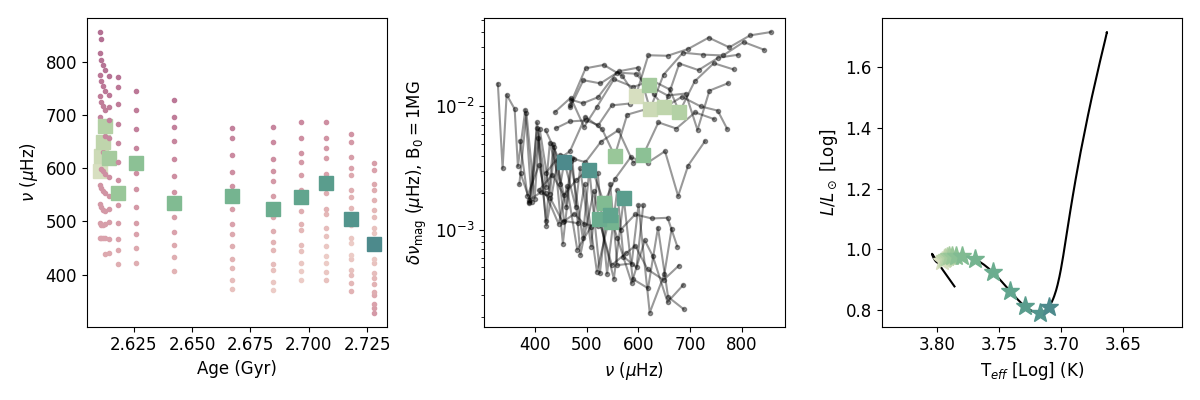}
    \caption{\textsl{Left:} The range of mixed mode frequencies are shown for different evolutionary stages along the subgiant phase (and early \rgb). Those individual stages are indicated on the evolutionary track of plotted in the right panel. Purple-coloured dots represent the individual frequencies of mixed modes, with the shade of the dots indicating the value of the mixed-mode order $n_{pg}$. Squares represent the frequency of the $\ell=1$ g-dominated mode closest to \numax{} for each evolutionary stage. We refer to Fig.~\ref{fig:nu_1g_def} for the determination of this frequency.
    \textsl{Middle:} Magnetic splitting of $\ell=1, m=1$ mixed modes located inside the range $[\nu_{\mathrm{max}}-7\Delta\nu:\nu_{\mathrm{max}}+7\Delta\nu]$ for each of the evolutionary stages denoted on the right panel. Squares represent the frequency of the $\ell=1$ g-dominated mode closest to \numax{} (see Fig.~\ref{fig:nu_1g_def} for further reference).
    \textsl{Right:} The \HR{} diagram of a $M_\star=1.5\mathrm{M_\odot}$, $Z=0.02$ star evolving on the subgiant phase. Coloured stars indicate the different evolutionary stages represented on the other panels.}
    \label{fig:evol_SG}
\end{figure*}

We now construct criteria similar to those in Sect.~\ref{sec:colormap} characterising the magnetic effect on the rotational triplet. By studying the $m=-1$ component of the triplet, one can compare the effect of rotation and magnetism on the frequency of the modes. Indeed, magnetism and rotation have opposite effects on the frequency of the $m=-1$ component. We conclude that when the stretched period of the $m=-1$ mode increases with the frequency of the mode, then magnetic effects dominate rotation effects. We identify three different regimes, depending upon the curvature of the $m=-1$ ridge of the stretched spectrum:
\begin{enumerate}
    \item When the green ridge representing $m=-1$ mode components is convex, the effect of magnetism is negligible relative to rotational effects. This scenario is evinced as the white areas in Fig.~\ref{fig:stretched}.
    \item When $\delta\nu_{\mathrm{mag}}$ approaches $\delta\nu_{\mathrm{rot}}$, the curvature of the $m=-1$ ridge reverses and becomes concave. At that point, magnetism and rotation have comparable effects on mixed-mode frequencies. This scenario corresponds to the orange areas in Fig.~\ref{fig:stretched}.
    \item When the magnetic field is even larger, its effects can dominate the rotational effects. Such instances are rendered as red areas in Fig.~\ref{fig:stretched}. One must take caution in this regime, as the magnetic field amplitudes may be large enough to violate the first-order approximation considered in this paper.
\end{enumerate}

We choose not to indicate the frequency values on the $y$-axes, because the position of the ridges of the stretched spectrum depends on the choice of the integration constant, and thus on the minimum frequency we considered, where we set all $\tau$ values to $0$.

\section{Magnetic effect on mixed mode frequencies of subgiants}
\label{sec:SG}

The magnetic effect on mixed modes inside subgiants is more complicated to study than those occurring on the \rgb, due to the transition from $p-m$ to $g-m$ modes that occurs during this evolutionnary stage. As a consequence, the driving terms listed in Appendix~\ref{sec:terms} are no longer fully represented by Eq.~\eqref{eq:mag_g}, and the mode inertia is no longer simply Eq.~\eqref{eq:massmode_g} during the subgiant stage \citep{Hekker2017a}, leading to variations in the magnetic-splitting patterns.

\subsection{Evolution of the magnetic splitting along the subgiant stage}

The subgiant phase is much shorter than the \rgb{}, especially for intermediate-mass stars. It lasts $\sim 0.1$ Gy for a $M_\star=1.5 \mathrm{M_\odot}$, $Z=0.02$ star, during which the frequency of maximum power varies approximately between $700$ and $500 $\,\si{\micro\hertz}. Figure~\ref{fig:evol_SG} represents the same three panels as Fig.~\ref{fig:evol} at this earlier stage, which is indicated by the position of the considered stars in the right panel. In the left panel, we observe that the value of $\nu_{\mathrm{1,g,max}}$ is no longer monotonously decreasing as it was the case during the \rgb. In the middle panel, the amplitude of the magnetic signature for modes contained in the [$\nu_{\mathrm{max}}-7\Delta\nu:\nu_{\mathrm{max}}+7\Delta\nu$] range are represented for all the considered evolutionary states by the black-dotted lines. Green squares indicate the position of $\nu_{\mathrm{1,g,max}}$. We observe that for evolved subgiants (and early red giants, e.g. stars older than $\sim 2.675$ Gy), the magnetic signature is consistent with the asymptotic theoretical pattern detailed in Sect.~\ref{sec:asympt_g}. However, younger red giants have larger magnetic frequency splittings at $\nu_{\mathrm{1,g,max}}$, and even more substantial signatures for $p-m$ modes, as seen in the middle panel of Fig.~\ref{fig:evol_SG}. We thus investigate the asymptotic regime where high-radial-order $p-m$ modes dominate the frequency spectrum.

\subsection{Analytic expression of the magnetic perturbations for high-radial-order $p-m$ modes}
\label{sec:asympt_p}
In the approximation of acoustic modes, the local radial displacement is much larger than the horizontal one ($\xi_h \ll \xi_r$). The high-radial-order modes have a small wavelength, allowing us to perform an asymptotic JWKB analysis, where terms with high-order $\xi_r$ derivatives dominate.
This approximation is valid for high-frequency, high-order mixed modes, for which the splitting components can be expressed as:
\begin{equation}
\langle \boldsymbol{\xi}_0,\boldsymbol{\xi}_0 \rangle_{p} \simeq  \int_0^R \rho r^2 |\xi_r|^2 \textrm{d}r, 
\label{eq:massmode_p}
\end{equation}

\begin{equation}
\langle \boldsymbol{\xi}_0,\boldsymbol{F_c}(\boldsymbol{\xi}_0) \rangle_{p} \simeq 4 m \omega_0 \int_0^R \rho r^2 \Omega \mathcal{R}e(\xi_r^* \xi_h) \textrm{d}r ,
\label{eq:corioliseffect_p}
\end{equation}

\begin{equation}
\langle \boldsymbol{\xi}_0,\boldsymbol{F_f}(\boldsymbol{\xi}_0) \rangle_{p} \simeq{-} m \omega_0 \int_0^R \rho r^2 \Omega |\xi_r|^2\textrm{d}r,
\label{eq:frameeffect_p}
\end{equation}

\begin{multline}
\langle \boldsymbol{\xi}_0,\boldsymbol{\delta F_L}(\boldsymbol{\xi}_0)/\rho \rangle_{p} \simeq 2 \pi B_0^2 \int_0^R r \xi_r^* \left(b_\theta \left(r\xi_r b_\theta\right)'' +  b_\phi \left(r\xi_r b_\phi\right)''\right) \textrm{d}r \\
\int_0^\pi\left( Y_l^m\right)^2 \sin^3{\theta} \textrm{d}\theta.
\end{multline}

\begin{figure}[t]
    \centering
    \includegraphics[width=.5\textwidth]{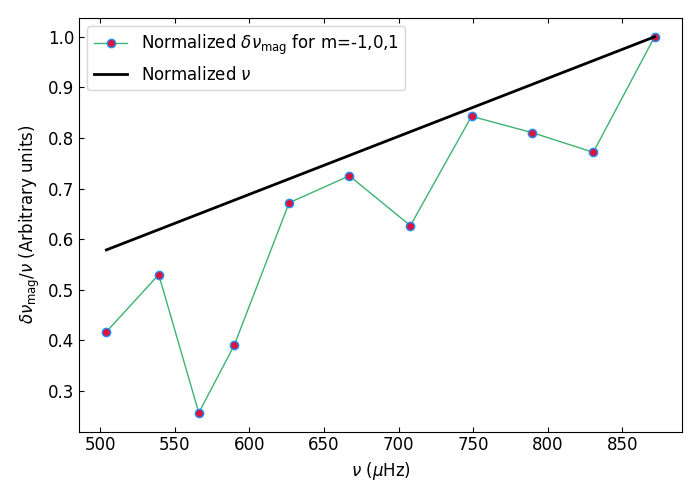}
    \caption{The normalised magnetic splittings of the $p-m$ modes are shown versus mixed-mode frequencies computed for a simulated $M_\star=1.5 \mathrm{M_\odot}$, $Z=0.02$, \numax$\simeq750 $\,\si{\micro\hertz} subgiant. The normalised power law following $\nu$ is superimposed in black.}
    \label{fig:asympt_p}
\end{figure}

\begin{figure*}[t]
    \centering
    \includegraphics[width=1.\textwidth]{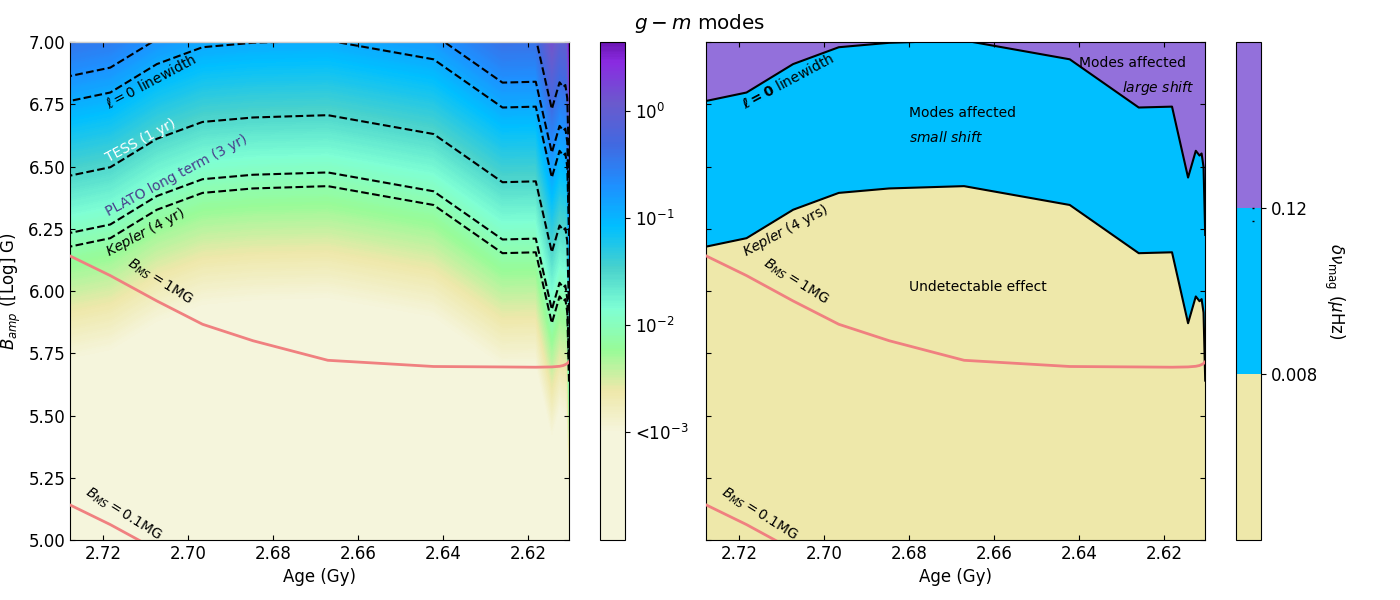}
    \includegraphics[width=1.\textwidth]{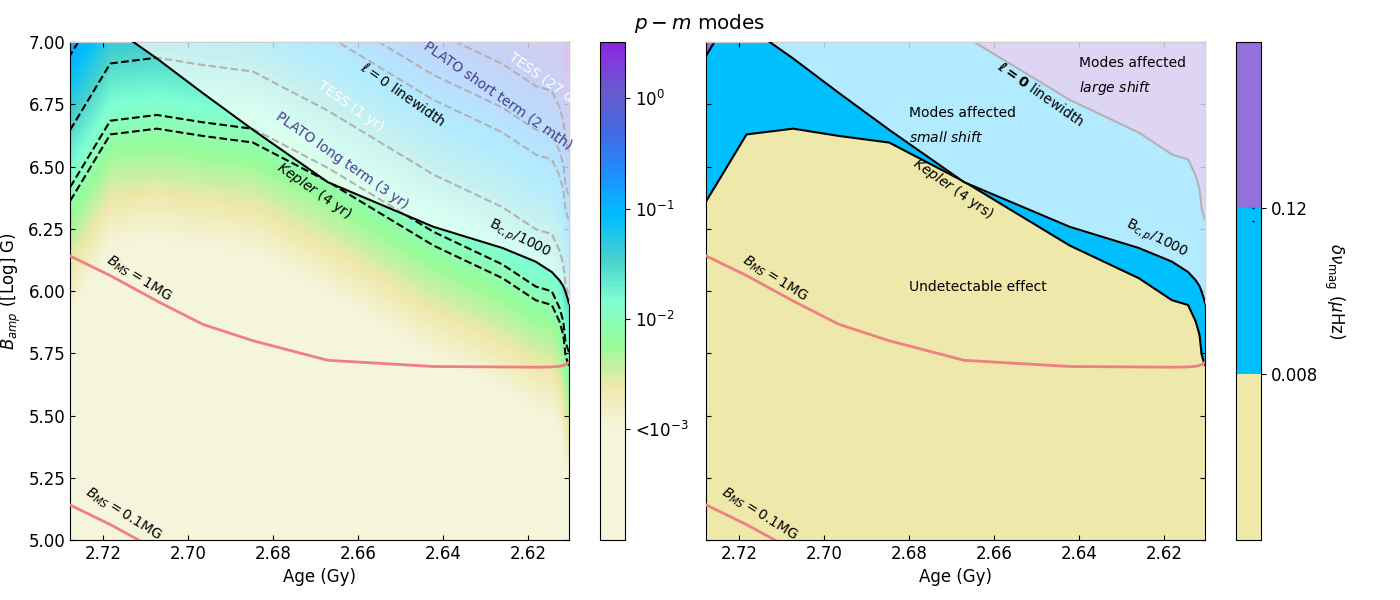}
    \caption{\textsl{Top panels:} Same as Fig.~\ref{fig:nu_1g_def}, for younger stars (\numax$\gtrsim 350 $\,\si{\micro\hertz}) with the abscissa changed to the age of the star (reversed to be consistent with the convention used in Fig.~\ref{fig:nu_1g_def}). \textsl{Bottom panels:} same as top panels but considering $p-m$ modes instead of $g-m$ modes.}

    \label{fig:splitting_SG}
\end{figure*}

\noindent In the case of high-order acoustic modes, the dominant magnetic ratio becomes:
\begin{equation}
    \mathcal{I}_{p}= \frac{\int_0^R{ \left( \left|(r b_\theta \xi_r)'\right|^2 + \left|(r b_\phi \xi_r)'\right|^2 \right) dr}}{\int_0^R{|\xi_r|^2 {\rho}r^2 dr}}.
    \label{eq:I_p}
\end{equation}

\noindent At high frequency, where $\omega_0 \gg S_l \gg N$, Eq.~(\ref{eq:oscillations}) may be written as
\begin{equation}
    \frac{d^2 \xi_r}{dr^2}=\frac{S_l^2}{c_s^2}\xi_r = -k_{\mathrm{r}}^2\xi_r,\,\mathrm{with}\, k_{\mathrm{r}}=-\frac{\omega_0}{c_s}.
\end{equation}
\bld{Taking $S_l=S_{l, \mathrm{min}}\times g(r)$, with $S_{l,\mathrm{min}}$ the minimum value of the Lamb frequency inside the acoustic cavity and $g(r)$ containing all radial dependencies of $S_l$}, the solution can be expressed as
\begin{equation}
    \xi_r \propto \frac{e^{i\int{k_{\mathrm{r}} dr}}}{k_{\mathrm{r}}} \propto \left(\frac{\omega_0^2}{S_{l, \mathrm{min}}^2}\right)^{-1/4}.
\end{equation}
We thus arrive at the following expression for the amplitude scaling of the radial displacement:
\begin{equation}
    \left|\xi_r\right|^2 \propto  \left(\frac{\omega_0^2}{S_{l, \mathrm{min}}^2}\right)^{-1/2}.
\end{equation}

\noindent Thus for a rapidly oscillating radial eigenfunction ($\xi_r' \gg \xi_r$), and $i\in\{\theta, \varphi\}$,
\begin{equation}
    |(r \xi_r b_i)'|^2\propto \left(\frac{\omega_0^2}{S_{l, \mathrm{min}}^2}\right)^{1/2}
\end{equation}

\noindent By employing the scaling for $\xi_r$ in Eq.~\eqref{eq:I_p}, we can see that

\begin{equation}
    \mathcal{I}_{p} \propto \frac{\omega_0^2}{S_{l, \mathrm{min}}^2}
\end{equation}

\noindent and 

\begin{equation}
\delta \omega_{\mathrm{mag,p}} \propto \frac{B_0^2}{S_{l, \mathrm{min}}^2} \omega_0.
\label{eq:asympt_p}
\end{equation}

For high-frequency $p-m$ modes, the magnetic splitting is thus proportional to the unperturbed frequency of the mode at first order. This asymptotic behaviour explains the average rise in the measured $\delta\nu_{\mathrm{mag, core},m}$ with $\nu_{\mathrm{1,g,max}}$ at high frequencies (early subgiants, see Fig.~\ref{fig:evol_SG}). As for red giants, we check this power law for a typical $M_\star-1.5\mathrm{M_\odot}$, $Z=0.02$, $\nu_{\mathrm{max}}=750$\,\si{\micro\hertz}. In Fig.~\ref{fig:asympt_p} the normalised magnetic splitting of the $\ell=1, m\in\{1,0,-1\}$ modes within the $[\nu_{\mathrm{max}}-7\Delta\nu:\nu_{\mathrm{max}}+7\Delta\nu]$ range is represented by the green-dotted line. The subgiant mixed-mode frequency space is less populated, as $p-m$ modes completely dominate the spectrum. We observe a $g-m$ mode among $p-m$ modes at $\sim565 $\,\si{\micro\hertz}. The black line corresponds to the normalised power law describing the $p-m$-mode magnetic signature from Eq.~\eqref{eq:asympt_p}, proportional to the mixed-mode unperturbed frequencies $\nu_0=\omega_0/(2\pi)$. We confirm that the modelled $p-m$ modes follow this $\nu_0$ asymptotic trend well at high frequency. We refer to the paper Mathis et al. (submitted) for a complete study of the asymptotic behaviour of $p-m$-mode frequencies in presence of magnetism.

 \subsection{Detectability of magnetic signature along the subgiant stage}
 
 As we did for stars on the \rgb{} (see Sect.~\ref{sec:colormap}), we now investigate the detectability of such magnetic signatures within the power spectrum density of subgiants. Since this evolutionary stage is a transition period between $p-m$ and $g-m$ modes, we choose to represent the detectability of the two asymptotic regimes $p-m$ and $g-m$ in Fig.~\ref{fig:splitting_SG}. In contrast with Fig.~\ref{fig:splitting_color}, in which we follow the ascension of the \rgb{} by using the $\nu_{\mathrm{1,g,max}}$ proxy, we choose to study the star during the \sg{} stage by following its age, as $\nu_{\mathrm{1,g,max}}$ does not evolve monotonously from consecutive subgiant stages. The top panels of Fig.~\ref{fig:splitting_SG} represent the amplitude of the magnetic splitting (not including rotational effects) affecting $\ell=1, m=1$ $g-m$ modes along the evolution of the star during the \sgb. The abscissa are reversed in order to retain the same orientation as used in Fig.~\ref{fig:splitting_color}. From right to left in the diagrams, we observe a decrease of the magnetic effect on $g-m$ modes as the star evolves along on the \sgb, followed by an increase of the magnetic signature as the star reaches the \rgb. The top-right panel delimits regions of detectability in the (Age, $B_0$) space. For a similar field amplitude, the magnetic signatures of the $g-m$ modes are harder to detect in the power density spectra of subgiants than in those of a red giant. We observe the same change of slope in the color map for $p-m$ modes (see the bottom panels of Fig.~\ref{fig:splitting_SG}), which are the majority among the mixed modes in subgiants. This change is due to the switch of asymptotic regime. In all panels, the red lines delineate the expected magnetic field amplitude evaluated by conservation of the magnetic flux inside the radiative interior from the end of the \ms. By comparing the relative position of these red lines with the frequencies characterising asteroseismic data, if such magnetic field amplitudes ($B_\lesssim1$ MG) are indeed present in the subgiants' radiative interiors, their effect on mixed-mode frequencies is not detectable with data from any current satellite. In order to reach the frequency resolution needed to be able to identify magnetic effects on mixed mode frequencies during the \sgb{} ($\delta f \lesssim 10^{-3}$\,\si{\micro\hertz}), the observation duration must be longer than about $30$ years. Therefore, we do not present further analysis of the magnetic signature on mixed-mode frequencies for subgiants.

\begin{figure*}[t]
    \centering
    \subfloat{{\includegraphics[width=0.49\textwidth]{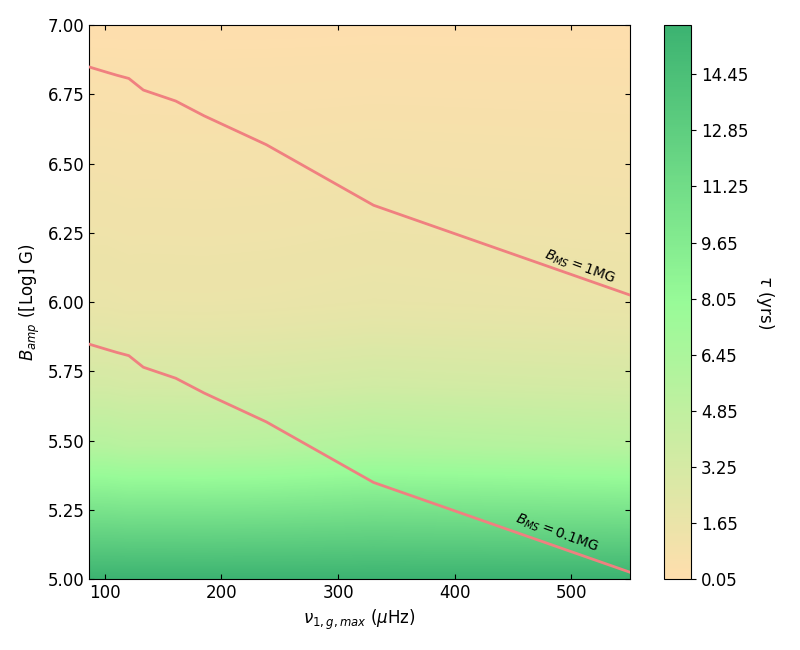} }}
    \subfloat{{\includegraphics[width=0.49\textwidth]{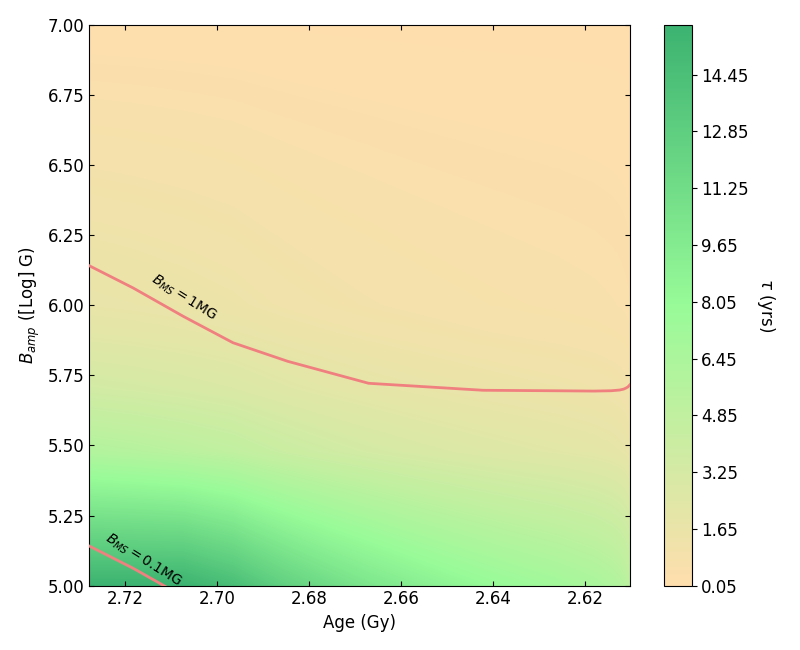}}}
    \caption{The characteristic time of angular momentum transport $\tau$ in years as expressed by Eq.~(\ref{eq:tau_transport}), depending upon the given evolutionary stages and the magnetic field amplitude. The magnetic field amplitude expected at each stage from the conservation of flux from the end of the \ms{} is indicated by red lines, for $B_{MS}=0.1$ and $1$ MG. \textsl{Left:} Transport along the \rgb\ with $\nu_{1,g,max}$ as the abscissa. \textsl{Right:} Transport along the subgiant branch with age as the abscissa.}
    \label{fig:transport}
    
\end{figure*}



\section{Angular momentum transport by \bld{fossil magnetic fields in evolved solar-like stars}}
\label{sec:AM}

The transport of angular momentum inside stars is a consequence of internal dynamical mechanisms. The understanding of the transport of angular momentum inside stars thus leads to an understanding of its global dynamical evolution. Current observational constraints on the transport inside stars begin with an estimation of the profile of their internal rotation rate \citep[e.g.][]{Beck2012}. {Constraining the core rotation rate also helps to estimate the surface rotation rate \citep{Gallet2013, Spada2020}, and the measure of the surface rotation rate provides an estimate of the age of the star during the \ms{} through gyrochronology \citep[e.g.][]{Barnes2003,Barnes2010,Angus2015}}. Stars with magnetic activity may possess signatures of their surface rotation rate in their \psd{} through the periodic reduction of their brightness by a few percent due to magnetised dark spots at the surface \citep[e.g.][]{Mcquillan2013a,Vansaders2013, Mathur2014,Garcia2014a,Ceillier2017,Santos2019}. However, internal rotation rate measurements are more difficult to obtain in general, for instance even the  rotation rate of the core of the Sun has yet to be measured, as it requires the presence of $g$ modes to efficiently probe the deepest layers of the radiative interior \citep[e.g.][]{Garcia2008a,Mathur2008}. In the case of subgiants and red giants that possess mixed modes, the internal rotation rate of a restricted sample of evolved stars has recently been measured \citep[e.g.][Deheuvels et al., \textsl{submitted}]{Deheuvels2012a, Mosser2012, Gehan2018}. As detailed in Sect.~\ref{sec:rotation}, the observed rotation-rate ratio ($\Omega_{\mathrm{core}}/\Omega_{\mathrm{env}}\simeq 5-10$) between the core and the envelope of evolved solar-like stars \citep{Deheuvels2012a, Deheuvels2014a, Mosser2015, Vrard2015, Gehan2018} is not consistent with the strong contraction of the core after the \ms{} that should lead to much higher ratio values \citep[e.g.][]{ Eggenberger2012abis, Eggenberger2019b, Ceillier2013,Cantiello2014a}. A dynamical process must be identified to transport angular momentum from the contracting core towards the envelope in order for this observation to be understood. \bld{In the recent literature that tackles this important question, most of the attention has been given to the potential effects of Maxwell stresses triggered by unstable fields in stably stratified radiative regions \citep[e.g.][]{fuller2019a,Eggenberger2020a,DenHartogh2020,Jouve2020}. Here we recall the potential strong efficiency of a stable axisymmetric field to redistribute angular momentum along poloidal field lines, a well-known result since \cite{Ferraro1937,Mestel1987}. We aim to provide a quantitative estimate of the characteristic time scale for such a field to redistribute angular momentum for field amplitude that could be detected in seismic data. As such, we are closing the loop: we are looking for seismic signatures of a potential axisymmetric fossil field as a potential candidate to explain the strong angular momentum transport revealed by the observed weak surface-to-the-core rotation contrast. Given the amplitude of the field that we predict to be able to detect, we compute the characteristic time scale on which it redistributes angular momentum. If it is very short compared to evolution time scales, as expected, this confirms that it is an excellent candidate for the observed strong extraction of angular momentum.}  

By considering a large-scale axisymmetric fossil magnetic field trapped inside the radiative interior of evolved solar-like pulsators, we investigate the impact of such magnetism on the internal rotation profile of the star \citep[see e.g.][for previous studies on the Sun]{Mestel1987, Charbonneau1993}. We consider the field lines to be closed, without any re-connection with the dynamo-generated field in its convection zone. Such configurations are depicted in Fig.~\ref{fig:field}, and following \cite{Mestel1987}, consider the following \bld{poloidal} current due to the \bld{toroidal component of the} axisymmetric magnetic field:

\begin{equation}
    \boldsymbol{j_p}=\frac{1}{4\pi r}\boldsymbol{\nabla}(rB_\varphi\boldsymbol{e_\varphi})\wedge \boldsymbol{e_\varphi}.
\end{equation} 
The toroidal component of the equation of induction can be written 
\begin{equation}
    \frac{\partial B_\varphi}{\partial t}=r\sin{\theta}(\boldsymbol{B_p}\cdot \boldsymbol{\nabla})\Omega,
\end{equation}
where $\boldsymbol{B_p}$ is the poloidal magnetic field vector. The current has a component perpendicular to $B_p$, so it exerts a torque that changes the rotation profile according to
\begin{equation}
\rho r^2 \sin{\theta}\frac{\partial \Omega}{\partial t}=r\boldsymbol{j_p}\wedge \boldsymbol{B_p}\cdot\boldsymbol{e}_\varphi.
\end{equation}
The rotation profile is thus affected in time according to the toroidal component of the momentum equation 
\begin{equation}
\rho r^2 \sin{\theta}\frac{\partial \Omega}{\partial t}=\frac{1}{4\pi}\boldsymbol{B_p}\cdot\boldsymbol{\nabla}(rB_\varphi).
\end{equation}
Assuming no turbulence, and given that the Ohmic diffusion time scale due to atomic processes is \bld{very large}, 
changes to the poloidal component of the magnetic field \bld{can be neglected at the leading order}. 
This leads to the partial differential equation \bld{that describes the redistribution of angular momentum along the poloidal field lines by Alfv\'en waves}:
\begin{equation}
    \frac{\partial^2\Omega}{\partial t^2}=\frac{1}{4\pi\rho r^2}\boldsymbol{B_p}\cdot\boldsymbol{\nabla}\left(r^2 \boldsymbol{B_p}\cdot\boldsymbol{\nabla}\right)\Omega.
\end{equation}
{If the variation of $\Omega$ along the magnetic field lines of $\boldsymbol{B_p}$, which define the coordinate $s$, is small with $r$ and with the scale of variation of $\boldsymbol{B_p}$, then}
\begin{equation}
    \frac{\partial^2\Omega}{\partial t^2}=\frac{{B_p}^2}{4\pi\rho}\frac{\partial^2\Omega}{\partial s^2}.\label{eq:omega_transport}
\end{equation}
From this equation, Alfv\'{e}n waves transport angular momentum 
leading to the Ferraro iso-rotation law $\left(B_p\cdot\nabla\Omega\right)=0$ where rotation becomes constant along the poloidal field lines since they are considered to be fixed in time. Given the defintion of our axisymmetric magnetic field, $\boldsymbol{B_p}$ scales with $B_0$, and as the magnetic field extent covers all the radiative interior delimited by $R_{\mathrm{rad}}$, the characteristic time for angular momentum transport derived from Eq.~\eqref{eq:omega_transport} is equal to the Alfv\'{e}n time:
\begin{equation}
    \tau=\frac{R_{\mathrm{rad}}}{v_A},
    \label{eq:tau_transport}
\end{equation}
with $v_A$ the Alfv\'{e}n speed defined by Eq.~\eqref{eq:Alfvén_speed}. As shown in \cite{Ferraro1937}, \cite{Mestel1953}, and \cite{Mestel1987}, this characteristic time to flatten the rotational profile of the radiative interior is very short.

Figure~\ref{fig:transport} evaluates the characteristic time for magnetism to flatten the rotational profile of the radiative interior of the reference star along the \rgb{} (left panel) and along the \sgb{} (right panel), as a function of $\nu_{\mathrm{1,g,max}}$ and age, respectively and as a function of the magnetic field amplitude ranging from $0.1$ to $10$ MG. As expected, a greater magnetic amplitude implies a shorter characteristic transport time. We do not observe any significant impact of \bld{the evolutionary stage of the star on the transport characteristic} time during the \sgb{} and \rgb. In any case, the characteristic time to flatten the radiative interior of the star is of the order of the year and thus such considered magnetic-field amplitudes (as represented by red lines in Fig.~\ref{fig:transport}) are very efficient to transport angular momentum inside the radiative region. With this order-of-magnitude analysis we reconfirm that magnetism is a very good candidate for rapidly transporting angular momentum after the \ms. It may actually be too powerful to maintain the slight differential rotation rate observed inside evolved solar-like stars \citep{Eggenberger2012abis, Eggenberger2017bis, Cantiello2014a}.

 The isolated-field scenario used for these calculation is however a strong assumption; this should be discussed. It has been proposed that the primordial magnetic field buried inside the radiative zone of the Sun inhibits the spread of the tachocline \citep[e.g.][]{Rudiger1997, Gough1998,Barnes1999}. Such confined magnetism can explain the quasi-uniform rotation rate of the radiative interior of the Sun. However, \cite{Brun2006a} and \cite{strugarek2011} showed that we may expect the confined magnetic field to spread by Ohmic diffusion towards the envelope, and to eventually reconnect with the convection zone dynamo field. Due to Ferraro's law of iso-rotation, the latitudinal differential rotation of the convection zone would then  imprint on the radiation zone. Such a phenomenon is however not observed in the radiative interior of the Sun. On the other hand, from a more theoretical standpoint, it has been argued that with a proper ordering of time scales of the dynamical processes, this spread can be prevented \citep[e.g.][]{Wood2011, Acevedo-Arreguin2013}. Indeed, this question of tachocline confinement and its role in global dynamics is still an open question; it is nevertheless still topical for the radiative interior of red giants. If the fossil field reconnects with the surface dynamo field, we expect the radiative interior to undergo a small differential rotation despite the strong magnetic field.

\section{Dependency of the magnetic splitting on stellar parameters}
\label{sec:dependency}

\begin{figure*}[t]
    \centering
    \includegraphics[width=1\textwidth]{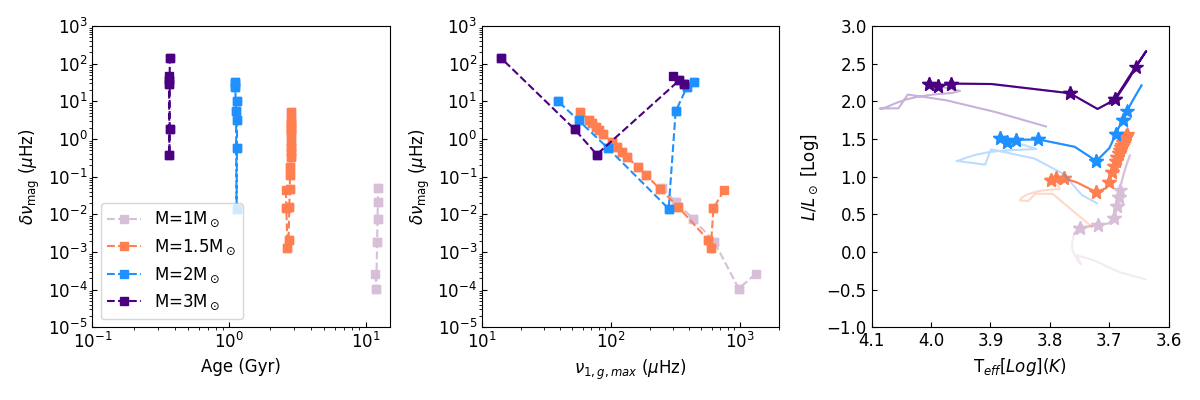}
     \caption{\textsl{Left:} Magnetic frequency splitting calculated at the central mode frequency $\nu_{\mathrm{1,g,max}}$ versus the age for different evolutionary stages reported on the right panel for Z=0.02, $M_\star\in\{1, 1.5, 2, 3\}\mathrm{M_\odot}$.
     \textsl{Middle:} Magnetic frequency splitting calculated at the central mode frequency $\nu_{\mathrm{1,g,max}}$ for different the different evolutionary stages and masses reported on the right panel. \textsl{Right:} Hertzsprung-Russell diagram corresponding to the colour-coded Z=0.02, $M_\star\in\{1, 1.5, 2, 3\}\mathrm{M_\odot}$ stars. The star symbols indicate the position of the star at which measurements are reported on left and middle panels.}
    \label{fig:mass_effect}
\end{figure*}

\begin{figure*}[h]
    \centering
    \includegraphics[width=1\textwidth]{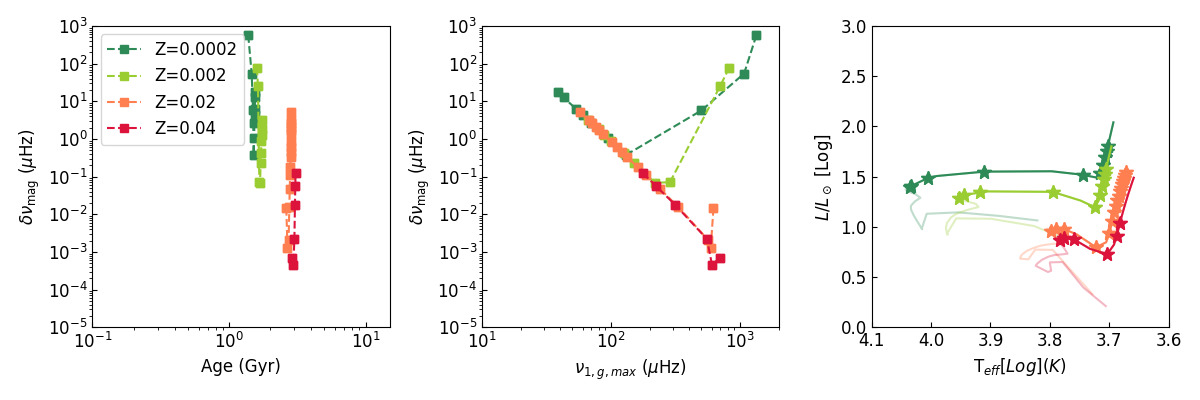}
     \caption{\bld{Same as figure~\ref{fig:mass_effect} but with varying metallicity; we represent $M_\star=1.5\mathrm{M_\odot}$, Z$\in\{0.04, 0.02, 0.002,0.0002\}$ stars.}}
    \label{fig:metal_effect}
\end{figure*}

In previous sections, we derived our study from mixed modes of a typical $M_\star=1.5 \mathrm{M_\odot}$, Z=0.02 star, during its evolution from the base of the subgiant stage towards the top of the \rgb. In the following section, we explore the impact of a change of mass and metallicity on previous results. From theoretical studies, stellar-evolution models, and observations, we know that massive stars evolve on shorter timescales than less massive ones. It is also well known that {metal-rich} stars live longer. We follow the evolution of stars with different initial masses ($M_\star\in\{1, 1.5, 2,3\} \mathrm{M_\odot}$) and different metallicities (Z$\in\{0.0002, 0.002, 0.02, 0.04\}$) in the Hertzsprung-Russell diagram in right panel of Figs.~\ref{fig:mass_effect}~and~\ref{fig:metal_effect}. In the left panels, for each mass we observe the evolution of the magnetic splitting $\delta\nu_{\mathrm{mag, core},m}$ of the central $g-m$ mode as the star evolves. 

\subsection{Mass dependency}

\bld{To investigate the effect of the mass of the star on the magnetic splitting, we consider low- and intermediate-mass solar-type oscillators from the subgiant towards the red giant phases, for which the convective envelope is thick enough for acoustic modes to be excited at the surface.} 
From the left panel of Fig.~\ref{fig:mass_effect}, it takes around $12.5$ Gy of the $M_\star=1 \mathrm{M_\odot}$ star to ascend the \rgb, whereas the $M_\star=3 \mathrm{M_\odot}$ reaches this stage in less than $1$ Gy. As a first result, significant magnetic signatures arise on frequency spectra at an earlier time for more intermediate-mass stars than for lighter ones. This first effect is due to the fact that intermediate mass stars evolve from the \sg{} to the \rg{} stage quicker than low-mass stars. 

\bld{We also represent in the middle panel the value of the magnetic splittings as a function of the dominant $g-m$ mode frequencies $\nu_{\mathrm{1,g,max}}$. Intermediate-mass stars show the transition between the $p-m$- and $g-m$-dominated mixed modes (see Fig.~\ref{fig:evol_SG}) at lower frequencies than low-mass stars. We interpret this mass dependency as follows: the frequency range inside which we can detect mixed-mode form is set by the frequency range of acoustic modes that propagate all over the star, strongly correlated with the size of the star. As a result, acoustic modes propagating inside intermediate-mass stars, bigger than low-mass stars, have the lowest eigenfrequencies for a given evolutionary stage. In other words, the frequency associated with the base of the \rgb{} where $g-m$ modes dominate is lower for intermediate-mass stars than for low-mass stars. The mass difference however does not impact much the magnetic frequency-splitting values for a given $g-m$ mode frequency. The small mass dependency seen at low frequency on the middle panel comes from the mixed-mode nature of the mode, which is more dominated by its acoustic nature for massive stars than for low-mass stars at a given frequency on the \rgb. The consequence is a slightly smaller effect of magnetism on mixed-mode frequencies for more massive stars during the \rgb.}


\subsection{Metallicity dependency}

In Fig.~\ref{fig:metal_effect} the same three panels as in Fig.~\ref{fig:mass_effect} are represented, with the change in mass replaced by a change in metallicity from Z=0.0002 to Z=0.04, for a fixed $M_\star=1.5 \mathrm{M_\odot}$. Without any surprise, the more metallic the star, the slower it evolves as seen in the left panel. The effect of metallicity on the magnetic signature at a given mixed-mode frequency of the stars is negligible during the \rgb{}. However, the transition from \sg{} to \rg{} arises at lower frequencies for low-metal stars. It leads to a large spreading of magnetic-signature values at a given frequency (see the middle panel of Fig.~\ref{fig:metal_effect}), similar to what is observed when varying the mass of the star shown in the middle panel of Fig.~\ref{fig:mass_effect}. 

\section{Comparison with low-amplitude dipolar mixed-modes critical field}
\label{sec:depressed}
The magnetic green house effect, as proposed by \cite{Fuller2015} and supported by the study of \cite{Lecoanet2017b}, consists in a complete transfer of energy of magnetised gravity waves towards Alfvén waves, geometrically trapped inside the core of the star. The resulting oscillation mode is then purely acoustic, with a great loss of power compared to the corresponding mixed mode due to the loss of the gravity wave energy trapped inside the core. Such mechanism implies the total disappearing of $g$ components in the $\ell=\{1,2\}$ modes regions in the \psd. \cite{Mosser2017} thus led an observational study, looking for signatures of mixed-mode residuals inside low-amplitude $\ell=1$ regions. The complete rotational mixed-mode frequency pattern is adjusted: it reconstructs well the observed $\ell=1$ pattern. The authors conclude that even for stars showing low-amplitude inside the $\ell=1$ region, the remaining oscillation modes have a mixed $p$ and $g$ nature. \cite{Loi2019} however set a warning about the physical conditions under which the suppression of mixed-mode amplitude may arise along with the disappearance of the $g$ components: the authors support that interactions between gravity modes and magnetic fields may result in various behaviours, depending on the configuration and strength of the magnetic field, and of the star's structure and stratification. This study brings perspectives concerning the validity of the theory proposed by \cite{Fuller2015}, which may be adapted for mixed-mode amplitude suppression by magnetism without loosing all the mixed nature of the modes. 

This topic being very controversial, we do not intend to decide whether or not the green house effect is the key to mode suppression. However, we provide the comparison of the critical magnetic-field amplitude needed for mode suppression as fixed by \cite{Fuller2015} with the minimum field amplitudes needed for the magnetic signature of mixed-mode frequencies to be detectable in observational data. Table~\ref{tab:comparison_amplitudes} contains the approximated values of the magnetic field corresponding to the minimum detection threshold inside data from Sect.~\ref{sec:colormap} ($B_{\mathrm{min}}$) and of the critical field needed for mode suppression to occur according to the theory by \cite{Fuller2015}. Results depend on the duration of observation, and on the considered frequency of the modes. In any case, we observe that as the star evolves (corresponding to a decrease of \numax) the two critical magnetic-field amplitudes are of the same order of magnitude, whereas the critical field for mode suppression is much larger for less evolved red giants. It means that if magnetic suppression by green house mechanism is at work inside \rg s, one should also observe magnetic signatures on mixed-mode frequencies, especially for young \rg s \citep[see also the work of][]{Rieutord2017}.

\begin{table*}[t]
    \centering
    \begin{tabular}{|l|c|c|c|c|c|}
    \hline
         Instrument&  time (yrs)& $M_\star$&\numax{} (\si{\micro\hertz})& $B_{\mathrm{min}}$ (MG) & $B_{c, \mathrm{low-amplitude}}$ (MG)\\
         \hline
         \hline
         \kepler&  4 & 1.5 & 300& 0.50& 3\\
         \kepler&  4 & 1.5 & 200& 0.25& 1\\
         \kepler&  4 & 1.5 & 100& 0.10& 0.1\\
         \hline
         PLATO&  3 & 1.5 & 300& 0.6& 3\\
         PLATO&  3 & 1.5 & 200& 0.35& 1\\      
         PLATO&  3 & 1.5 & 100& 0.11& 0.1\\   
         \hline
         TESS&  1 & 1.5 & 300& 1.2& 3\\
         TESS&  1 & 1.5 & 200& 0.55& 1\\      
         TESS&  1 & 1.5 & 100& 0.20& 0.1\\
         \hline
    \end{tabular}
    \caption{Comparison of the critical field $B_{c, \mathrm{low-amplitude}}$ needed  for mode suppression by the green-house effect \citep{Fuller2015} and the minimum magnetic field amplitude leading to frequency shifts of about the frequency resolution inside data. }
    \label{tab:comparison_amplitudes}
\end{table*}

\section{Discussion \& Perspectives}
\label{sec:discussion}

In this study, we investigate the effects of an axisymmetric mixed poloidal and toroidal fossil field aligned with the rotation axis of the star on mixed-mode frequencies during the \sg{} and \rg{} evolutionary stages. We concentrate on the fossil field scenario, in which a buried magnetism is present inside the radiative interior of the stars. This field results from the relaxation of a magnetic field generated from previously active dynamos. By using different force and energy balance regimes, we show that the expected fossil magnetic-field amplitude stabilised at the end of the \ms{} should be in the range of $[0.1-1]$ MG. This range is independent of the mass of the star, and of the presence or not of a sustainable convective core during the \ms. Such field amplitudes are large enough for our study of the effects of magnetism on mixed-mode frequencies to be interesting. Indeed, our perturbative analysis of mixed-mode frequencies of evolved solar-like stars provide constraints for the expected field amplitudes inside evolved stars. Such field amplitudes during the \rgb{} may permit the detection of magnetic signatures on $g-m$ mixed-mode frequencies. Specifically, the amplitude of magnetically induced frequency shifts is of about $\sim0.1 $\,\si{\micro\hertz} during the \rgb{} when considering field amplitudes between $0.1$ and $1$ MG, and this shift is larger than the frequency resolution of long-term observations by the \kepler, TESS, and PLATO missions. Depending on the nature ($p-m$ or $g-m$) of the mixed mode and of its frequency, the magnetically induced frequency shift may as well exceed the typical linewidth of the mode. Along the \rgb, the effect of magnetism on $g-m$ modes, which propagate inside the magnetised radiative interior, is larger than on $p-m$ modes, which also propagate inside the convective envelope. This difference between $g-m$ and $p-m$ modes is also visible when evaluating the effect of the rotation of the star, where $p-m$ modes probing the more slowly-rotating envelope are less affected by rotation than $g-m$ modes that primarily provide information about the more rapidly rotating core. Our results thus mostly concern the effect of magnetism on $g-m$ modes, despite the fact that they are more difficult to observe than $p-m$ modes due to their relatively low amplitude in current data sets. 

During the \sg{} stage however, $p-m$ modes dominate the observed spectra, and they are more affected by magnetism than $g-m$ modes. It can be understood by the large extent of the radiative interior: $R_{\mathrm{rad}}\rightarrow R_\star$ during the \sg{} phase while $R_{\mathrm{rad}}\rightarrow 0$ during the \rgb. As a result, $p-m$ modes probe more magnetised plasma than $g-m$ modes during this phase. Fossil magnetism resulting from active dynamos before or during the main-sequence is however not strong enough throughout the \sgb{} to produce detectable changes neither on their $p-m$ nor $g-m$ mixed-mode frequencies. We do not exclude the possibility of having stronger fields than the one obtained by the balanced regimes.  For example, \cite{Fuller2015} proposed a mechanism that uses the frequency of magneto-gravity waves to estimate field amplitudes as large as 10 MG in KIC8561221 during the \sg{} phase. 

The two $g-$ and $p-$ dominated mode regimes correspond to two different asymptotic cases: $\omega\ll N\ll S_l$ for $g-m$ modes, and $N\ll S_l \ll \omega$ for $p-m$ modes. Asymptotic power laws corresponding to the magnetic effect on $p-m$ and  $g-m$ mode frequencies are provided, where low-frequency $g-m$-mode magnetic splittings behave as $\delta\omega_{\mathrm{mag},g}\sim1/\omega^{-3}$ whereas $p-m$-mode magnetic splittings behave as $\delta\omega_{\mathrm{mag},p}\sim\omega$. These asymptotic power laws possess amplitude scaling factors, and correlate with the $\zeta$ mode-coupling function. Complete expressions for these are derived in the paper Mathis et al., (\textsl{submitted}). Considering the typical balanced field amplitudes, we argue that, as for rotational perturbations, these first-order expressions suffice to ascertain the effect of magnetism on mixed-mode frequencies. As the star evolves on the \rgb, the effect of magnetism on mixed-mode frequencies gets larger, where very evolved red giants or \agb{} stars with $\nu_{\mathrm{max}}\lesssim100 $\,\si{\micro\hertz} may be non-perturbatively impacted by magnetism. For such evolved stars, we refer to the study of \cite{Loi2020}.

We show that the considered axisymmetric magnetism, aligned with the rotation axis of the star, acts as a new perturbation of the already present rotational multiplet. The magnetised mixed multiplet with order $n_{pg}$ is made asymmetric by the presence of a magnetic field: all $\ell=1$ and $\ell=2$ multiplet components are shifted towards higher frequencies, with the amplitude of the shifts depending on the azimuthal order $m$ of the mode. We argue that this asymmetry can be distinguished from other asymmetry sources such as non-degenerate effects. Adjustment methods such as described in \cite{Vrard2015},  \cite{Mosser2015}, and \cite{Gehan2018} may be adapted to investigate magnetic asymmetries. If the magnetic effect is small compared to the rotational impact on mixed-mode frequencies, the multiplet is simply shifted towards higher frequencies. However, we may observe crossings of the components between successive $n_{pg}$ mixed-mode multiplets if the magnetic field amplitude is sufficiently large, yet a strong core rotation also yields similar crossings.

If such isolated fossil fields exist inside the radiative interior of evolved stars, they would also affect the rotational profile of the radiative zone: we show that within the assumptions of \cite{Mestel1987}, the rotation is frozen to poloidal field lines within a few years, eventually leading to an almost flat rotational profile inside the radiative interior. A reconnection of the fossil and dynamo-generated magnetic fields in the convection zone at the tachocline may however permit a small differential rotation to persist inside the radiative interior. 

Under the hypothesis that $g$ modes may be trapped inside the core of the star as described by \cite{Fuller2015}, we also show that magnetic signatures may be detectable in the star's frequency pattern before the complete suppression of the mode amplitudes, especially in the case of young red giants. This result is of great interest since the magnetic signature within mixed-mode frequencies may appear in the \psd{} with $\ell=1$ modes of normal amplitude.




\begin{acknowledgements}
    \bld{We thank the referee for very useful and detailed comments that allow to improve the quality of our study and the article.} L. Bugnet, V. Prat, S. Mathis, A.A stoul, and K. Augustson acknowledge support from the European Research Council through ERC grant SPIRE 647383. All CEA members acknowledge support from GOLF and PLATO CNES grants of the Astrophysics Division at CEA. S. Mathur acknowledges support by the Ramon y Cajal fellowship number RYC-2015-17697. L. Amard acknowledges funding from the European Research Council (grant agreement No. 682393 AWESoMeStars). We made great use of the megyr python package for interfacing MESA and GYRE codes.In honour of our dear friend and colleague Michael J. Thompson. À la plus belle étoile de mon ciel.
\end{acknowledgements}

\bibliographystyle{aa}  
\bibliography{references.bib}
\newpage
\begin{appendix}
\section{Magnetic scaling-law regimes}
\label{sec:appendix_magnetostrophy}

By considering the fluid in the rotating frame under the effect of magnetism (via the Lorentz force) and rotation (via the Coriolis acceleration), assuming stationarity and neglecting viscosity \bld{and the centrifugal acceleration}, the Navier-Stokes equation in a convection zone may be written as:

\begin{equation}
 \underbrace{\left(\boldsymbol{v}\cdot \boldsymbol{\nabla}\right){\boldsymbol{v}}}_{\text{Advection}} = -\underbrace{\frac{\boldsymbol{\nabla}P}{\rho}}_{\text{Fluid pressure}} - \underbrace{2\boldsymbol{\Omega}\wedge\boldsymbol{v}}_{\text{Coriolis}} +\underbrace{\frac{1}{4\pi\rho}(\boldsymbol{\nabla}\wedge\boldsymbol{B})\wedge \boldsymbol{B}}_{\text{Lorentz force}}  + \underbrace{\boldsymbol{g}}_{\text{Gravity}},
 \label{eq:NS}
\end{equation}
with $\boldsymbol{g}$ the gravitational acceleration.

For this equation several force balances \bld{allow us to evaluate the amplitude of the magnetic field in different regimes. The balance between forces is a partitioning that depends upon the convective Rossby number. It is more subtle than a typical asymptotic analysis because it is meant to encompass three convective Rossby number regimes smoothly:
\begin{itemize}
    \item At high Rossby number (Ro$\gg 1$) flows are weakly rotationally constrained but possess a strong dynamo, 
    \item For Ro closes to one flows are modestly rotationally constrained, again with a strong dynamo,
    \item For low Ro (Ro$\ll 1$), the magnetostrophic regime is reached, corresponding to rotationnally constrained and magnetized flows.
\end{itemize}
  It is found in \cite{Augustson2019}, that the partitioning of forces is dominated by combinations of the inertial terms (Reynolds stresses), the Coriolis acceleration, and the Lorentz force (or Maxwell stresses) for all Rossby numbers. However, depending upon Rossby number regime, the relative influence of each of these three strongest forces changes. }

\bld{The force balance described in \cite{Augustson2019} fits the available simulation data well when the force balance is described as }

\begin{equation}
    I + C + L \approx 0,
\end{equation} 

\noindent \bld{where $I$ is the magnitude of the inertial forces, $C$ that of the Coriolis force, and $L$ that of the Lorentz force. Using a scaling argument about the dominant length scales, velocity and magnetic field amplitudes at that scale, one can show that in this case}

\begin{equation}
    \mathrm{ME}/\mathrm{KE} \approx a + b/\mathrm{Ro},
\end{equation}

\noindent \bld{where Ro $\approx v_l / (2 \Omega_0 l$) and $l$ is that dominant length scale. $a$ describes the efficiency of a small-scale local dynamo that depends only on the helical turbulence generated by convection, and $b$ refers to how close to perfect magnetostrophy the system is in. In this framework, three specific regimes can be reached:}

\begin{itemize}
    \item The magnetostrophic regime \bld{(low Rossby number)} is reached when the Lorentz force balances the Coriolis acceleration and other forces are much smaller in magnitude. This regime can be achieved for a sufficiently high rotation rate. From Eq.~\eqref{eq:NS}, we deduce that 
    \begin{equation}
        B\simeq\sqrt{8\pi\Omega\rho R_{\mathrm{conv}} v_{\mathrm{conv}}},
    \end{equation}
    with $R_{\mathrm{conv}}$ the size of the convective zone, and $v_{\mathrm{conv}}$ the convective velocity. This can be rewritten in terms of convective Rossby number $\mathrm{Ro}=v_{\mathrm{conv}}/(2\Omega R_{\mathrm{conv}})$ and kinetic energy density associated with convective motions $\mathrm{KE}=\rho v_{\mathrm{conv}}^2/2$ as:
    
    \begin{equation}
        B\simeq \sqrt{\frac{8\pi \mathrm{KE}}{\mathrm{Ro}}}.
    \end{equation}
    
    \item Dynamo action converts a fraction of the kinetic energy of the convective motions into magnetic energy. In the equipartition regime \bld{(high Rossby number regime where the Coriolis acceleration can be neglected)}, the assumption is thus that the dynamo is efficient enough such that the magnetic energy density $\mathrm{ME}=B^2/(8\pi)$ balances the convective kinetic energy density $\mathrm{KE}$ of the fluid:
    \begin{equation}
        B\simeq\sqrt{8\pi\mathrm{KE}}.
    \end{equation}
    
    \item \bld{In the modest Rossby number regime, all three forces have roughly the same magnitude.}
\bld{So, ultimately, in the Rossby number regime that is close to unity, all three forces play nearly equal roles. }By neglecting inertial forces, considering constant density, and with $\boldsymbol{\nabla}.\boldsymbol{v}=0$, the curl of Eq.~\eqref{eq:NS} can be written
    \begin{equation}
\left(\boldsymbol{\Omega}.\boldsymbol{\nabla}\right)\boldsymbol{v} =\frac{1}{4\pi\rho}\boldsymbol{\nabla}\wedge\left((\boldsymbol{\nabla}\wedge\boldsymbol{B})\wedge \boldsymbol{B} \right) + \boldsymbol{\nabla}\wedge\boldsymbol{g}.
\label{eq:curl}
    \end{equation}
    
    Comparing the three terms composing Eq.~\eqref{eq:curl}, the balance is reached for \begin{equation}
    B\simeq \sqrt{\frac{8\pi\mathrm{KE}}{\sqrt{\mathrm{Ro}}}}.
\end{equation}


We refer to the \cite{Augustson2019} and \cite{Astoul2019} studies for additional details on the estimation of the buoyancy dynamo field strength. This field has an intermediate value between the equipartition and magnetostrophy regimes, which set the upper and lower boundaries of the expected magnetic amplitude during episodes of internal convective dynamo action.
\end{itemize}

  \section{Magnetic field stable topology}
  \label{appendix:mag_topo}

 \begin{figure*}[ht]
     \centering
     \includegraphics[width=0.7\textwidth]{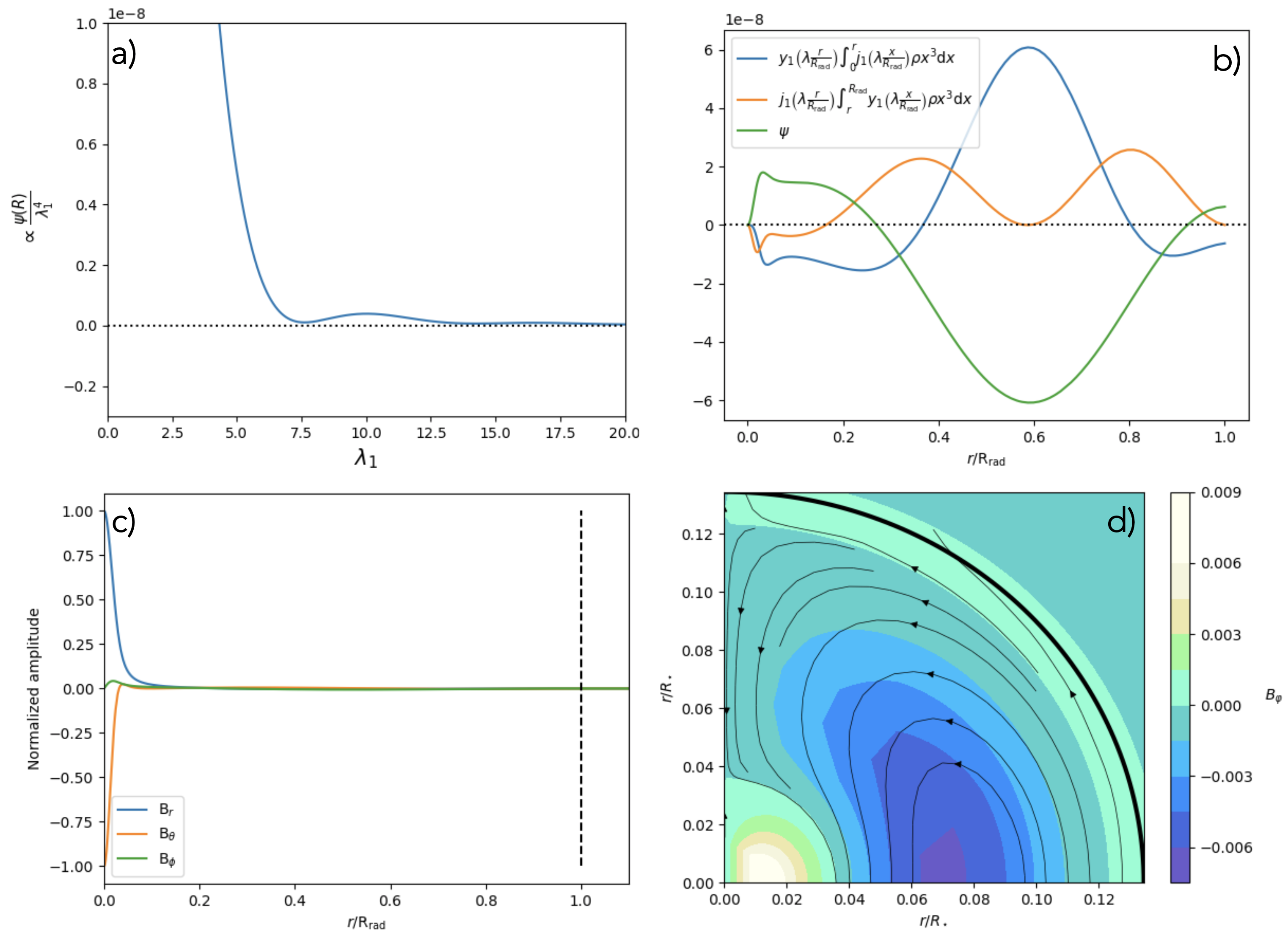}
     \caption{\bld{Attempt to find the first zero of $\int_0^r j_1\left(\lambda \frac{x}{R_{\mathrm{rad}}} \right)\rho x^3 \textrm{d}x$ at the radiative/convective boundary. We consider the typical red giant with $M=1.5\mathrm{M}_\odot$, $Z=0.02$ on the \rgb. \textsl{Panel a):} Trend of the evolution of the $\Psi$ function (normalized by $\lambda_1^4$ for convenience) with the first eigenvalue $\lambda_1$. \textsl{Panel b):} Values of the functions with $\lambda_1\simeq7.62$. \textsl{Panel c):} Magnetic field components with $\lambda_1\simeq7.62$. \textsl{Panel d):} Resulting magnetic field topology with $\lambda_1\simeq7.62$.}}
     \label{fig:wrong}

\vspace*{\floatsep}
     \centering
     \includegraphics[width=0.7\textwidth]{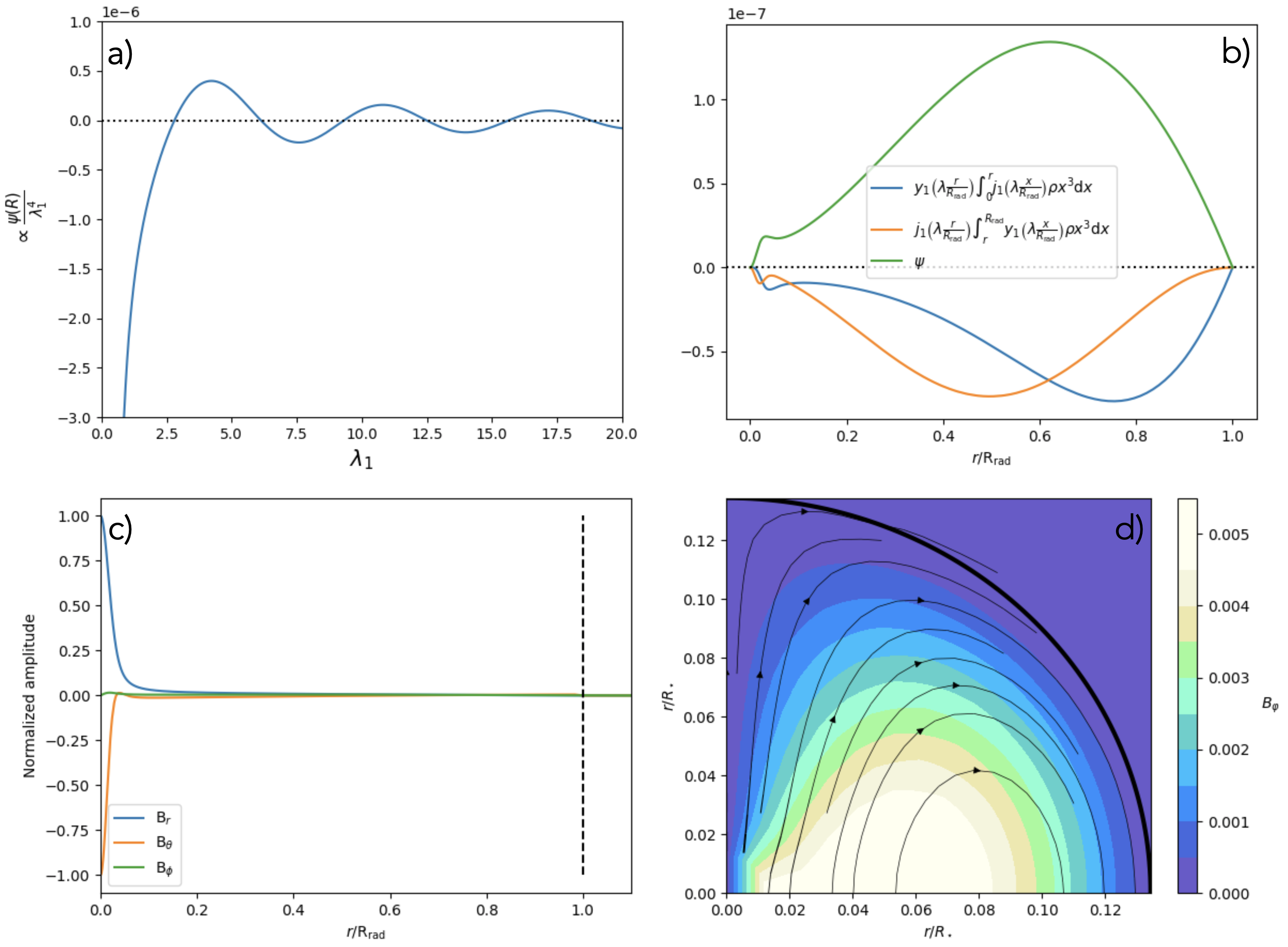}
     \caption{\bld{Same legend than fig.~\ref{fig:wrong} for the first zero of $y_1\left(\lambda \frac{r}{R_{\mathrm{rad}}} \right)$ instead ($\lambda_1\simeq 2.80$). }}
     \label{fig:good}
 \end{figure*}

\bld{We remind the stable configuration theoretical expression \citep{Duez2010} that is used in our study to represent fossil fields aligned with the rotation axis of the star:} 
\bld{\begin{equation}
    \boldsymbol{B}=\begin{cases}
    \displaystyle
    \frac{1}{r \sin{\theta}}\left( \boldsymbol{\nabla} \psi(r,\theta) \wedge \boldsymbol{e_\varphi} + \lambda \frac{\psi(r,\theta)}{R_{\mathrm{rad}}} \boldsymbol{e_\varphi}\right) \mathrm{\,if\,} r<R_{\mathrm{rad}}\, ,\\
    \displaystyle
    0 \mathrm{\,if\,} r>R_{\mathrm{rad}}
    \end{cases}
\end{equation}}

\bld{\noindent where $\psi$ is the stream function:}
\bld{\begin{equation}
    \psi(r,\theta)=\mu_0 \alpha \lambda \frac{A(r)}{R_{\mathrm{rad}}}\sin^2{\theta}\, ,
\end{equation} with $\mu_0$ the vacuum magnetic permeability, $\alpha$ a normalisation constant, $\lambda$ the eigenvalue of the problem that fixes the shape of the magnetic configuration, $R_{\mathrm{rad}}$ the radius of the radiative cavity, and 
\begin{multline}
A(r)=-r j_1\left(\lambda \frac{r}{R_{\mathrm{rad}}} \right) \int_r^{R_{\mathrm{rad}}} y_1\left(\lambda \frac{x}{R_{\mathrm{rad}}} \right)\rho x^3 \textrm{d}x\\
-r y_1\left(\lambda \frac{r}{R_{\mathrm{rad}}} \right) \int_0^r j_1\left(\lambda \frac{x}{R_{\mathrm{rad}}} \right)\rho x^3 \textrm{d}x,
\label{eq:A2}
\end{multline}
\noindent with $j_1$ (resp. $y_1$) the first-order spherical Bessel function of the first (resp. second) kind \citep{Abramowitz1972}.}

\bld{In order for the field to be confined inside the radiative interior of evolved solar-like stars, $\Psi(r)$ (and thus $A(r)$) should go to zero at the radiative/convective boundary located by $R_{\mathrm{rad}}$. There are two options to cancel $\Psi(R_{\mathrm{rad}})$:
\begin{itemize}
    \item Cancelling $\rho\int_0^r j_1\left(\lambda \frac{x}{R_{\mathrm{rad}}} \right) x^3 \textrm{d}x$ allows to confine $b_r$ and $b_\theta$ inside the radiative interior and to keep the field in its more stable configuration \citep{Woltjer1959, duez2010a}.
    \item Cancelling $y_1\left(\lambda \frac{r}{R_{\mathrm{rad}}} \right)$ allows only $b_r$ to go to zero and the radiative boundary. Non-zero $B_\theta$ generates an azimuthal current sheet that potentially creates instabilities \citep{Duez2010}. 
\end{itemize}}
 \bld{We observe on panel \textsl{a} of fig.~\ref{fig:wrong} that the $\Psi$ function fails to reach zero at the edge of the convective interior for any value of $\lambda_1$ when trying to annulate $\rho\int_0^r j_1\left(\lambda \frac{x}{R_{\mathrm{rad}}} \right) x^3 \textrm{d}x$. The first minimum of the function $\Psi(\lambda_1)$ is found at $\lambda_1\simeq7.62$, and leads to the $\Psi(r)$ function represented on panel \textsl{b} and to the magnetic field topology represented on panels \textsl{c} and \textsl{d} of fig.~\ref{fig:wrong}, for which the field is not trapped inside the radiative interior. Such magnetic-field configuration is not the most stable one according to the studies of \citep{Braithwaite2008a} and \cite{Duez2010}.}\\
 
\bld{By performing instead the analysis with a constant density profile, (i.e. searching for zeros of the function $\rho\int_0^r j_1\left(\lambda \frac{x}{R_{\mathrm{rad}}} \right) x^3 \textrm{d}x$) allows the search for $\lambda_1$ to converge. We thus conclude that the steep density profile inside red giants prevents the $\int_0^r j_1\left(\lambda \frac{x}{R_{\mathrm{rad}}} \right) x^3 \rho\textrm{d}x$ integral to converge to zero easily.}\\

\bld{The only remaining option to find the eigenvalue $\lambda_1$ that cancels $\Psi$ at the radiative boundary is thus to search for the zeros of $y_1\left(\lambda \frac{r}{R_{\mathrm{rad}}} \right)$ instead of those of the integral in Eq.~\ref{eq:A}. Results are represented on fig.~\ref{fig:good}, with $\lambda_1\simeq 2.80$ the eigenvalue used in this study. This method allows $b_r$ to cancel at the radiative boundary, but does not constrain the horizontal component of the field. As shown by panel \textsl{c} of Fig.~\ref{fig:good}, angular components are still very close to zero at the radiative/convective boundary. As a result, this magnetic field topology is stable \citep{Duez2010} and can be considered as trapped inside the radiative interior along the evolution.}
 


\section{Non-zero-average terms of the Lorentz work}
\label{sec:terms}

In this section, we present all the non-zero terms composing the work of $\boldsymbol{\delta F_L}\cdot\boldsymbol{\xi}^*/\rho$. They either involves only poloidal components of the magnetic field, or toroidal components only. The prime symbol ($'$) indicates a total derivative, either radial or latitudinal depending on the considered variable.

\subsection{Poloidal terms}

\bld{By using $A = [(rb_\theta)'+b_{\rm r}]$, the terms {of the work of $\boldsymbol{\delta F_{L, j+t}}\cdot\boldsymbol{\xi}^*/\rho$} involving poloidal components write:}
\begin{align*}
    &-m^2\frac{\xi_{\rm h}b_\theta A\xi_{\rm r}^*}{r^2}\left(Y_l^{m}\right)^2
    +\frac{(r\xi_{\rm r}b_\theta)'A\xi_{\rm r}^*}{r^2}\left(Y_l^{m}\right)^2\sin^2\theta\\
&    -\frac{(r\xi_{\rm h}b_{\rm r})'A\xi_{\rm r}^*}{r^2}Y_l^m \partial_\theta Y_l^m\sin\theta\cos\theta
    +\frac{\xi_{\rm r}b_\theta A\xi_{\rm h}^*}{r^2}\partial_\theta Y_l^m (Y_l^m\sin^2\theta)'\\
   & -\frac{Ab_{\rm r}|\xi_{\rm h}|^2}{r^2}\partial_\theta Y_l^m(\partial_\theta Y_l^m\sin\theta\cos\theta)'
    +m^2\frac{Ab_{\rm r}|\xi_{\rm h}|^2}{r^2\sin{\theta}}\partial_\theta Y_l^m Y_l^m\cos\theta\\
    &-m^2\frac{(\xi_{\rm h}b_\theta)'b_\theta\xi_{\rm r}^*}{r} \left(Y_l^{m}\right)^2 
    +\frac{(r\xi_{\rm r}b_\theta)''b_\theta\xi_{\rm r}^*}{r}\left(Y_l^{m}\right)^2\sin^2\theta\\
    &-\frac{(r\xi_{\rm h}b_{\rm r})''b_\theta\xi_{\rm r}^*}{r}Y_l^m \partial_\theta Y_l^m\sin\theta\cos\theta\\
    &+\frac{b_\theta^2|\xi_{\rm r}|^2}{r^2}Y_l^m\sin\theta\left[\frac{(Y_l^m\sin^2\theta)'}{\sin\theta}\right]'\\
    &-\frac{\xi_{\rm h}b_{\rm r}b_\theta\xi_{\rm r}^*}{r^2}Y_l^m\sin\theta\left[\frac{(\partial_\theta Y_l^m\sin\theta\cos\theta)'}{\sin\theta}\right]'\\
    &+m^2\frac{\xi_{\rm h}b_{\rm r}b_\theta\xi_{\rm r}^*}{r^2}Y_l^m\sin\theta\left(Y_l^m\frac{\cos\theta}{\sin^2\theta}\right)'
    +m^2\frac{(\xi_{\rm h}b_\theta)'b_{\rm r}\xi_{\rm h}^*}{r\sin\theta}\partial_\theta Y_l^m Y_l^m\cos\theta\\
    &-\frac{(r\xi_{\rm r}b_\theta)''b_{\rm r}\xi_{\rm h}^*}{r}Y_l^m \partial_\theta Y_l^m\sin\theta\cos\theta
    +\frac{(r\xi_{\rm h}b_{\rm r})''b_{\rm r}\xi_{\rm h}^*}{r}\left(\partial_\theta Y_l^m\right)^2\cos^2\theta\\
    &-\frac{\xi_{\rm r}b_{\rm r}b_\theta\xi_{\rm h}^*}{r^2}\partial_\theta Y_l^m\cos\theta\left[\frac{(Y_l^m\sin^2\theta)'}{\sin\theta}\right]'\\
    &+\frac{b_{\rm r}^2|\xi_{\rm h}|^2}{r^2}\partial_\theta Y_l^m\cos\theta\left[\frac{(\partial_\theta Y_l^m\sin\theta\cos\theta)'}{\sin\theta}\right]'\\
    &-m^2\frac{b_{\rm r}^2|\xi_{\rm h}|^2}{r^2}\partial_\theta Y_l^m\cos\theta\left(Y_l^m\frac{\cos\theta}{\sin^2\theta}\right)'\\
    &+m^2\frac{(r\xi_{\rm h}b_{\rm r})'b_\theta\xi_{\rm h}^*}{r^2\sin\theta}Y_l^m(Y_l^m\cos\theta)'\\
    &+m^2\frac{b_\theta^2|\xi_{\rm h}|^2}{r^2\sin^2\theta}Y_l^m[\sin\theta(Y_l^m)']'
    -m^4\frac{b_\theta^2|\xi_{\rm h}|^2}{r^2\sin^2\theta}\left(Y_l^{m}\right)^2\\
    &+m^2\frac{(r\xi_{\rm r}b_\theta)'b_\theta\xi_{\rm h}^*}{r^2}\left(Y_l^{m}\right)^2
    -m^2\frac{(r\xi_{\rm h}b_{\rm r})'b_\theta\xi_{\rm h}^*}{r^2\sin\theta}\partial_\theta Y_l^m Y_l^m\cos\theta\\
    &-m^2\frac{\xi_{\rm r}b_\theta b_{\rm r}\xi_{\rm h}^*}{r^2}Y_l^m\frac{\cos\theta}{\sin^3\theta}(Y_l^m\sin^2\theta)'\\
    &+m^2\frac{b_{\rm r}^2|\xi_{\rm h}|^2}{r^2}Y_l^m\frac{\cos\theta}{\sin^3\theta}(\partial_\theta Y_l^m \sin\theta\cos\theta)'\\
    &-m^4\frac{b_{\rm r}^2|\xi_{\rm h}|^2}{r^2}\left(Y_l^{m}\right)^2\frac{\cos^2\theta}{\sin^4\theta}
    +m^2\frac{(r\xi_{\rm h}b_{\rm r})''b_{\rm r}\xi_{\rm h}^*}{r\sin^2\theta}\left(Y_l^{m}\right)^2\cos^2\theta\\
    &+m^2\frac{(\xi_{\rm h}b_\theta)'b_{\rm r}\xi_{\rm h}^*}{r\sin\theta}Y_l^m\cos\theta(Y_l^m)'
\end{align*}

{\bld{\noindent and the work of $\boldsymbol{\delta F_{L, c}}\cdot\boldsymbol{\xi}^*/\rho$ is composed of:}
\begin{align*}
&-\frac{A b_\theta}{r}\frac{\left(\rho r^2 \xi_{\rm r} \right)'}{\rho}\xi_{\rm r}^*\left(Y_l^{m}\right)^2\sin^3\theta
-b_\theta A \xi_{\rm r}^*\xi_{\rm h}\partial_\theta\left(\partial_\theta Y_l^m \sin\theta\right)\sin^2\theta Y_l^{m}\\
&+m^2 A b_\theta \xi_{\rm r}^*\xi_{\rm h} \left(Y_l^{m}\right)^2\sin\theta
+\frac{A b_{\rm r}}{r} \frac{\left(\rho r^2 \xi_{\rm r} \right)'}{\rho} \xi_{\rm h}^* Y_l^m \partial_\theta Y_l^m \sin^2\theta \cos\theta\\
&-m^2 A b_{\rm r} \xi_{\rm h}^2 Y_l^m \partial_\theta Y_l^m \cos\theta.
\end{align*}}

\subsection{Toroidal terms}
\label{sec:tor}

\bld{The terms that involve the toroidal component of the field are for the work of $\boldsymbol{\delta F_{L, j+t}}\cdot\boldsymbol{\xi}^*/\rho$:}
\begin{align*}
    &2\frac{(r\xi_{\rm r}b_\varphi)'b_\varphi\xi_{\rm h}^*}{r^2}Y_l^m \partial_\theta Y_l^m\sin\theta\cos\theta\\
    &+2\frac{b_\varphi^2|\xi_{\rm h}|^2}{r^2}\partial_\theta Y_l^m\cos\theta(\partial_\theta Y_l^m\sin\theta)'\\
    &+2m^2\frac{b_\varphi^2|\xi_{\rm h}|^2}{r^2\sin\theta}\partial_\theta Y_l^m Y_l^m\cos\theta
    +\frac{(r\xi_{\rm r}b_\varphi)''b_\varphi\xi_{\rm r}^*}{r}\left(Y_l^{m}\right)^2\sin^2\theta\\
    &+\frac{(\xi_{\rm h}b_\varphi)'b_\varphi\xi_{\rm r}^*}{r}Y_l^m\sin\theta(\partial_\theta Y_l^m\sin\theta)'\\
    &+\frac{(r\xi_{\rm r}b_\varphi)'b_\varphi\xi_{\rm h}^*}{r^2}\partial_\theta Y_l^m(Y_l^m\sin^2\theta)'\\
    &+\frac{b_\varphi^2|\xi_{\rm h}|^2}{r^2}\partial_\theta Y_l^m[\sin\theta(\partial_\theta Y_l^m\sin\theta)']'
    -m^2\frac{b_\varphi^2|\xi_{\rm h}|^2}{r^2}\left(\partial_\theta Y_l^m\right)^2\\
    &+\frac{(r\xi_{\rm r}b_\varphi)'(rb_\varphi)'\xi_{\rm r}^*}{r^2}\left(Y_l^{m}\right)^2\sin^2\theta\\
    &+\frac{\xi_{\rm h}b_\varphi(rb_\varphi)'\xi_{\rm r}^*}{r^2}Y_l^m\sin\theta(\partial_\theta Y_l^m\sin\theta)'\\
    &+m^2\frac{\xi_{\rm r}b_\varphi(rb_\varphi)'\xi_{\rm h}^*}{r^2}\left(Y_l^{m}\right)^2
    -m^2\frac{b_\varphi^2|\xi_{\rm r}|^2}{r^2}\left(Y_l^{m}\right)^2.
\end{align*}\noindent \bld{and:}
\bld{{\begin{align*}
&-\frac{(r b_\varphi)'b_\varphi}{r}\frac{\left(\rho r^2 \xi_{\rm r} \right)'}{\rho}\xi_{\rm r}^*\left(Y_l^{m}\right)^2\sin^3\theta\\
& -\left(r b_\varphi\right)'b_\varphi \xi_{\rm r}^*\xi_{\rm h}\partial_\theta\left(\partial_\theta Y_l^m \sin\theta\right)\sin^2\theta Y_l^{m}\\
&+m^2 (r b_\varphi)'b_\varphi \xi_{\rm r}^*\xi_{\rm h} \left(Y_l^{m}\right)^2\sin\theta\\
&-\frac{2b_\varphi^2}{r}\frac{\left(\rho r^2 \xi_{\rm r}\right)'}{\rho} \xi_{\rm h}^* Y_l^m \partial_\theta Y_l^m \sin^2\theta \cos\theta\\
&+2m^2b_\varphi^2\xi_{\rm h}^2 Y_l^m \partial_\theta Y_l^m \cos\theta
\end{align*}}
\noindent for the work of $\boldsymbol{\delta F_{L, c}}\cdot\boldsymbol{\xi}^*/\rho$}.

\section{Stretched spectrum spacing $\Delta\tau_m$ in the presence of magnetism}
\label{sec:delta_tau_with_magnetism_appendix}
In order to investigate the internal rotation rate of evolved solar-like stars possessing mixed modes, \cite{Mosser2015a} concentrates on the rotational splitting on $g-m$ modes. \bld{We assume that solar-like stars are slow rotators, and thus neglect the centrifugal acceleration and other second order rotational effects which scales as $\Omega^2$.} From the study of \cite{Goupil2013a} the global rotational splitting of a mixed mode is written as 

\begin{equation}
    \delta\nu_{\mathrm{rot},m}= \delta\nu_{\mathrm{rot,g}}\zeta + \delta\nu_{\mathrm{rot,p}}(1-\zeta).
\end{equation}

For $g-m$ modes the $\zeta$ function is very close to 1, so $\delta\nu_{\mathrm{rot},m}\simeq \delta\nu_{\mathrm{rot,g}}\zeta \simeq \delta\nu_{\mathrm{rot,core}}\zeta$, providing a measure of the core rotation rate of the star. Therefore, the unperturbed frequency of $g-m$ modes can be written as $\nu_g=\nu-\delta\nu_{\mathrm{rot}, g}\zeta$ with $\nu$ the measured frequency in the presence of rotation.

For a rotating and magnetised star, we rewrite the unperturbed frequency as a function of the observed frequency through:

\begin{equation}
    \nu_g=\nu-\left(\delta\nu_{\mathrm{rot}, g}+ \delta\nu_{\mathrm{mag},g}\right)\zeta,
\end{equation}
with $\delta\nu_{\mathrm{mag},g}$ the frequency perturbation due to magnetism only. The validity of this expression comes from the asymptotic study contained in the paper (Mathis et al., \textsl{submitted}). As a consequence, the period of mixed modes {of azimuthal component $m$} varies as

\begin{equation}
    \frac{\mathrm{d} P_m}{\mathrm{d} \nu}=-\frac{1}{\nu_g^2},
\end{equation}
\noindent leading to
\begin{equation}
    {\mathrm{d}P_m}=-\frac{\mathrm{d}\nu}{\left(\nu- \left(\delta\nu_{\mathrm{rot, g},m}+\delta\nu_{\mathrm{mag, g},m}\right)\zeta\right)^2}.
\end{equation}

\noindent With the use of a limiting case where $\left(\delta\nu+\delta\nu_{\mathrm{mag, core},m}\right)/\nu \rightarrow0$, one obtains:

\begin{equation}
    {\mathrm{d}P_m}=-\frac{\mathrm{d}\nu}{\nu^2}
    \left(1+2\zeta\frac{\delta\nu_{\mathrm{rot, g},m}+\delta\nu_{\mathrm{mag, g},m}}{\nu}\right).
    \label{eq:dPm}
\end{equation}

\bld{The period spacing between two consecutive mixed modes is thus written as:}

\begin{equation}
    {\Delta P_m}=\Delta P
    \left(1+2\zeta\frac{\delta\nu_{\mathrm{rot, g},m}+\delta\nu_{\mathrm{mag, g},m}}{\nu}\right),
    \label{eq:DPm}
\end{equation}
\noindent \bld{with $\Delta P$ the period spacing between two consecutive $m=0$ axisymmetric mixed modes. Noting that $\Delta \tau_m=\Delta P_m/\zeta$ and by using the fact  $\Delta P$ can be written as $\Delta P=\zeta  \Delta\Pi_1$, Eq.~\eqref{eq:DPm} becomes:}

\begin{equation}
    {\Delta \tau_m}=\Delta\Pi_1
    \left(1+2\zeta\frac{\delta\nu_{\mathrm{rot, g},m}+\delta\nu_{\mathrm{mag, g},m}}{\nu}\right).
    \label{eq:Dtm}
\end{equation}

\section{MESA inlist}
\label{sec:MESA_inlist}
In this appendix we report the MESA inlist used to calculate the stellar evolution
models of the $1.5\mathrm{M}_\odot$, $Z=0.02$ star:

\begin{Verbatim}
&star_job

  ! begin with a pre-main sequence model
  create_pre_main_sequence_model = .true.

/ !end of star_job namelist


&controls

  ! starting specifications
    initial_mass  = 1.5 ! in Msun units
    initial_z     = 0.02
    use_Type2_opacities = .true.
    Zbase        = 0.02
    max_model_number    = 900

  !———————————— MISC
  profile_interval = 10
  history_interval = 10
  max_num_profile_models = 1000

  calculate_Brunt_N2 = .true.
  star_history_name = 'history.data'
  profile_data_prefix = 'profile'
  profiles_index_name = 'profiles.index'

  set_min_D_mix = .true.
  min_D_mix = 1d1


  !———————————— Output pulse files for GYRE
  pulse_data_format = 'GYRE'
  write_pulse_data_with_profile = .true.
  add_center_point_to_pulse_data = . true.
  add_double_points_to_pulse_data = . true.


  !———————————— WIND
  cool_wind_RGB_scheme = 'Reimers'
  cool_wind_AGB_scheme = 'Blocker'
  RGB_to_AGB_wind_switch = 1d-4
  Reimers_scaling_factor = 0.2
  Blocker_scaling_factor = 0.5
  use_accreted_material_j = .true.
  accreted_material_j = 0

  !————————————— OVERSHOOTING
  overshoot_scheme(1) = 'exponential'
  overshoot_zone_type(1) = 'any'
  overshoot_zone_loc(1) = 'any'
  overshoot_bdy_loc(1) = 'any'
  overshoot_f(1) = 0.015
  overshoot_f0(1) = 0.004

  !————————————— MESH
  mesh_delta_coeff = 0.7
  varcontrol_target = 0.7d-3
  predictive_mix(1) = .true.
  predictive_superad_thresh(1) = 0.005
  predictive_avoid_reversal(1) = 'he4'
  predictive_zone_type(1) = 'any'
  predictive_zone_loc(1) = 'core'
  predictive_bdy_loc(1) = 'top'
  dX_div_X_limit_min_X = 1d-4
  dX_div_X_limit = 5d-1
  dX_nuc_drop_min_X_limit = 1d-4
  dX_nuc_drop_limit = 1d-2

/ ! end of controls namelist
\end{Verbatim}

\end{appendix}
\end{document}